\documentclass[a4paper,11pt]{article}
\usepackage{jheppub}
\usepackage[T1]{fontenc}
\usepackage[utf8]{inputenc}
\usepackage{amsmath,amssymb,amsfonts}
\usepackage[titletoc]{appendix}
\usepackage{braket}
\usepackage{graphicx}
\usepackage{stmaryrd}
\usepackage[dvipsnames,svgnames,x11names]{xcolor}
\usepackage{hyperref}
\usepackage{bm}
\usepackage{float}
\usepackage{comment}
\usepackage{enumerate}
\usepackage{romannum}
\usepackage{subfigure}
\usepackage{longtable}
\usepackage{mathtools}
\usepackage[normalem]{ulem}
\DeclareMathAlphabet{\mathbbold}{U}{bbold}{m}{n}

\DeclareMathOperator{\tr}{tr}
\DeclareMathOperator{\Tr}{Tr}

\newcommand{\Rmnum}[1]{\expandafter\@slowromancap\romannumeral #1@}

\newcommand{\be}{\begin{equation}}
\newcommand{\ee}{\end{equation}}

\newcommand{\lb}{\left (}
\newcommand{\rb}{\right )}

\makeatletter
\newcommand{\fixed@sra}{$\vrule height 2\fontdimen22\textfont2 width 0pt\shortrightarrow$}
\newcommand{\shortarrow}[1]{%
\mathrel{\text{\rotatebox[origin=c]{\numexpr#1*45}{\fixed@sra}}}
}
\makeatother

\DeclareRobustCommand{\rchi}{{\mathpalette\irchi\relax}}
\newcommand{\irchi}[2]{\raisebox{\depth}{$#1\chi$}}

\begin{document}

\title{\centering {\huge To gauge or to double gauge?}
\\ {\Large Matrix models, global symmetry, and black hole cohomologies} 
}

\author[a]{Adwait Gaikwad,}
\author[a,c]{Tanay Kibe,}
\author[a,b,c]{Sam van Leuven}
\author[a]{and Kayleigh Mathieson}

\affiliation[a]{Mandelstam Institute for Theoretical Physics, School of Physics, University of the Witwatersrand, Johannesburg 2050, South Africa
}
\affiliation[b]{DSI-NRF Centre of Excellence in Mathematical and Statistical Sciences (CoE-MaSS),  South Africa}
\affiliation[c]{National Institute for Theoretical and Computational Sciences (NITheCS), Gauteng, South Africa}

\emailAdd{adwaitgaikwad@gmail.com}
\emailAdd{tanay.kibe@wits.ac.za}
\emailAdd{sam.vanleuven@wits.ac.za}
\emailAdd{kayleigh.mathieson.wits@gmail.com}

\date{\today}

\abstract{We study the structure of the Hilbert space of gauged matrix models with a global symmetry.
In the first part of the paper, we focus on bosonic matrix models with $U(2)$ gauge group and $SO(d)$ global symmetry, and consider singlets under both the gauge and global symmetry. 
We show how such ``double--gauged'' matrix models can be described in terms of a simpler $SO(3)$ single--matrix model.
In the second part of the paper, we consider the so-called BMN subsector of the  $\mathcal{N}=4$ $SU(N)$ super Yang--Mills theory, which is closely related to the BMN matrix model. Among the 1/16 BPS operators in this sector, ``non-graviton'' operators were recently discovered, which are expected to relate to the microstates of supersymmetric $AdS_5$ black holes. We show that a double gauging of this model, where one projects onto $SU(3)_R$ $R$--symmetry singlets, considerably simplifies the analysis of the non--graviton spectrum. In particular, for low values of $N$, we show that (almost) all graviton operators project out of the spectrum, while important classes of non-graviton operators remain. In the $N=3$ case, we obtain a closed form expression for the superconformal index of singlet non-gravitons, which reveals structural features of their spectrum.}

\maketitle 

\newpage

\section{Introduction}

Matrix models are among the simplest quantum mechanical models that have been proposed to describe black holes and, more generally, to provide a non-perturbative definition of theories of quantum gravity.
Probably the most famous example is the Banks-Fischler-Shenker-Susskind (BFSS) matrix model, which in the large $N$ limit is conjectured to describe M--theory in flat space \cite{Banks:1996vh}.\footnote{Further early developments, including matrix models for string theories, can be found in \cite{Banks:1996vh,Dijkgraaf:1997vv,Seiberg:1997ad,Itzhaki:1998dd,Aoki:1998vn}, see also \cite{Polchinski:1999br,Taylor:2001vb} and more recently \cite{Maldacena:2023acv,Lin:2025iir} for reviews.}

The BFSS model can be viewed as the worldline quantum mechanics of $D0$ branes in a decoupling limit of type IIA string theory \cite{Banks:1996vh,Itzhaki:1998dd}.
In this limit, the effective dynamics can be described as the dimensional reduction of $10d$ $\mathcal{N}=1$ (maximally) supersymmetric Yang-Mills theory to $0+1$ dimensions.
Consequently, the degrees of freedom comprise nine bosonic $U(N)$ matrices and their superpartners.
A well--studied generalization of the BFSS model is the Berenstein-Maldacena-Nastase (BMN) matrix model \cite{Berenstein:2002jq}, which can be viewed as a mass deformation of the former. 

For sufficiently large $N$ and strong coupling, matrix models have a large number of degrees of freedom with rich and complicated dynamics.
Over the years, a variety of methods, both analytical and numerical, have been developed to study these models, see \cite{Lin:2025iir} for a recent overview and a comprehensive list of references.
Instead of tackling the BFSS or BMN models head--on, one can also consider toy versions of these models using various types of simplifications.
For example, one can study models with a smaller number of matrices $d<9$, small values of $N$, purely bosonic or classical dynamics, singlet sectors of a global symmetry (\textit{e.g.}, states with vanishing angular momentum), or any combinations thereof \cite{Sethi:1997pa,Hubener:2014pfa,Aoki:2015uha,Gur-Ari:2015rcq,Berenstein:2016zgj,Anous:2017mwr,Buividovich:2018scl,Asano:2020yry,Kovacik:2020cod,Bergner:2021goh}.

In this paper, we use a blend of these simplifications, but the main role will be played by the global symmetry singlet projection.
Since the singlet sector is the only sector which survives upon gauging a symmetry, we will refer to such models as \emph{double--gauged matrix models}. It should be contrasted with the \emph{To gauge or not to gauge?} paper \cite{Maldacena:2018vsr}.
In this work, the authors compare the gauged and ungauged BFSS model at strong coupling and observe that gauge non-singlets decouple, effectively demonstrating the equivalence of the two models in the gravitational regime (see also \cite{Berkowitz:2018qhn} for further justifications). 
Clearly, in the ungauged model one does not have to worry about gauge invariance, which makes it easier to study numerically and avoids the complications of trace relations in determining the structure of the Hilbert space.
As we will demonstrate in this paper, at least for the matrix models we study, the structure of the Hilbert space and dynamics also simplify in the double--gauged theory.

We focus on two sets of models. 
The first involves bosonic matrix models with $U(2)$ gauge symmetry and a global $SO(d)$ symmetry. 
The second set comprises supersymmetric matrix models, in particular, the so-called BMN sector of the $\mathcal{N} = 4$ super Yang--Mills (SYM) theory with $SU(N)$ gauge group.
In this model, we focus on a 1/16 BPS subsector which has a global $SU(3)_{R}$ $R$-symmetry.
In the remainder of the introduction, we motivate the study of these specific models in turn.

\subsection*{Bosonic matrix models}\label{ssec:interacting-ham}

Consider a model with $d$ Hermitian matrices $X_{i}$, $i=1,2,\hdots d$, transforming in the adjoint of the $U(2)$ gauge group, with the Lagrangian
\begin{equation}\label{Eq:BFSSbosonicLag}
    L =\frac{1}{2g^2}\tr \lb \sum_{i=1}^d (D_t X_i)^2-\frac12 \sum_{i\neq j}[X_i,X_j]^2\rb.
\end{equation}
Here, the gauge covariant derivative is $D_t X_i = \partial_t X_i -i[A_t,X_i]$, with $A_t$ being the $U(2)$ gauge field. When $d=9$, this model corresponds to the bosonic part of the BFSS matrix model. 
The diagonal components of the matrices can be understood as the positions of two $D0$ branes, and the off-diagonals correspond to strings stretching between the branes. 
The Lagrangian splits into a free $U(1)$ part, corresponding to the trace components of $X_i$, which describes the centre--of--mass motion, and an $SU(2)$ part, which contains the degrees of freedom associated with the relative motion.
Correspondingly, the Hilbert space of the theory decomposes as a tensor product $\mathcal{H}_{U(1)}\otimes\mathcal{H}_{SU(2)}$.

As mentioned above, we will be interested in the singlet sector of this model under its global $SO(d)$ symmetry, corresponding to the zero angular momentum sector in target space.
The (chaotic) dynamics of this model has been studied before, see, \textit{e.g.}, \cite{Hubener:2014pfa,Asano:2015eha,Gur-Ari:2015rcq,Berenstein:2016zgj,Buividovich:2022jgv,Hoppe:2023qem}. 
In particular, it was observed that (the $SU(2)$ part of) the Hamiltonian simplifies significantly upon gauge fixing both $SU(2)$ \emph{and} $SO(d)$ \cite{Sethi:1997pa,Hoppe:2023qem}
\begin{equation}\label{Eq:d=3ham}
    H = \frac12\lb p_x^2+p_y^2+p_z^2\rb+ \frac12 \lb x^2y^2+y^2z^2+z^2x^2\rb\,.
\end{equation}
Perhaps surprisingly, the model only involves three bosonic degrees of freedom. 
If one studies the $U(2)$ model instead, one can impose zero angular momentum either on the relative and centre-of-mass motions separately, or only on the total combined motion.
In the latter case, the $SO(d)$ singlet projection does not respect the $\mathcal{H}_{U(1)}\otimes\mathcal{H}_{SU(2)}$ tensor product structure of the Hilbert space (see Sec.~\ref{Sec:Bmolinweyl}).
In the remainder of this work, we will mostly consider the latter case and therefore treat the $U(2)$ and $SU(2)$ models separately. 

In this work, we show in detail how the structure of the gauge-invariant Hilbert space simplifies in the $SO(d)$ singlet sector, which is related to the simplification of the dynamics. 
As a tool we use the $SO(d)$ singlet partition functions of bosonic matrix models with the $U(2)$ and $SU(2)$ gauge groups, where the interactions are turned off and a mass deformation is introduced, \textit{i.e.}, matrix harmonic oscillators.
We interpret these expressions by identifying the complete set of gauge and $SO(d)$ invariant operators, accounting for trace relations. 
Our analysis reveals that the $SO(d)$ singlet sector of $SU(2)$ models with $d\geq4$ can effectively be described by an $SO(3)$ gauged matrix model of a single (real symmetric) matrix $B$. 
For $d=3$, there is an additional invariant that squares to ${\rm det}(B)$.
Similarly, the $d\geq5$  $U(2)$ bosonic models in the $SO(d)$ singlet sector are described in terms of an $SO(3)$ matrix model of a single real symmetric matrix, a vector and a scalar, which can be combined into a single matrix $\mathcal{B}$. 
For $d=4$, there is an additional invariant that squares to ${\rm det}(\mathcal{B})$.  
Although our partition functions are computed for the matrix harmonic oscillators, the analysis of the invariants relies only on \textit{kinematic} trace relations, and should therefore extend to interacting models.

\subsection*{BMN sector of the $\mathcal{N} = 4$ theory}\label{ssec:intro-bmn}

The BMN sector of the $4d$ $\mathcal{N}=4$ $SU(N)$ SYM theory refers to a truncation of the theory where only the lowest lying modes of the fields on $S^3$ are retained.
At the classical level, this truncation turns out to be equivalent to the BMN matrix model, hence its name, although this equivalence does not extend beyond one-loop corrections \cite{Kim:2003rza}.
We focus on a 1/16 BPS subsector of this model, for which the higher loop corrections conjecturally vanish \cite{Grant:2008sk}.\footnote{Recently, some doubt has been cast on this conjecture, which we will return to in due course.}
This sector has recently played an important role in the identification of potential operators, at finite $N$, which could correspond to the microstates of supersymmetric $AdS_5$ (quantum) black holes \cite{Chang:2022mjp,Choi:2022caq,Choi:2023znd,Choi:2023vdm,Chang:2024zqi,deMelloKoch:2024pcs,Gadde:2025yoa}.\footnote{Technically, recent works have identified $Q$--cohomology classes of operators, which are in one--to--one correspondence with BPS operators. See Section \ref{ssec:rev-bmn} for further details.}

This progress relies on a number of earlier results.
Perhaps the oldest result, dating back to the very start of the AdS/CFT correspondence, is the identification of a set of 1/16 BPS multiplets in the $\mathcal{N}=4$ theory which map, at large $N$ and strong Yang--Mills coupling, onto short supergravity multiplets in $AdS_5$ \cite{Witten:1998qj}.
These are the famous chiral primary multiplets, also called $S_n$ multiplets in \cite{Kinney:2005ej}, and we will loosely refer to such multiplets as graviton multiplets, or simply gravitons for short.
Various follow-ups have attempted to construct more general multiplets, which could map onto non-perturbative objects in $AdS_5$, most importantly black holes, but did not succeed \cite{Kinney:2005ej,Grant:2008sk,Chang:2013fba}.
This program recently gained new impetus from results in \cite{Cabo-Bizet:2018ehj,Choi:2018hmj,Benini:2018ywd}, which showed that the superconformal index of the $\mathcal{N}=4$ theory, which counts 1/16 BPS operators (with signs), exhibits $\mathcal{O}(e^{N^2})$ growth.
This growth cannot be accounted for by graviton multiplets \cite{Kinney:2005ej}.
As is common in the recent literature, we will refer to the responsible operators interchangeably as non-graviton, fortuitous or black hole.

A key tool underlying the identification of non-gravitons is the superconformal index \cite{Kinney:2005ej}, which, crucially, is independent of the Yang--Mills coupling.
It can therefore be exactly evaluated in the free theory, where one can explicitly enumerate the full 1/16 BPS spectrum.
The resulting formula may, however, also be interpreted in terms of the strongly coupled spectrum.
Since the graviton multiplets are known and defined also at finite $N$ in the $\mathcal{N}=4$ theory, one may also evaluate an index over this subset of (finite $N$) multi-graviton operators.
The idea is then to evaluate the difference between the full and ``graviton index''.
A discrepancy between the two provides clues as to the charge combinations at which non-gravitons arise \cite{Murthy:2020scj,Agarwal:2020zwm,Chang:2022mjp}.

An important feature of the BMN sector is that the index restricted to this sector takes on a relatively simple closed form.
This makes it easier to deduce various structural features of the BPS spectrum in this sector \cite{Choi:2023znd,Choi:2023vdm,Gadde:2025yoa}.
Even better is when the graviton index is also known in closed form. 
In this case, one can compute the difference in closed form, such that structural features of the non-graviton spectrum can also be deduced.
As it turns out, the graviton index is difficult to compute in closed form at finite $N$ (see, \textit{e.g.}, \cite{Kinney:2005ej,Chang:2013fba} for a closed form at large $N$).
This difficulty is due to trace relations, which complicate the explicit construction of the space of states.
As far as we are aware, a closed form at finite $N$ exists only for the BMN subsector when the gauge group is $SU(2)$ \cite{Choi:2023znd} (see also \cite{Grant:2008sk} for closed forms in other sectors).

In the second part of our paper, we show that a double--gauging of the BMN sector simplifies this problem.
In particular, the 1/16 BPS sector transforms under an $SU(3)_R$ $R$-symmetry, which is a subgroup of the full $SU(4)_R$ $R$-symmetry.
Somewhat surprisingly, we find that important classes of known non-graviton operators are $SU(3)_R$ singlets, whereas most graviton operators, at least for $N\leq 4$, are non--singlets.
For example, in the $SU(2)$ case, we find that the singlet graviton index is trivial,
which suggests that the graviton operators are eliminated entirely in the $SU(3)_R$ singlet sector. 
We prove this through explicit construction, demonstrating that potential singlet gravitons always vanish. 

In the singlet sector of the theory with $SU(3)$ gauge group, we also find a closed form for the graviton index.
In this case, we find only two $SU(3)_R$ singlet gravitons. 
This provides a closed-form for the $SU(3)_R$ singlet \emph{non-graviton} index, which allows us to study structural features of the non-graviton spectrum for $SU(3)$ gauge group.
Finally, we comment on extensions to larger values of $N$, making a start with $SU(4)$ gauge group.

\subsection*{Organization}

The rest of this paper is organized as follows. In Sec.~\ref{Sec:molienweyl}, we introduce the Molien-Weyl formula, which we use to calculate the (double--gauged) $SO(d)$ singlet partition functions for bosonic matrix models with the $U(2)$ and $SU(2)$ gauge groups. These partition functions are then interpreted in Sec.~\ref{sec:struct-loop-singlet}, where we identify a complete set of operators in the singlet sector by accounting for trace relations. 
This is followed by the second part of the paper in Sec.~\ref{sec:non-grav}, where we generalize our analysis to the BMN sector of the $\mathcal{N}=4$ theory. In particular, we study the global $SU(3)_R$ singlet projection for $SU(N)$ gauge group with $N=2,3,4$.  We conclude in Sec.~\ref{Sec:conclusions} with a summary and future directions. 

In Appendix~\ref{app:A} we provide details on the Molien-Weyl projection formulas and a compendium of partition functions for bosonic matrix models. We analyse the partition functions and the invariant theory for bosonic models in non-singlet sectors in Appendix~\ref{Sec:loopspaceJneq0}. In Appendix~\ref{App:gravsinglets}, we provide details for our procedure to enumerate $SU(3)_R$ singlet gravitons in the BMN sector of the $\mathcal{N}=4$ theory with $SU(3)$ gauge group. Finally, in Appendix~\ref{App:SU4P300} we record the expression for the superconformal index for $SU(3)_R$ singlets in the BMN sector of the $SU(4)$ $\mathcal{N}=4$ theory.

\section{Partition functions in bosonic matrix models}\label{Sec:molienweyl}

In this first part of the paper, we study the Hilbert space of bosonic multi-matrix models composed of $d$ matrices that transform in the adjoint of the $U(2)$ or $SU(2)$ gauge group.\footnote{We distinguish between $U(2)$ and $SU(2)$ for reasons explained in Section \ref{ssec:interacting-ham}.} Our focus is on models with a global $SO(d)$ symmetry. This provides a decomposition of the Hilbert space into a direct sum over the irreducible representations (irreps) of $SO(d)$. 
Its structure is determined by the ring of $U(2)$ or $SU(2)$ invariants, which are generated by (multi-)traces of the $d$ matrices, that transform in a given fixed irrep of the global $SO(d)$ symmetry. As described in the introduction, our main focus will be on the $SO(d)$ singlet sector of the Hilbert space. However, in Appendix \ref{Sec:loopspaceJneq0}, we also consider more general representations. 

One can use (free) multi-matrix harmonic oscillator partition functions as a tool to identify a generating set of invariants \cite{deMelloKoch:2025ngs}. The completeness of this set can be proven using trace relations, which are purely kinematic relations that are not modified by interactions. Hence, using the partition function of the free matrix model as a crutch, we can develop an understanding of the gauge-invariant Hilbert space of the interacting theory. 

The multi-matrix harmonic oscillator is a model of $d$ Hermitian matrices $X_{1,\ldots,d}$, transforming in the adjoint representation of $U(2)$ or $SU(2)$, with Hamiltonian
$$H=\frac12 \sum_{i=1}^d\tr \lb {P}_i^2+ X_i^2\rb,$$
where $P_i$ is the momentum matrix that is canonically conjugate to $X_i$, and the trace ensures gauge invariance. The spectrum can be determined by defining the set of creation and annihilation operators as
\begin{equation}
    a_i^\dagger \coloneqq X_i -i P_i.
\end{equation}
By acting with traces of the creation operators on the vacuum $\ket{0}$, one can construct the gauge invariant Hilbert space with energy eigenstates 
\begin{equation}
    \tr(a^\dagger_1 a^\dagger_2 \hdots a_n^\dagger) \ket{0}, \quad \tr(a^\dagger_1 a^\dagger_2) \tr(a_3^\dagger \hdots a_n^\dagger) \ket{0}, \quad \hdots.
\end{equation}
Therefore, the Hamiltonian simply counts the number of creation operators in a given (multi-)trace structure (invariant). 
However, to avoid constructing an overcomplete basis for the Hilbert space, one needs to carefully account for trace relations \cite{PROCESI1976306}. 
In general, this is a complicated problem (see \cite{deMelloKoch:2025ngs,deMelloKoch:2025qeq,deMelloKoch:2025rkw,deMelloKoch:2025eqt} for analyses of the problem).
At the level of the partition function, though, there is a simple method to avoid the overcounting without the need to explicitly consider trace relations.
The basic idea is to first construct an extended Hilbert space, containing all matrix degrees of freedom, and evaluate the partition function over this Hilbert space.
One then projects onto gauge singlets by integrating the extended partition function over the gauge group with the Haar measure, as we review in Appendix \ref{app:A}.
Schematically, this procedure can be expressed as:
\begin{equation}\label{eq:singlet-projection}
    Z(\beta)=\text{Tr}_{\mathcal{H}}\,e^{-\beta H}= \int_G d\mu_G\, \text{Tr}_{\mathcal{H}_{\text{ext}}}\,e^{-\beta H}\,,
\end{equation}
with $d\mu_G$ the Haar measure on the gauge group $G$.
This was referred to as the Molien-Weyl formula in \cite{Gray:2008yu} and we will stick to that nomenclature in this paper as well.

\subsection{Molien-Weyl formulas}\label{Ssec:Molien-Weylformulas}

For a free multi-matrix model with $d$ (bosonic) matrices in the adjoint representation of $U(2)$, the Molien-Weyl formula leads to the following partition function
\begin{equation}\label{eq:mol-weyl-U(2)-d}
    Z^{U(2)}(x_i)=\frac{1}{\prod^d_{i=1}(1-x_i)^2}\oint_{|t|=1} \frac{dt}{2\pi i t}\frac{(1-t)}{\prod^d_{i=1}(1-x_i t)(1-x_i t^{-1})},
\end{equation}
where 
\begin{equation}
    x_i = e^{-\beta E_i},\quad i=1,2,\hdots,d,
\end{equation}
with $E_i,\, i =1,2,\hdots, d$ the energies of the $d$ matrices. See Appendix \ref{app:A} for details. 

The contour integral in \eqref{eq:mol-weyl-U(2)-d} can be evaluated by summing over residues.
If we close the contour inside the unit circle and assume $|x_i|<1$, we find contributions from poles at
\begin{equation}
    t=x_i,\quad i=1,...,d\,.
\end{equation}
The residue sum can thus be expressed as \cite{deMelloKoch:2025ngs}
\begin{equation}\label{eq:2mol-weyl-U(2)-d}
    Z^{U(2)}(x_i)= \frac{1}{\prod_{i=1}^d(1-x_i)^2} \sum_{l=1}^d\left(\frac{(1-x_l)\ x_l^{d-1}}{\left[\prod_{j=1}^d(1 - x_j x_l)\right] \left[\prod_{k \neq l} (x_l-x_k)\right]}\right)\,.
\end{equation}
We can find the corresponding $SU(2)$ partition function by dividing by the $U(1)$ partition function, which is given by 
\begin{equation}\label{eq:2mol-weyl-U(1)-d}
    Z^{U(1)}(x_i)= \frac{1}{\prod_{i=1}^d(1-x_i)}\,.
\end{equation}

As mentioned before, we will be interested in sectors of the matrix model for a fixed irrep of the global $SO(d)$ symmetry and, in particular, the singlet or trivial representation. 
Sometimes, we also refer to the $SO(d)$ singlet sector as the zero angular momentum sector, or for short $\vec{J}=0$ sector, which is the target space interpretation.

The partition function in the double--gauged (singlet) sector can again be found applying the formula \eqref{eq:singlet-projection}, where now we integrate over the global $SO(d)$ group.
As reviewed in Appendix \ref{app:A}, when $d=2n$ we have
\begin{align}\label{eq:mol-weyl-so(even)}
    \begin{split}
        Z^{U(2)/SU(2),J=0}_{d=2n}(x)=\prod^n_{i=1}\oint_{|s_i|=1}\frac{ds_i}{2\pi i s_i}&\prod_{1\leq j<k\leq n}(1-s_js_k)(1-s_js_k^{-1})\\ &\times Z^{U(2)/SU(2)}(xs_1,xs_1^{-1},\ldots , xs_n,xs_n^{-1}),
    \end{split}
\end{align}
where the $s_i$ parametrize the eigenvalues of the $SO(2n)$ matrix and we again use a reduced measure to compute the singlet projection \cite{Hanany:2008sb}. Similarly, for $d=2n+1$, we have 
\begin{align}\label{eq:mol-weyl-so(odd)}
    \begin{split}
         Z^{U(2)/SU(2),J=0}_{d=2n+1}(x)=\prod^n_{i=1}\oint_{|s_i|=1}&\frac{ds_i}{2\pi i s_i}(1-s_i)\prod_{1\leq j<k\leq n}(1-s_js_k)(1-s_js_k^{-1})\\ 
         &\times Z^{U(2)/SU(2)}(xs_1,xs_1^{-1},\ldots , xs_n,xs_n^{-1},x).
    \end{split}
\end{align}
In the $J\neq0$ sectors, we will focus only on models with $d=2,3$ matrices and $SO(2)$ or $SO(3)$ global symmetry. For $SO(2)$, we use the character of the spin-$J$ representation to obtain the projection formula (see Appendix \ref{app:A} for details)
\begin{equation}\label{eq:mol-weyl-so(2)Jneq0}
    Z^{U(2)/SU(2),J}_{d=2}(x)=\oint \frac{ds}{2\pi i s} \frac{s^J+s^{-J}}{2}Z^{U(2)/SU(2)}_{d=2}(xs,xs^{-1}).
\end{equation}
Here, we note that the symmetric projection $\frac{s^{J}+s^{-J}}{2}$ is equivalent to a projection with $s^J$, due to the symmetry of the integrand under $s\to s^{-1}$.
In particular,
\begin{equation}
   Z_{d=2}^{U(2)/SU(2)}(x s,x s^{-1})=Z_{d=2}^{U(2)/SU(2)}(x s^{-1},x s).
\end{equation}
It follows that for $J\neq0$, we automatically project onto states of even parity under the reflection $X_1\leftrightarrow X_2$.
This is different from the $J=0$ projection \eqref{eq:mol-weyl-so(even)} 
\begin{equation}
    Z^{U(2)/SU(2),J=0}_{d=2}(x)=\oint \frac{ds}{2\pi i s} Z^{U(2)/SU(2)}_{d=2}(xs,xs^{-1}),
\end{equation}
which counts all $SO(2)$ invariants, irrespective of their parity under the reflection. This subtle distinction between how the $J=0$ and $J\neq0$ sectors include/exclude parity odd invariants will be important in Sec.~\ref{Sec:invd2j0}.

For $SO(3)$ fixed spin-$J$ sectors, we obtain
\begin{equation}\label{eq:mol-weyl-so(3)Jneq0}
    Z^{U(2)/SU(2),J}_{d=3}(x)=\oint \frac{ds}{2\pi i s}\lb1+\sum_{j=1}^J s^j +\sum_{j=1}^{J}s^{-j} \rb (1-s) Z^{U(2)/SU(2)}_{d=3}(xs,xs^{-1},x).
\end{equation}
These fixed $J$ projections, namely \eqref{eq:mol-weyl-so(2)Jneq0} and \eqref{eq:mol-weyl-so(3)Jneq0} can be similarly evaluated using residues.

One can bring the partition functions in \eqref{eq:2mol-weyl-U(2)-d}, \eqref{eq:mol-weyl-so(even)} and \eqref{eq:mol-weyl-so(odd)} into the so-called Hironaka form \cite{Sturmfels1993AlgorithmsII,Grinstein:2023njq}
\begin{equation}\label{eq:hironaka-gen-form}
    Z(x) = \frac{1+\sum_j c^s_j x^j}{\prod_i(1-x^i)^{c_i^p}}.
\end{equation}
This form encodes various features of the ring of invariants.
In particular, the denominator indicates the number of \emph{primary} invariants $c_i^p$, which involve $i$ matrices. And the numerator indicates the number of \emph{secondary} invariants $c_j^s$, which involve $j$ matrices. The primary invariants are algebraically independent and act freely in the ring, whereas the secondary invariants are quadratically reducible, as we will explain in more detail in Sec.~\ref{Sec:Hironaka}. Finally, we will refer to the number of matrices that appear in an invariant as its order. For example, the $SU(2)$ invariants $\tr(ABCD)$ and $\tr(A^2) \tr(CD)$ both have order 4.


\subsection{Examples}\label{Ssec:examplesmolien-weyl}

We will now evaluate the partition functions for the $SO(d)$ singlet sectors of $U(2)$ and $SU(2)$ matrix models in turn for $d=2$, $d=3$ and finally $d\geq 4$, using the Molien-Weyl formulas \eqref{eq:mol-weyl-so(even)} and \eqref{eq:mol-weyl-so(odd)}.

\subsubsection{$d=2$ matrix model}\label{Ssec:d=2matrixmodel}

We first consider a model with two Hermitian matrices transforming in the adjoint representation of $U(2)$. 
Simplifying \eqref{eq:2mol-weyl-U(2)-d} with $d = 2$, we obtain the following partition function for the free matrix model
\begin{equation}\label{eq:res-sum-U(2)-2}
    Z^{U(2)}_{d=2}(x_1,x_2) = \frac{1}{(1 - x_{1})(1 - x_{2})(1 - x_{1}^2)(1 - x_{2}^2)(1 - x_{1} x_{2})}.
\end{equation}
This form indicates the presence of five primary invariants, which can be identified as the following single trace operators \cite{deMelloKoch:2025ngs} 
\begin{equation}\label{Eq:u2invs}
    m_1=\tr(X_1), \quad m_2=\tr(X_2), \quad m_3=\tr(X_1^2), \quad m_4=\tr(X_1X_2) ,\quad m_5=\tr(X_2^2) ,
\end{equation}
where $X_{1,2}$ are the two matrices corresponding to the two fugacities $x_{1,2}$ in \eqref{eq:res-sum-U(2)-2}.
We also note that this ring of invariants does not contain any secondary invariants. Then the partition function for the $SO(2)$ singlet sector can be found by applying \eqref{eq:mol-weyl-so(even)} to \eqref{eq:res-sum-U(2)-2} as
\begin{align}\label{Eq:U2d2J0}
    \begin{split}
         Z^{U(2),J = 0}_{d=2} (x) =& \frac{1}{(1 - x^2)} \oint \frac{ds}{2 \pi i s} \frac{1}{(1 - x s^{\pm})(1 - x^2 s^{\pm 2})} \\
    =& \frac{1 + x^{4}}{(1 - x^2)^{2} (1 - x^4)^{2}}\,,
    \end{split}   
\end{align}
where we use the notation that $(1 - x s^{\pm })=(1 - x s)(1 - x s^{-1 })$.
Comparing with \eqref{eq:hironaka-gen-form}, this partition function indicates the presence of two primary invariants at both order $2$ and order $4$, and one secondary invariant at order $4$. 
We will identify these invariants in Sec.~\ref{Sec:invd2j0}.

Similarly, the $SU(2)$ partition function can easily be obtained from \eqref{eq:res-sum-U(2)-2} by excluding the generators $\tr(X_1), \tr(X_2)$, which vanish for $SU(2)$ matrices, and takes the form
\begin{equation}\label{eq:res-sum-SU(2)-2}
    Z^{SU(2)}_{d=2}(x_i)=\frac{1}{(1-x_1^2)(1-x_2^2)(1-x_1x_2)}.
\end{equation}
Projecting onto $SO(2)$ singlets, we then find
\begin{equation}\label{eq:su2so2j0}
    Z^{SU(2),J=0}_{d=2}(x)=\frac{1}{(1-x^2)}\oint \frac{ds}{2\pi i s}\frac{1}{(1-x^2s^{2})(1-x^2s^{-2})}=\frac{1}{(1-x^2)(1-x^4)},
\end{equation}
which indicates one primary invariant at order $2$, and one at order $4$.
As we will see in Sec.~\ref{Sec:invd2j0}, the invariant theory underlying this partition function can be obtained from the $U(2)$ case by setting the matrices to be traceless.

\subsubsection{$d=3$ matrix model}\label{Sec:threeormore}

The partition function for the $d=3$, $U(2)$ matrix model can be obtained from \eqref{eq:2mol-weyl-U(2)-d}.
Upon simplification, we can express the partition function in Hironaka form as follows
\begin{equation}
            Z^{U(2)}_{d=3}(x_i)=\frac{1+x_1x_2x_3}{\prod^3_{i=1}(1-x_i)(1-x_i^2)\prod_{i<j}(1-x_ix_j)}\,.
\end{equation}
Then performing the $SO(3)$ singlet projection, we obtain
\begin{equation}\label{Eq:U2d3j0}
    Z^{U(2),J = 0}_{d=3} (x) = \frac{1 + x^{9}}{(1 - x^2)^2 (1 - x^3) (1 - x^4)^2 (1 - x^6)},
\end{equation}
which indicates a secondary invariant at order $9$ and six primary invariants at various orders.

In comparison, for the $SU(2)$ partition function we have
\begin{equation}
    Z^{SU(2)}_{d=3}(x_1,x_2,x_3)=\frac{1+x_1x_2x_3}{(1-x_1^2)(1-x_2^2)(1-x_3^2)(1-x_1x_2)(1-x_2x_3)(1-x_1x_3)},
\end{equation}
which indicates six primary invariants at order $2$ and one secondary invariant at order $3$. 
The $SO(3)$ singlet projection yields
\begin{equation}\label{eq:su2so3j0p}
     Z^{SU(2),J=0}_{d=3}(x)=\frac{1+x^3}{(1-x^2)(1-x^4)(1-x^6)}.
\end{equation}
Here, we see that at order 2, only a single invariant is retained. At order 3, we retain a secondary invariant, and we find primary invariants at orders 4 and 6. We will identify these invariants in Sec.~\ref{Sec:Bmolinweyl}.

\subsubsection{$d\geq 4$ matrix model}\label{Ssec:dgeq4matrixmodel}

The Hironaka forms of the partition functions of $d\geq4$ $U(2)$ matrix models contain large numerators and are therefore kept in Appendix \ref{App:partfnexpr}.

We find a particularly simple form for $SO(d)$ singlet partition function of $SU(2)$ matrix models with arbitrary $d\geq 4$, which is given by
\begin{equation}\label{Eq:SU2d4plusj0}
   Z^{SU(2),J=0}_{d\geq 4}(x) =\frac{1}{(1 - x^2)(1 - x^4)(1 - x^6)}.
\end{equation}
Similar to the $d=3$ case, these partition functions reflect the presence of just three primary invariants. 
This saturation of the partition function in the singlet sector can be understood from the fact that this sector, in terms of the target space, is like an $s$-wave sector and only sees ``radial'' degrees of freedom.
This is directly related to the simplification of the Hamiltonian \eqref{Eq:d=3ham} in this sector, as we will discuss in more detail in Sec.~\ref{Sec:Btheory} and Sec.~\ref{Sec:Bmolinweyl}.

On the other hand, for the $U(2)$ gauge group, the partition functions are somewhat more complicated.
We find that in the $SO(d)$ singlet sector for $d=4$, the corresponding partition function can be expressed as
\begin{equation}\label{Eq:U2d4j0}
    Z^{U(2),J = 0}_{d=4} (x)=\frac{1+x^4 + x^{12}+x^{16}}{(1 - x^2)^2 (1 - x^4)^2 (1 - x^6)^2(1-x^8)}\,,
\end{equation}
with 7 primary invariants and three secondary invariants at order $4,12$ and $16$. 
When $d\geq 5$, the partition function for the $SO(d)$ singlets saturates as well and takes the form
\begin{equation}\label{Eq:U2d5plusj0}
    Z^{U(2),J = 0}_{d\geq 5} (x)=\frac{1 + x^{12}}{(1 - x^2)^2(1 - x^4)^2(1 - x^6)^2 (1 - x^8)}\,,
\end{equation}
with 7 primary invariants, as in the $d=4$ case above. The count of the primary invariants up to $d=5$ has a straightforward explanation in terms of the number of $SO(d)$ constraints. However, the count for $d\geq6$ is more subtle, since the $SO(d)$ constraints cannot be imposed independent of the $U(2)$ gauge constraints. 
We return to this point, and an interpretation of the above partition functions, in Sec.~\ref{Sec:Btheory} and Sec.~\ref{Sec:Bmolinweyl}.

\section{\texorpdfstring{$SO(d)$}{SO(d)} singlet sector of bosonic matrix models}\label{sec:struct-loop-singlet}

The $SO(d)$ singlet projections of the partition functions for the $U(2)$ or $SU(2)$ bosonic matrix models, discussed in Sec.~\ref{Sec:molienweyl}, can be interpreted as the Hilbert series for $U(2)\times SO(d)$ or $SU(2)\times SO(d)$ invariants, constructed out of the Hermitian matrices $X_{1,\ldots,d}$. 
In this section, we interpret these expressions by determining the set of so-called primary and secondary generators of the ring of invariants.
For the $d=2$ matrix model, we are able to derive the full structure of the ring of invariants, making use of $U(2)$ trace relations.
Then for the $d>2$ models, we observe that the $SO(d)$ singlet constraint effectively reduces the $SU(2)$ multi-matrix model to a theory of a single real symmetric $3\times 3$ matrix. 
Similarly, for $d>4$, the $U(2)$ multi-matrix model reduces to a theory of a real symmetric and anti-symmetric $3\times 3$ matrix and an additional (real) scalar. 

The partition functions for $U(2)$ and $SU(2)$ models in $SO(d)$ \emph{non}-singlet sectors, and the associated invariant theory, may be found in Appendix~\ref{Sec:loopspaceJneq0}.

\subsection{Hironaka decomposition}\label{Sec:Hironaka}

The set of invariants forms a graded ring, which, for linearly reductive groups\footnote{A group is linearly reductive if all of its finite dimensional representations can be decomposed as a direct sum of irreps.} over the complex numbers, such as $U(2)$, $SU(2)$ and $SO(d)$, is guaranteed to be Cohen–Macaulay and admits a Hironaka decomposition due to the Hochster–Roberts theorem \cite{HOCHSTER1974115}. The Hironaka decomposition of such invariant rings $\mathcal{R}$ takes the form
\begin{equation} \label{Eq:hironakagen}
    \mathcal{R} = \bigoplus_{m}s_m K[p_1,p_2, \hdots, p_l],
\end{equation}
where $p_{n}$ are called the primary invariants, $s_m$ are called the secondary invariants and $K[p_1,p_2,\hdots, p_l]$ is the ring freely generated by the primary invariants. Primary invariants are algebraically independent, whereas secondary invariants are quadratically reducible through relations of the form \cite{deMelloKoch:2025ngs}
\begin{equation}
    s_m s_n =\sum_l s_l f^l_{mn}, \quad f^{l}_{mn} \in K[p_1,p_2,\hdots, p_l],
\end{equation}
where we define $s_0=1$. When a ring has the structure \eqref{Eq:hironakagen} it is said to behave as a free module over a single sub-ring, the ring of primary invariants $K[p_1,p_2,\hdots p_l]$.

As seen in Sec.~\ref{Sec:molienweyl} and below, the partition functions and invariant rings for the $U(2)\times SO(d)$ and $SU(2)\times SO(d)$ invariants admit a Hironaka decomposition. However, we find that the invariant rings for the $SO(d)$ non-singlet sectors do not have a Hironaka decomposition. This does not contradict the Hochster-Roberts theorem, as explained in Appendix \ref{Sec:loopspaceJneq0}, where we investigate the invariant theory of these non-singlet sectors.


\subsection{Invariants in the \texorpdfstring{$d=2$}{d=2} matrix model} \label{Sec:invd2j0}

As shown in Sec.~\ref{Ssec:d=2matrixmodel}, the partition function of the $SO(2)$ singlet sector in the $d=2$ $U(2)$ matrix model \eqref{Eq:U2d2J0} indicates the presence of two primary invariants at order 2 and order 4, respectively.
One can easily see that the two primaries at order $2$ are
\begin{equation}
    p_1 = \tr(X_i X_j) \delta_{ij}, \quad p_2= \tr(X_i)\tr(X_j) \delta_{ij},
\end{equation}
where the sum over repeated indices is implied. Here, the invariants $p_1$ and $p_2$ are algebraically independent, since there are no trace relations at order $2$ for $2\times2$ matrices. 
We note that while the ring of $U(2)$ invariants is generated by single traces \eqref{Eq:u2invs}, the global symmetry singlet sector can have primary generators, such as $p_2$ above,  which are multi-trace from the gauge group point of view. This feature will also be encountered in the second part of the paper in the context of the BMN matrix model.

We can contract $SO(2)$ indices with the Kronecker delta $\delta_{ij}$ or the antisymmetric tensor $\epsilon_{ij}$. This implies that the $SO(2)$ singlets must have even order, and so the next set of invariants should occur at order $4$. There are five order $4$ invariants, which can be constructed using Kronecker delta contractions of the $SO(2)$ indices and they are listed as
\begin{align}
    &q_1 = \tr(X_l X_l X_m X_m),\quad q_2=\tr(X_lX_mX_lX_m),\\
    &q_3= \tr(X_lX_m)\tr(X_lX_m), \quad q_4= \tr(X_l) \tr(X_lX_mX_m),\\
    &q_5=\tr(X_lX_m)\tr(X_l)\tr(X_m).
\end{align}
Only two out of these five invariants turn out to be independent due to trace relations.
To see this, let us note that all $U(2)$ trace relations can be obtained from an identity satisfied by arbitrary $2\times2$ matrices $A$, $B$ and $C$
\begin{align}
    \begin{split}
        0=T_2(A,B,C) &= \tr(A) \tr(B)\tr(C)-\tr(AB)\tr(C)-\tr(AC)\tr(B)-\tr(A)\tr(BC)\\
        &+\tr(ABC)+\tr(ACB).
    \end{split}
\end{align}
This identity can be derived from the Cayley-Hamilton theorem, and it is the only relation generating equation for two-dimensional matrices. Using $T_2(A,B,C)$ we can derive trace relations for our matrices $X_{1,2}$ by choosing $A,B$ and $C$ appropriately.
At order 4, we find the following relations between the order 2 and order 4 invariants $p_m$ and $q_n$:
\begin{equation}\label{Eq:J0d2U2quartconst}
    q_5-q_3-2q_4+q_1+q_2=0,\quad p_1p_2-p_1^2-2q_4+2q_1=0,\quad p_2^2-2q_5-p_1p_2+2q_4=0\,.
\end{equation}
We choose $q_3$ and $q_5$ as the two independent order 4 invariants, which turns out to be convenient in the proof of completeness in Sec.~\ref{Sec:completed2j0}.

Let us explain why \eqref{Eq:J0d2U2quartconst} generates all trace relations among the invariants $p_m$ and $q_n$. Firstly, there is only one possible partition (up to permutations) of four matrices $X_{i,j,k,l}$ into three non-empty sets, which corresponds to $A=X_i$, $B=X_j$, $C=X_kX_l$ in $T_2(A,B,C)=0$. Secondly, the $T_2$ relation is symmetric under permutations of $A,B,C$. Finally, to ensure $SO(2)$ invariance, we must contract all the indices that appear. This implies that permutations of the choice of $A,B,C$ above, and of the indices $i,j,k,l$, do not generate new trace relations. There are two possible contractions of the indices in $T_2(X_i,X_j,X_k X_l)$, namely $\delta_{ik}\delta_{jl}$, which gives the first trace relation in \eqref{Eq:J0d2U2quartconst}, and $\delta_{ij}\delta_{kl}$, which gives the second relation in \eqref{Eq:J0d2U2quartconst}. The final trace relation in \eqref{Eq:J0d2U2quartconst} is obtained from
$$\delta_{ij} \delta_{kl}\tr(X_i) T_2(X_j,X_k,X_l)=0.$$ 
Note that we have not used an order $3$ trace relation yet, so this is an independent relation. All possible Kronecker delta contractions of this equation yield the same relation, which is the final equality in \eqref{Eq:J0d2U2quartconst}. 

So far, we have explained the presence of the two order 2 invariants, and two order 4 invariants.
To find the third order 4 invariant, we note that order 4 invariants can also be constructed using the two-dimensional anti-symmetric tensor $\epsilon_{ij}$. As explained in Sec.~\ref{Sec:molienweyl}, even though invariants with a single $\epsilon$ contraction are parity odd under the $X_1\leftrightarrow X_2$ exchange, they survive the $SO(2)$ singlet projection. 
The $q_{1,2,3,4,5}$ invariants are an exhaustive list of the possible non-trivial trace and index contractions at order 4. We simply need to replace one or both of the $\delta_{ij}$ contractions in these invariants with $\epsilon_{ij}$ contractions to generate the further order 4 invariants. The set of non-vanishing invariants obtained this way are
\begin{align}
    & q_1^{\epsilon,\epsilon} = \tr(X_i X_j X_kX_l) \epsilon_{ij} \epsilon_{kl} = q_2-q_1,\\
    &q_3^{\epsilon,\epsilon} = \tr(X_iX_j)\tr(X_kX_l)\epsilon_{ik}\epsilon_{jl} =p_1^2-q_3,\\
    &q_4^{\epsilon,\delta} = \tr(X_i)\tr(X_jX_k X_k)\epsilon_{ij},\\
    &q_5^{\epsilon,\delta} = \tr(X_iX_j)\tr(X_k)\tr(X_l)\epsilon_{ik}\delta_{jl},\\
    &q_5^{\epsilon,\epsilon} = \tr(X_iX_j)\tr(X_k)\tr(X_l)\epsilon_{ik}\epsilon_{jl}\,.
\end{align}
Since the first two invariants can be expressed in terms of the previously identified generators, we find only three new invariants.

The trace relations for this new set of invariants can be obtained as before. We can choose $A=X_i$, $B=X_j$, $C=X_kX_l$ in $T_2(A,B,C)=0$, and consider various possible contractions with $\delta_{ij}$ and $\epsilon_{ij}$. There are only two non-vanishing contraction structures that yield linearly independent trace relations. These are the $\epsilon_{ik} \delta_{jl}$ contraction, which gives
\begin{equation}\label{eq:rel1d=2}
    q_5^{\epsilon,\delta}+q_4^{\epsilon,\delta}=0,
\end{equation}
and the $\epsilon_{ik}\epsilon_{jl}$ contraction, which gives
\begin{equation}\label{eq:rel2d=2}
    q_5^{\epsilon,\epsilon}-q_3^{\epsilon,\epsilon} +q_2-q_1=0.
\end{equation}
By imposing these two relations, we can reduce two of these remaining invariants, leaving us with only one linearly independent invariant. In these relations, one can choose either $q_4^{\epsilon,\delta}$ or $q_5^{\epsilon,\delta}$ as our linearly independent invariant. We select $q_5^{\epsilon,\delta}$ as the independent generator and list the set of generators as
\begin{equation}\label{eq:gens-d2-U2-J0}
    \mathcal{G}=\{p_1,p_2, q_3,q_5;\, q_5^{\epsilon,\delta}\},
\end{equation}
which contains the expected five generators seen in the partition function \eqref{Eq:U2d2J0}.
In particular, four of the generators are primary, \textit{i.e.}, their action in the ring is free, while $q_5^{\epsilon,\delta}$ is a secondary invariant, since it is quadratically reducible by the following relation\footnote{Alternatively, we could have kept $q_5^{\epsilon,\delta}$ as a primary generator in which case $q_5$ would be secondary.}
\begin{align}
    (q_5^{\epsilon,\delta})^2= -\frac14 p_1^4-\frac14 p_1^2 p_2^2+\frac12 p_1^3 p_2 -\frac12 p_1 p_2^3+\frac 12p_2^2 q_3-2p_1^2 q_5+2 p_1p_2 q_5+\frac32 p_2^2 q_5-4 q_5^2,
    \label{Eq:SecondaryJ0U2d2}
\end{align}
which can be verified numerically. We have also numerically\,\footnote{In particular, we first list all possible monomials made from our five generators at a given order. If we have $k$ monomials at order $m$ we generate a $k\times k$  matrix, where each row is the $k$ monomials evaluated using random Hermitian $2\times2$ matrices $X_1$ and $X_2$. The null-space (kernel) of this $k\times k$ matrix allows us to identify relations between the monomials.} verified that there are no further relations among the generators in \eqref{eq:gens-d2-U2-J0}, up to order 20. Furthermore, in Sec.~\ref{Sec:completed2j0}, we will prove that this set of generators is complete.

Note that for the $SU(2)$ theory, we have $p_2=q_4=q_5=q_4^{\epsilon,\delta}=0$, and the relations \eqref{Eq:J0d2U2quartconst} imply that $q_1$ and $q_3$ are not independent generators. Therefore, only $p_1$ and $q_2$ are independent in this case.  Lastly, the syzygy \eqref{Eq:SecondaryJ0U2d2} for the $SU(2)$ theory is 
\begin{equation}
    0=-\frac14 p_1^4 -q_1^2+q_1 p_1^2,
\end{equation}
which is just the square of the second quartic order constraint in \eqref{Eq:J0d2U2quartconst}. This enables us to conclude that we do not have any secondary invariants in the $SU(2)$ theory.
It follows that the generators for $SU(2)$ are all primary and given by
\begin{equation}\label{eq:gens-d2-SU2-J0}
    \mathcal{G}_{S}=\{p_1,q_2\},
\end{equation}
This explains the form of the partition function as calculated in \eqref{eq:su2so2j0}.

\subsubsection{Completeness of $\mathcal{G}$} \label{Sec:completed2j0}

Above, we have derived the set of generators $\mathcal{G}$ of the ring of invariants associated with the $SO(2)$ singlet sector of the $d=2$, $U(2)$ matrix model.
Using trace relations, we also proved that these are the only independent invariants up to order 4. Below, using the Cayley-Hamilton theorem, we show that this set of generators is complete, \textit{i.e.}, they generate all possible higher order ($\geq6$) $U(2)\times SO(2)$ invariants.

Using trace relations and induction, it has already been proven that any $U(2)$ invariant can be generated by the following single trace invariants \cite{deMelloKoch:2025ngs}
\begin{equation}
    m_1=\tr(X_1), \quad m_2=\tr(X_2), \quad m_3=\tr(X_1^2), \quad m_4=\tr(X_1X_2) ,\quad m_5=\tr(X_2^2).
\end{equation}
Our $U(2) \times SO(2)$ invariants form a sub-ring of the full ring generated by the above. This sub-ring is generated by $SO(2)$ invariant polynomials in $m_{1,2,3,4,5}$. Such polynomials can be constructed by first identifying how $m_{1,2,3,4,5}$ transform under $SO(2)$ and then taking $SO(2)$ invariant combinations. The pair $(m_1,m_2)^T$ transforms as a column vector under $SO(2)$, which we denote as $Z_i$, with $Z_{1,2}=m_{1,2}$. One can easily verify that the remaining $m_{3,4,5}$ invariants transform as a bi-vector under $SO(2)$. We combine them into the symmetric matrix
\begin{equation}
    W = \begin{pmatrix}
        m_3&m_4\\
        m_4&m_5
    \end{pmatrix},
\end{equation}
which transforms under $SO(2)$ as $W\to g Wg^T$, where $g\in SO(2)$ is a $2\times2$ orthogonal matrix, with ${\rm det}(g)=1$. This enables us to rewrite the generators $\mathcal{G}$ in terms of the $m_{1,2,3,4,5}$ invariants as
\begin{align}
    &p_1=\tr(W), \quad p_2=Z^T\cdot Z,\\
    &q_3=\tr(W^2),\quad q_5=Z^T\cdot W\cdot Z,\quad q_5^{\epsilon,\delta}=Z_iW_{jk}Z_l \epsilon_{ij}\delta_{kl}.
\end{align}

Now we need to show that any higher order $SO(2)$ invariants that can be composed of $W$ and $Z$ are in the ring with generators $\mathcal{G}$. Let us first consider higher order ($\geq6$) $SO(2)$ invariants that are made from $W$ only. These can only take the form $\tr(W^n)$ or ${\rm det}(W^n)$, or their products. Leveraging the Cayley-Hamilton theorem, we can express our $2 \times 2$ matrix $W$ in terms of the following characteristic equation 
\begin{equation}\label{Eq:CH2}
    W^2-\tr(W)W-{\rm det}(W) I_{2\times2}=0,
\end{equation}
where $I_{2\times 2}$ is the two-dimensional identity matrix. This provides a way to express ${\det}(W)$ in terms of $\tr(W)$ and $\tr(W^2)$. By induction, we can express $\tr(W^n)$ for $n\geq3$ in terms of $\tr(W)$ and $\tr(W^2)$. Furthermore, ${\det}(W^n) = {\rm det}(W)^n$. Therefore, any higher order $SO(2)$ invariant made only from $W$ can be expressed in terms of our generating set $\mathcal{G}$.

Next, we consider higher order $SO(2)$ invariants that are made from both $W$ and $Z$. They take the form of non-vanishing index contractions of $Z_i (W^n)_{jk} Z_l$. For the $\delta_{ij} \delta_{kl}$ contraction, by \eqref{Eq:CH2}, we have
\begin{equation}
    \delta_{ij} \delta_{kl} Z_i (W^2)_{jk} Z_l=\tr(W)Z^T\cdot W\cdot Z+{\rm det}(W) Z^T\cdot Z,
\end{equation}
where $Z^T\cdot W\cdot Z=q_5$ and $Z^T\cdot Z=p_2$. It is then straightforward to see that $\delta_{ij} \delta_{kl}Z_i (W^n)_{jk} Z_l$ can be written in terms of our generators $\mathcal{G}$ using \eqref{Eq:CH2} and induction on $n$. Similarly, for the $\epsilon_{ij}\delta_{kl}$ contraction, from \eqref{Eq:CH2}, we get
\begin{equation}
    \epsilon_{ij} \delta_{kl} Z_i (W^2)_{jk} Z_l=\tr(W)\epsilon_{ij}\delta_{kl}Z_iW_{jk}Z_l = p_1 q_5^{\epsilon,\delta}.
\end{equation}
Using induction on $n$, it then follows that $\epsilon_{ij} \delta_{kl}Z_i (W^n)_{jk} Z_l$ can be expressed in terms of our generators $\mathcal{G}$. Finally, for the $\epsilon_{ij}\epsilon_{kl}$ contraction, from \eqref{Eq:CH2}, we get 
\begin{equation}
    \epsilon_{ij} \epsilon_{kl} Z_i (W^2)_{jk} Z_l=\tr(W)\epsilon_{ij}\epsilon_{kl}Z_iW_{jk}Z_l+{\rm det}(W)\delta_{jk}\epsilon_{ij}\epsilon_{kl}Z_iZ_l = p_1 q_5^{\epsilon,\epsilon}-2{\rm det}(W) p_2.
\end{equation}
In Sec.~\ref{Sec:invd2j0} we have already shown that $q_5^{\epsilon,\epsilon}=\epsilon_{ij}\epsilon_{kl}Z_iW_{jk}Z_l$ can be expressed in terms of our generators $\mathcal{G}$. Using induction on $n$, it follows that $\epsilon_{ij} \epsilon_{kl}Z_i (W^n)_{jk} Z_l$ can be expressed in terms of $\mathcal{G}$. 

The only contraction structure that remains is contractions of $W_{ij} W_{kl} \epsilon_{jk}$. Since $W$ is a symmetric matrix, we have
\begin{equation}
    W_{ij} W_{kl} \epsilon_{jk} = {\rm det}(W)\epsilon_{il}.
\end{equation}
As we have shown, any $\epsilon_{il}$ contractions of any combination of $W$ and $Z$ are reducible in terms of $\mathcal{G}$.
We can therefore deduce that any further contractions of the above with $W$ and $Z$ can similarly be expressed in terms of $\mathcal{G}$. 

Having exhausted all of the non-vanishing combinations and contractions, we conclude our proof of the completeness of the set of $U(2)\times SO(2)$ generators $\mathcal{G}$. As discussed in Sec.~\ref{Sec:molienweyl}, the partition functions for the free multi-matrix models are used merely as a crutch to identify the generators of the invariant ring. Our proof relies on trace relations, which are purely kinematic constraints, and are unaffected by interactions. Therefore, the set of invariants $\mathcal{G}$ also generates the ring of invariants, or equivalently, the gauge-invariant Hilbert space of when interactions are turned on.

In  Sec.~\ref{Sec:Bmolinweyl}, we will identify the generating set of invariants for $d\geq3$ based on the form of the partition functions in the singlet sector calculated in Sec.~\ref{Sec:threeormore}, and prove their completeness using a different approach.

\subsection{Counting primary invariants in \texorpdfstring{$d>2$}{d>2} matrix models} \label{Sec:Btheory}

It has been observed that the number of degrees of freedom in matrix models, after gauge fixing, matches with the number of primary invariants \cite{deMelloKoch:2025qeq,deMelloKoch:2025cec}.\footnote{Another way to determine the count of the independent primary invariants is by using the Krull dimension of the invariant ring \cite{Herstein1970Notes}.}
A simple formula for the number of secondary invariants is not known, but their number is expected to scale exponentially with the (square of the) rank of the gauge group $N$ \cite{deMelloKoch:2025ngs,deMelloKoch:2025qeq,deMelloKoch:2025cec}.
In this section, we show how the count of primary invariants for the $SO(d)$ singlet sector of $d>2$, $U(2)$ and $SU(2)$ matrix models, as deduced from the partition functions in Sec.~\ref{Sec:threeormore}, can be reproduced from a count of the gauge invariant degrees of freedom.

For a model with $d$ Hermitian matrices that transform in the adjoint of $U(N)$, we use the following gauge fixing procedure \cite{deMelloKoch:2025ngs}. Firstly, we can use the $U(N)$ transformation to diagonalize one of the $d$ matrices, say $X_1$. This leaves $(d-1)N^2$ (real) degrees of freedom in the matrices $X_i$, $i=2,3,\hdots d$, and $N$ degrees of freedom in $X_1$. However, there is a residual gauge symmetry that corresponds to the adjoint action of a diagonal $N\times N$ unitary matrix, which preserves $X_1$. Only $N-1$ out of the $N$ non-zero entries (phases) in this diagonal unitary matrix are meaningful, since an overall phase, which leaves all $X_i$ matrices unchanged, can be removed. These $N-1$ real variables can be used to set the $(j,j+1)^{\rm th}$, $j=1,2\hdots,N-1$, entries of one of the matrices, say $X_2$, to be real. Thus, the total number of degrees of freedom that remain after a complete gauge fixing is
\begin{equation}\label{Eq:NpUN}
    N_p^{U(N)} = (d-1)N^2+N-(N-1)=(d-1)N^2+1.
\end{equation}
Similarly for $SU(N)$ we get
\begin{equation}\label{Eq:NpSUN}
    N_p^{SU(N)} = N_p^{U(N)}-d= (d-1)(N^2-1),
\end{equation}
where the subtraction is due to tracelessness. These counts match with the number of primary invariants, as determined from the number of denominators in the Hironaka form of the partition functions for $U(2)$ and $SU(2)$ singlets without the $SO(d)$ projection. See \eqref{eq:res-sum-U(2)-2}, \eqref{eq:res-sum-SU(2)-2}, and  Appendix \ref{App:partfnexpr}.

A naive count of the primary invariants in the $SO(d)$ singlet sector can be obtained by subtracting the number of $SO(d)$ constraints, which is $\frac{d(d-1)}{2}$, from the above. 
We will focus on $U(2)$ and $SU(2)$ models, for which we get
\begin{equation}
    N_p^{U(2),J=0} = 4 d-3 -\frac{d(d-1)}{2}, \quad N_p^{SU(2),J=0} = 3(d-1)- \frac{d(d-1)}{2}\,.
\end{equation}
These numbers can indeed be seen to match with the number of primary invariants, as deduced from the partition functions computed in Sec.~\ref{Ssec:examplesmolien-weyl}, for $U(2)$ gauge group when $d\leq 5$ and for $SU(2)$ gauge group when $d\leq 4$.
Beyond these values, however, there is a mismatch. 
In fact, for sufficiently large $d$, the naive count above does not just underestimate the number of primary invariants, but gives a negative answer.
Instead, as seen in the partition functions \eqref{Eq:SU2d4plusj0} and \eqref{Eq:U2d5plusj0}, the number of primary invariants in the singlet sector saturates for sufficiently large $d$. 

Let us reproduce this count for the $SU(2)$ matrix models with $d\geq3$ matrices, by performing a more careful gauge fixing in the $SO(d)$ singlet sector. 
First, note that each of the $X_i$ matrices can be expressed as a three dimensional vector $Q^i_a$, where $i=1,2,\hdots d$ labels the $d$ matrices and $a=1,2,3$ labels their components in the Pauli $\sigma_{1,2,3}$ basis. 
We can therefore organize the configuration space of the model into a real $3\times d$ matrix. The $SU(2)$ transformations act on $Q_a^i$ as $3\times 3$ $SO(3)$ matrices from the left and the $SO(d)$ transformations act as $d\times d$ matrices from the right. We describe this with the general transformation
\begin{equation}
    Q^i_a \to g^b_a Q^j_b h^i_j, \label{Eq:qtransformation}
\end{equation}
where $g$ and $h$ are elements of $SO(3)$ and $SO(d)$, respectively. Since we are concerned with $SU(2)\times SO(d)$ singlets, both of these transformations $g$ and $h$ can be thought of as gauge transformations. Using these transformations, we can convert the matrix $Q^i_a$ into a very simple form \cite{Sethi:1997pa}:
\begin{equation}\label{Eq:diagonalmatsu2d}
    Q^i_a \to 
    \begin{pmatrix}
        x&0&0&0&\cdots&0\\
        0&y&0&0&\cdots&0\\
        0&0&z&0&\cdots&0
    \end{pmatrix}.
\end{equation}
The above diagonal form of $Q_a^i$ follows from the singular value decomposition. Or more generally, any $n\times m$ real matrix $Q$, with $n\leq m$, has the singular value decomposition
\begin{equation}
    Q = U D {V}^T,
\end{equation}
where $U,V$ are $n\times n$ and $m\times m$ orthogonal matrices, respectively. Above, $D$ is a \textit{diagonal} $n\times m$ matrix with $n$ non-zero entries, as in \eqref{Eq:diagonalmatsu2d} for $n=3$ and $m=d\geq3$. 
Strictly speaking, the singular value decomposition only guarantees that $U \in O(3) $ and $V \in O(d)$. 
However, in cases where the determinants of $U$ and/or $V$ are $-1$, we can simply absorb the sign in a row of $D$, ensuring the $U/V$ matrix has unit determinant.
 
The above shows that there are only three degrees of freedom in the $SO(d)$ singlet sector of $SU(2)$ matrix models with $d\geq3$, which is consistent with the number of primary invariants as read off from the partition functions in \eqref{eq:su2so3j0p} and \eqref{Eq:SU2d4plusj0}.

A similar argument can be made for the number of primary invariants in a theory of $d$, $U(2)$ adjoint matrices. We can construct a $4\times d$ matrix $\mathcal{Q}_a^i$, where now $a=0,1,2,3$ labels apart from the Pauli matrices also the identity matrix, corresponding to the non-zero trace of $X_i$. 
The general $U(2)\times SO(d)$ transformation acts as
\begin{equation}
    \mathcal{Q}_a^i \to g_a^b \mathcal{Q}_b^j h_j^i, \quad h\in SO(d),
\end{equation}
and $g\in U(2)$ are $4\times 4$ matrices of the form $g = 1\oplus R$, $R\in SO(3)$, which leave the traces of the $d$ matrices unchanged. The matrix $\mathcal{Q}_a^i$ takes the form
\begin{equation}\label{Eq:u2gaugefixmat1}
    \begin{pmatrix}
        t_1&t_2&t_3&t_4&t_5&\cdots&t_d\\
        x_1&x_2&x_3&x_4&x_5&\cdots&x_d\\
        y_1&y_2&y_3&y_4&y_5&\cdots&y_d\\
        z_1&z_2&z_3&z_4&z_5&\cdots&z_d,
    \end{pmatrix}
\end{equation}
where the first row corresponds to the traces of the matrices. As for the $SU(2)$ case, we can use the singular value decomposition of the $3\times d$ block to convert to three degrees of freedom on the diagonal.
Using a further $SO(d-3)$ transformations on the upper row, the gauge fixed matrix can be written as
\begin{equation}\label{Eq:u2gaugefixmat2}
  \mathcal{Q}_a^i \to  \begin{pmatrix}
        \tilde{t}_1&\tilde{t}_2&\tilde{t}_3&\tilde{t}_4&0&\cdots&0\\
        \tilde{x}_1&0&0&0&0&\cdots&0\\
        0&\tilde{y}_2&0&0&0&\cdots&0\\
        0&0&\tilde{z}_3&0&0&\cdots&0
    \end{pmatrix}.
\end{equation}
We thus remain with $7$ degrees of freedom, which matches with the number of primary invariants in the $SO(d)$ singlet sector of $U(2)$ matrix models with $d\geq4$, as seen in the partition functions \eqref{Eq:U2d4j0} and \eqref{Eq:U2d5plusj0}.

\subsection{Invariants in \texorpdfstring{$d>2$}{d>2} matrix models}\label{Sec:Bmolinweyl}

In this section, we will describe the full ring of invariants in the $SO(d)$ singlet sector of the $SU(2)$ and $U(2)$ matrix models when $d>2$.

\subsubsection{The $SU(2)$ case}\label{Sec:BtheorySU2}

An elegant way to implement the gauge fixing \eqref{Eq:diagonalmatsu2d} for the $SU(2)$ theory is by first constructing an $SO(d)$ invariant (real) symmetric $3\times 3$ matrix \cite{Hoppe:2023qem}\footnote{One could similarly construct a $d\times d$ symmetric matrix of $SO(3)$ invariants, $\widetilde{B}^{ij}= Q_a^iQ^j_b\delta^{ab}$. However, $\widetilde{B}$ parametrizes an over-complete set of invariants, since the number \eqref{Eq:NpSUN} of primary $SO(3)$ invariants made from $d$ adjoint matrices scales linearly with $d$, which is smaller than $\frac{d(d+1)}{2}$ for sufficiently large $d$.}
\begin{equation}
    B= Q Q^T \quad \text{or}\quad  B_{ab}= Q^i_aQ^j_b\delta_{ij}\,,
\end{equation}
where $a,b$ are $SO(3)$ vector indices and, as introduced in the previous section, the $SO(3)$ vectors $Q^i_a$ replace the adjoint $SU(2)$ matrices $X_i$ for the remainder of this section.
The fact that $B$ only contains three degrees of freedom, just like $Q_a^i$, follows from the residual $SO(3)$ gauge symmetry, which can be used to diagonalize $B$:
\begin{equation}\label{eq:B-gauge-fixed-su2}
    B\to g B g^T=\begin{pmatrix}
        x^2 & 0 & 0\\
        0 & y^2 & 0\\
        0 & 0 & z^2
    \end{pmatrix}, \quad g\in SO(3)\,,
\end{equation}
where we used the same notation for the eigenvalues, as in \eqref{Eq:diagonalmatsu2d}.
For $d\geq 4$, the ring of invariants of the $SO(d)$ singlet sector of the $SU(2)$ matrix model is freely generated by three traces of the single $3\times 3$ matrix $B$:
\begin{equation}\label{eq:B-gens-su2}
    \{\tr(B), \tr(B^2),\tr(B^3)\}\,,
\end{equation}
since $\tr(B^n)$ for $n\geq 4$ can be expressed in terms of these generators through trace relations.
This also explains the form of the partition function calculated in \eqref{Eq:SU2d4plusj0}, which we repeat here for convenience:
\begin{equation}
  {Z}^{SU(2),J=0}_{d\geq 4}(x)  = \frac{1}{(1-x^2)(1-x^4)(1-x^6)}\,.
\end{equation}
Each factor in the denominator is clearly identified  with one of the generators in \eqref{eq:B-gens-su2}.
We conclude that the recasting of the singlet sector of the $d\geq 4$ matrix models into a model of a single matrix trivialises the proof of completeness of the above $SU(2)\times SO(d)$ invariants for $d\geq4$.

Let us note, however, that for $d=3$ the above argument does not quite work.
The reason for this is that in this case, $Q^i_a$ is a $3\times3$ matrix and ${\rm det}(Q)$ represents an additional independent invariant, which cannot be expressed in terms of traces of $B$.
However, this invariant is secondary, since it is quadratically reducible in terms of $B$ via:
\begin{equation}
  {\rm det}(Q)^2= {\rm det}(B)= \frac 16\lb{\rm tr}(B^3)-3{\rm tr}(B^2){\rm tr}(B)+2{\rm tr}(B)^3\rb\,.
\end{equation}
This secondary invariant together with the three primary invariants in \eqref{eq:B-gens-su2} account for the factors in the $d=3$ partition function \eqref{eq:su2so3j0p}, which we repeat here for convenience:
\begin{equation}
    Z^{SU(2),J=0}_{d=3}(x) = \frac{1+x^3}{(1-x^2)(1-x^4)(1-x^6)}\,.
\end{equation}
So, we conclude that the $SO(d)$ singlet sector of the Hilbert space of even an \emph{interacting} $SU(2)$ matrix model can be understood in terms of an effective $SO(3)$ single--matrix model. 
For example, the Hamiltonian in the $SO(d)$ singlet sector of the bosonic part of the BFSS model takes the simple form 
\begin{equation}
    H = \frac12\lb p_x^2+p_y^2+p_z^2\rb+ \frac12 \lb x^2y^2+y^2z^2+z^2x^2\rb\,,
\end{equation}
where the commutator squared potential is reduced to a potential in terms of the eigenvalues of $B$ \cite{Hoppe:2023qem}.
Indeed, using the gauge fixed matrices \eqref{eq:B-gauge-fixed-su2}, it is easy to see that the potential can be expressed in terms of $\tr(B)$ and $\tr(B^2)$. 
Furthermore, we note that $\tr(B) = x^2+y^2+z^2=\tr(X_i X_j) \delta_{ij}$ can be thought of as the radial (target space) distance between the two $D0$ branes of the $SU(2)$ BFSS model, and the other two primary invariants, $\tr(B^2)$ and $\tr(B^3)$, are its moments.

\subsubsection{The $U(2)$ case}\label{Sec:BtheoryU2}

We now analyze the ring of invariants for the $SO(d)$ singlet sector of the $U(2)$ matrix model for $d\geq 4$.
As in Section \ref{Sec:Btheory}, we first convert the $d$ $U(2)$ matrices into a $4 \times d$ matrix $\mathcal{Q}^i_a$, which transforms under $U(2) \times SO(d)$ as  
\begin{equation}
    \mathcal{Q}_a^i \to g_a^b \mathcal{Q}_b^j h_j^i, \quad h\in SO(d)\,,
\end{equation}
where $g$ is of the form $g = 1\oplus R$, $R\in SO(3)$. 
As for the $SU(2)$ case, we can define the $4\times 4$ matrix $\mathcal{B} =\mathcal{Q}\mathcal{Q}^T$ that transforms under $U(2)$ as
\begin{equation}
    \mathcal{B}\to g \mathcal{B} g^T =\begin{pmatrix} 1 & 0 \\ 0 & R \end{pmatrix}
\begin{pmatrix} S & \mathbf{v}^T \\ \mathbf{v} & T \end{pmatrix}
\begin{pmatrix} 1 & 0 \\ 0 & R^T \end{pmatrix}\,,
\end{equation}
where we parametrized $\mathcal{B}$ in terms of a $3 \times 3$ symmetric matrix $T$, a 3-dimensional vector $\mathbf{v}$, and a scalar $S$.
Using the isomorphism between the vector and asymmetric two-tensor representations of $SO(3)$, we can repackage $\mathcal{B}$ into a $7\times 7$ block diagonal matrix as
\begin{equation}
 \bar{\mathcal{B}}=   \begin{pmatrix} 
   S & 0 & 0 \\ 
   0 & A & 0 \\
   0 & 0 & T
   \end{pmatrix}\,,
\end{equation}
where $A$ is an antisymmetric $3 \times 3$  matrix related to $\mathbf{v}$ as
\begin{equation}
    A_{ab} = \epsilon_{abc}\, \mathbf{v}_c \, .
\end{equation}
Now recall the partition function calculated for the $d\geq 5$ models in \eqref{Eq:U2d5plusj0}:
\begin{equation}
    Z^{U(2),J = 0}_{d\geq 5} (x)=\frac{1 + x^{12}}{(1 - x^2)^2(1 - x^4)^2(1 - x^6)^2 (1 - x^8)}\,.
\end{equation}
In terms of the components of $\mathcal{B}$, we identify the seven primary invariants with
\begin{equation}
    \{S, \tr(T),\tr(T^2),\tr(T^3),\tr(A^2), \tr(TA^2), \tr(T^2A^2)\}\,.
\end{equation}
The first four primary invariants are as expected for a single $3\times 3$ matrix $T$.
Since $A$ is an anti-symmetric matrix, its trace is only non-zero for even powers. 
Moreover, since it is $3\times 3$, only $A^2$ appears.
Finally, we note that $\tr(T^2 A^2)$ is a generator involving four matrices.
Even though the matrices are $3\times 3$, this generator is not reducible by trace relations due to the fact that there is an additional invariant involving four matrices: $\tr(TATA)$. 
There is a single trace relation that features this invariant along with $\tr(T^2A^2)$, and we have chosen the latter to be the independent invariant.

The partition function also suggests the presence of a secondary invariant in terms of 6 matrices, which we identify as
\begin{equation}
    \tr(TAT^2A^2).
\end{equation}
We have numerically verified that this is an independent generator at order $6$, but that it is also quadratically reducible in terms of the primary generators listed above.\footnote{As before, we list all possible order 12 monomials that can be made from the $7$ primary and one secondary invariant. There are $591$ such monomials. We then generate $591$ random instances of these monomials to construct a $591\times 591$ matrix. The rank of this matrix is $590$, up to machine precision, and its null space gives us the desired syzygy. We do not provide the explicit relation here since it is too long.}

As in the $d=3$ $SU(2)$ case, the above argument does not quite work for $d=4$.
In this case, this is because $\mathcal{Q}^i_a$ is now a $4\times4$ matrix and ${\rm det}(\mathcal{Q})$ represents an additional independent invariant, which cannot be expressed in terms of traces of $\mathcal{B}$.
However, this invariant is secondary, since it is quadratically reducible in terms of $\mathcal{B}$ via:
\begin{equation}
  {\rm det}(\mathcal{Q})^2= {\rm det}(\mathcal{B})\,,
\end{equation}
and one can easily check that ${\rm det}(\mathcal{B})$ can be written in terms of the primary invariants listed above.
This additional secondary invariant in the $d=4$ case accounts for the additional two terms in the numerator of the $d=4$ partition function calculated in \eqref{Eq:U2d4j0}:
\begin{equation}
    Z^{U(2),J = 0}_{d=4} (x)=\frac{1 +x^4+ x^{12}+x^{16}}{(1 - x^2)^2 (1 - x^4)^2 (1 - x^6)^2(1-x^8)}.
\end{equation}
We stress that this is an exceptional case, because the determinant only exists at $d=4$, when $\mathcal{Q}^i_a$ is a $4\times4$ matrix. 
Thus, we see how also the $U(2)$ matrix model in the $SO(d)$ singlet sector reduces to a $SO(3)$ matrix model, but unlike the $SU(2)$ case, apart from a symmetric matrix, this model also contains an anti-symmetric matrix (or vector) and a scalar.
This demonstrates how the $SO(d)$ singlet projection significantly simplifies the original multi--matrix model. 

Let us end this section by returning to a point mentioned in Sec.~\ref{ssec:interacting-ham}.
Compared to an $SU(2)$ theory, a theory of $d$ $U(2)$ matrices corresponds to including the centre of mass motion of the two $D0$ branes. 
Specifically, for the commutator square potential, the centre of mass decouples and behaves like a free particle. This can be seen by computing the commutators using our gauge fixed matrix \eqref{Eq:u2gaugefixmat2}. It follows that the relative dynamics of the bosonic BFSS matrix model is not affected by the inclusion of the centre of mass motion. 
However, the dynamics in the $SO(d)$ singlet sector of the matrix models with other interaction potentials will generically be different for $U(2)$ vs. $SU(2)$ gauge groups. 
This is because the centre of mass itself carries angular momentum. In the $J=0$ sector of the $SU(2)$ theory we are setting the angular momentum of only the relative coordinates to zero. On the other hand, in the $J=0$ sector of the $U(2)$ theory, we set the combined angular momentum of the centre of mass and the relative degrees of freedom to zero.

\section{\texorpdfstring{$SU(3)_R$}{SU(3)R} singlets in the BMN sector of the \texorpdfstring{$\mathcal{N}=4$}{N=4} theory}\label{sec:non-grav}

In this section, we investigate a global symmetry singlet projection in the context of the $4d$ $\mathcal{N}=4$ SYM theory with $SU(N)$ gauge group.
More precisely, we will be interested in singlets of an $SU(3)_R$ (subgroup of) $R$-symmetry in the so-called BMN sector of the $\mathcal{N}=4$ theory.
This is a subsector of the 1/16 BPS sector, which we review below.
It was recently shown to harbour ``non-graviton'' operators, which are expected to be related to the microstates of supersymmetric $AdS_5$ black holes.
We will demonstrate that the restriction to $R$-symmetry singlets simplifies the analysis of their spectrum.
In particular, we show how a closed form expression for the index of singlet non-graviton operators can be obtained, and work out the detailed form for $SU(2)$ and $SU(3)$ gauge groups.

\subsection{Review of the BMN sector}\label{ssec:rev-bmn}

The 1/16 BPS sector of the $\mathcal{N}=4$ theory is defined with respect to a single complex supercharge (and its Hermitian conjugate) in the full $\mathcal{N}=4$ superconformal algebra \cite{Kinney:2005ej}.
We follow the convention of \cite{Grant:2008sk} and take the supercharge to be $Q\equiv Q^4_{-}$, and $Q^{\dagger}\equiv S^-_{4}$ its superconformal partner.
The supercharge $Q$ satisfies the following anti-commutator
\begin{equation}
    2\{Q,Q^\dagger\} = E-J_1-J_2-R_1-R_2-R_3\,,
\end{equation}
where $E$ is the dilatation operator on $\mathbb{R}^4$, $J_{1,2}$ generate rotations in the two planes $\mathbb{R}^2\subset \mathbb{R}^4$, and $R_{1,2,3}$ are the Cartan generators of $SU(4)_R$.
Since a 1/16 BPS operator is, by definition, annihilated by $Q$ and $Q^\dagger$, it follows that its energy, in radial quantization, is related to its angular momenta and $R$-charges through $E=J_1+J_2+R_1+R_2+R_3$.
In the free theory, it is easy to construct 1/16 BPS operators (modulo trace relations).
Indeed, one can list all components of the Lagrangian fields that satisfy the BPS condition, which are commonly referred to as the BPS letters.
Any gauge invariant word made up of the letters corresponds to a 1/16 BPS operator. 
For convenience, we list the BPS letters in Table \ref{tab:n=4-letters}.

\begin{table}[tbp]\label{tab:n=4-letters}
\begin{centering}
\bgroup
\def\arraystretch{1.3}
\begin{tabular}{|c|c|c|c|c|c|c|c|}
\hline
Letters & $ E$ & $J_1$ & $J_2$ & $R_1$ & $R_2$ & $R_3$  & $\mathcal{I}_{\text{1-loop}}(y_{i};p, q)$ \tabularnewline
  \hline
  $  f_{++}$ & $2$ & $1$ & $1$ & $0$ & $0$ & $0$  &  $y_1y_2y_3$ \tabularnewline
  \hline
  $  \bar{\lambda}_{\dot{\pm}}$  & $  \frac{3}{2}$ & $ \pm \frac{1}{2}$ & $ \mp \frac{1}{2}$ & $ \frac{1}{2}$ & $ \frac{1}{2}$  & $\frac{1}{2}$ &  $-p$, $-q$ \tabularnewline
  \hline
  $  \psi_{1+}$ & $  \frac{3}{2}$ & $  \frac{1}{2}$ & $ \frac{1}{2}$ & $- \frac{1}{2}$ & $ \frac{1}{2}$ & $\frac{1}{2}$  & $-y_2y_3$ \tabularnewline
  \hline
  $  \psi_{2+}$ & $  \frac{3}{2}$ & $  \frac{1}{2}$ & $ \frac{1}{2}$ & $ \frac{1}{2}$ & $ -\frac{1}{2}$ & $\frac{1}{2}$  & $-y_1y_3$ \tabularnewline
  \hline
  $  \psi_{3+}$ & $  \frac{3}{2}$ & $  \frac{1}{2}$ & $ \frac{1}{2}$ & $ \frac{1}{2}$ & $ \frac{1}{2}$ & $-\frac{1}{2}$  & $-y_1y_2$ \tabularnewline
  \hline
  $  \bar{\phi}^{1}$ & $1$ & $0$ & $0$ & $1$ & $0$ & $0$  & $y_1$   \tabularnewline
  \hline
   $  \bar{\phi}^{2}$ & $1$ & $0$ & $0$ & $0$ & $1$ & $0$  & $y_2$   \tabularnewline
  \hline
$  \bar{\phi}^3$ & $1$ & $0$ & $0$ & $0$ & $0$ & $1$  & $y_3$   \tabularnewline
  \hline
  $  \partial_{+\dot{+}}$ & $1$ & $1$ & $0$ & $0$ & $0$ & $0$   & $p$ \tabularnewline
   \hline
  $  \partial_{+\dot{-}}$ & $1$ & $0$ & $1$ & $0$ & $0$ & $0$   & $q$ \tabularnewline
\hline
$  Q^1_+$ & $  \frac{1}{2}$ & $\frac{1}{2}$ & $ \frac{1}{2}$ & $ \frac{1}{2}$ & $ -\frac{1}{2}$  & $-\frac{1}{2}$ & $y_1$  \tabularnewline
  \hline
  $  Q^2_+$ & $  \frac{1}{2}$ & $\frac{1}{2}$ & $ \frac{1}{2}$ & $- \frac{1}{2}$ & $ \frac{1}{2}$  & $-\frac{1}{2}$ & $y_2$  \tabularnewline
  \hline
  $  Q^3_+$ & $  \frac{1}{2}$ & $\frac{1}{2}$ & $ \frac{1}{2}$ & $ -\frac{1}{2}$ & $ -\frac{1}{2}$  & $\frac{1}{2}$ & $y_3$  \tabularnewline
  \hline
  $  \partial_{+\dot{\alpha}}  \bar{\lambda}^{\dot \alpha} =0$ & $  \frac{5}{2}$ & $\frac{1}{2}$ & $ \frac{1}{2}$ & $ \frac{1}{2}$ & $ \frac{1}{2}$  & $\frac{1}{2}$ & $pq$  \tabularnewline
  \hline
\end{tabular}
\egroup
\par\end{centering}
\caption{The 1/16 BPS letters of the free $\mathcal{N}=4$ theory, all of which transform in the adjoint of the $SU(N)$ gauge group. All letters except for derivatives and supercharges have $U(1)_Y$ charge $1$. The final line contains the gaugino equation of motion, which contributes to the index as a constraint. The expression for $\mathcal{I}_{\text{1-loop}}$ follows from \eqref{eq:ref-index-defn}.}
\end{table}

It is expected that the 1/16 BPS spectrum at strong coupling contains operators corresponding to the microstates of supersymmetric $AdS_5$ black holes \cite{Kinney:2005ej,Grant:2008sk}.
The determination of the spectrum at strong or even non-zero coupling is non-trivial, due to the fact that BPS operators in the free theory can recombine into long multiplets and thereby lift from the BPS spectrum.
Progress has so far relied on two simplifications \cite{Grant:2008sk}. 
First, a BPS operator, annihilated by both $Q$ and $Q^\dagger$, is in one-to-one correspondence with a $Q$-cohomology class of operators, \textit{i.e.}, the class of operators annihilated by $Q$ modulo the addition of $Q$-exact operators  $O'=\left[Q,O\right\rbrace$.
Furthermore, one relies on the conjecture that the BPS spectrum is exact at one-loop, \textit{i.e.}, that there is no further lifting of operators when higher loop (or non-perturbative) corrections are taken into account.\footnote{\label{fn:1-loop-conj}This conjecture was recently shown to be false in the context of $\mathcal{N}=4$ theories with $SO(2N+1)$ and ($S$-dual) $USp(N)$ gauge groups \cite{Gadde:2025yoa,Chang:2025mqp,Choi:2025bhi}. At present, the status is unclear for $SU(N)$ gauge groups. In the following, we will indicate to what extent our arguments depend on this conjecture.}
Assuming this, it follows that the BPS spectrum at strong coupling is in one-to-one correspondence with the \emph{classical} (or half-loop) $Q$-cohomology of the $\mathcal{N}=4$ theory.

Determining the classical 1/16 BPS cohomology of the $\mathcal{N}=4$ theory is still a highly non-trivial problem, even for an $SU(2)$ gauge group, see, \textit{e.g.}, \cite{Grant:2008sk,Chang:2013fba,Chang:2022mjp,Choi:2022caq} for partial progress.
The problem can be simplified by restricting to the so-called BMN subsector, as studied recently in \cite{Choi:2023znd,Choi:2023vdm,deMelloKoch:2024pcs,Gadde:2025yoa}, which is so named due to its connection with the BMN matrix model \cite{Berenstein:2002jq,Kim:2003rza}.\footnote{As shown in \cite{Kim:2003rza}, the BMN matrix model is classically equivalent to a consistent truncation of the four-dimensional $\mathcal{N}=4$ super Yang--Mills theory on a three-sphere, where one keeps only the lowest lying modes of the fields. However, as also shown in \cite{Kim:2003rza}, this equivalence does not extend to the full quantum theory and, in particular, breaks down at two-loops.}
The BMN sector is comprised by a subset of the letters in Table \ref{tab:n=4-letters}, namely: three scalars $\bar{\phi}^a$, three (positive chirality) fermions $\psi_{a+}$, and a component of the field strength $f_{++}$.
We note that the scalars and fermions transform respectively in a fundamental and anti-fundamental representation of $SU(3)_R\subset SU(4)_R$, which we indicate with an upper and lower index, respectively.
We will henceforth refer to the fermions simply as $\psi_a$ and to the field strength as $f$. 
These letters form a closed subset under the action of the (classical) supercharge $Q$, which acts as \cite{Grant:2008sk}
\begin{equation}\label{Eq:SCalgebra}
    [Q,\bar{\phi}^a]=0,\quad \{Q,\psi_a\}=-ig_{\rm YM}\epsilon_{abc}[\bar{\phi}^b,\bar{\phi}^c],\quad [Q,f]=-ig_{\rm YM}[\psi_a,\bar{\phi}^a]\,,
\end{equation}
where $g_{\rm YM}$ is the Yang-Mills coupling.
Restrictions of this type were termed letter-based truncations in \cite{Gadde:2025yoa}.
In the above references, it was verified that this sector still harbours interesting, ``non-graviton'' cohomologies, a concept we will define momentarily.

A subset of the (classical) $Q$-cohomology classes in the $\mathcal{N}=4$ $SU(N)$ theory have as representatives operators in the chiral primary multiplets $S_{n\leq N}$ \cite{Kinney:2005ej}.
At large $N$, these multiplets were matched with short (Kaluza-Klein) supergravity multiplets on $AdS_5\times S^5$ \cite{Witten:1998qj,Kinney:2005ej,Grant:2008sk,Chang:2013fba}, which shows that at least these operators do not lift from the BPS spectrum at strong coupling.
At finite $N$, we will refer to these cohomology classes somewhat loosely as \emph{graviton operators}.
In the BMN sector, the single trace generators for the $S_n$ multiplet are given by \cite{Choi:2023znd,Choi:2023vdm}
\begin{align}
    &(u_n)^{a_1,\hdots, a_n}=\Tr\lb\bar{\phi}^{(a_1}\cdots \bar{\phi}^{a_n)}\rb, \label{Eq:gravlet1}\\
    &(v_n)^{a_1,\hdots, a_{n-1}}_b= \Tr\lb\bar{\phi}^{(a_1}\cdots \bar{\phi}^{a_{n-1})}\psi_b\rb -{\rm trace},\label{Eq:gravlet2}\\
    &(w_n)^{a_1,\hdots, a_{n-1}}= \Tr\lb f \bar{\phi}^{(a_1}\cdots \bar{\phi}^{a_{n-1})}+\frac 12 \sum_{p=1}^{n-1} \bar{\phi}^{(a_1}\cdots \bar{\phi}^{a_{p-1}}\epsilon^{a_pbc} \psi_b\bar{\phi}^{a_{p+1}}\cdots \bar{\phi}^{a_{n-1})}\psi_c\rb,\label{Eq:gravlet3}
\end{align}
The antisymmetric parts of the scalars in the first line and the traces in the second line, with all possible upper and lower index contractions, have been subtracted out because they are $Q$-exact, as can be seen from \eqref{Eq:SCalgebra}. 
The letters that make up the chiral primaries in \eqref{Eq:gravlet1} can therefore be treated as simultaneously diagonalizable. 
The other generators in \eqref{Eq:gravlet2} and \eqref{Eq:gravlet3} are superconformal descendants.
They are obtained from the chiral primaries by the action of the supercharges $Q_+^a\,, a=1,2,3$ in the commutant of $Q$ 
(see, \textit{e.g.}, the appendix of \cite{Grant:2008sk}). 
In what follows we will refer to these supercharges simply as $Q^a$.

As observed in \cite{Choi:2023vdm}, the (classical) supercharges $Q^a$ act linearly on the BMN letters, which implies that, for the purpose of enumerating cohomologies, all letters appearing in the graviton operators can be treated as diagonal.
This also implies that for $SU(N)$ gauge group, the multiplets $S_{n>N}$ are reducible to the $S_{n\leq N}$ multiplets through trace relations.
Furthermore, graviton trace relations imply that particular polynomial combinations of single trace graviton vanish up to $Q$-exact terms \cite{Choi:2023vdm}.
We will encounter examples of such relations in the following sections.

We can now simply define the \emph{non-graviton operators}, \textit{i.e.}, non-graviton $Q$-cohomology classes, as the set of operators which remain BPS at strong coupling and are not of the above graviton type.
To find explicit representatives of such classes, a key tool is the superconformal index \cite{Kinney:2005ej}.
This index counts $Q$-cohomology classes (with signs) and, crucially, is independent of the Yang--Mills coupling.
It can therefore be exactly evaluated in the free theory, where one knows the full 1/16 BPS spectrum, but can just as well be interpreted in terms of the strongly coupled spectrum.
The idea is then to evaluate this index in the BMN sector, and subtract a ``graviton index'', \textit{i.e.}, the index for the set of (multi-)graviton operators.
A discrepancy between the two provides clues as to the charge combinations at which non-graviton cohomologies arise, which survive at strong coupling \cite{Murthy:2020scj,Agarwal:2020zwm,Chang:2022mjp}.
These results were instrumental in the recent, explicit construction of such operators \cite{Choi:2022caq,Choi:2023znd,Choi:2023vdm}.

\paragraph{BMN index and $SU(3)_R$ singlet projection:}

As noted in \cite{Choi:2023znd}, the BMN index cannot be obtained as a limit from the standard superconformal index.
However, a refinement of the index may be defined at one-loop due to the presence of an emergent  $U(1)_Y$ ``bonus'' symmetry 
\cite{Chang:2013fba,Gadde:2025yoa}.
The charge associated to this symmetry counts the number of letters in an operator and is preserved by the one-loop Hamiltonian.
As such, this refined ``one-loop index'' should strictly be viewed as an object that does not change between the free BPS spectrum and the one-loop BPS spectrum.
Assuming the one-loop exactness of the BPS spectrum, though, this would provide a (signed) count of the BPS spectrum at strong coupling.
However, given footnote \ref{fn:1-loop-conj}, we should consider the possibility that this assumption is incorrect.
Therefore, we interpret the BMN index with some care, as some states may lift from the spectrum upon including further (non-)perturbative corrections in $g_{\text{YM}}$. 
Note that this issue only occurs due to the refinement of the index with the $U(1)_Y$ symmetry; the ordinary index does not have this problem.

Keeping these issues in mind, the $U(1)_Y$ refined index is defined as
\begin{equation}\label{eq:ref-index-defn}
    \mathcal{I}_{\text{1-loop}}(y_a;p,q)=\text{tr}\,(-1)^F p^{E-J_2-Y}q^{E-J_1-Y}y_1^{Y-R_2-R_3}y_2^{Y-R_1-R_3}y_3^{Y-R_1-R_2}\,,
\end{equation}
which turns into the ordinary index for $y_1y_2y_3=pq$.
From Table \ref{tab:n=4-letters}, it should be clear that the BMN sector is captured by the limit $p,q\to 0$ of this index.
The single letter index is then easily written down:
\begin{equation}\label{eq:bmn-single-letter-index}
    i^{SU(N)}_{\text{BMN }} (y_a;U)=\left(y_1y_2y_3-\sum_{a<b}y_ay_b+\sum_a y_a\right)\chi_{\text{adj}}(U)\,,
\end{equation}
with $\chi_{\text{adj}}(U)$ the adjoint $SU(N)$ character.
Plethystic exponentiation of the single letter index together with a gauge singlet projection through integration over $SU(N)$, with the Haar measure, then yields the complete $SU(N)$ BMN index (as discussed in Appendix \ref{app:A}):
\begin{align}\label{eq:bmn-index-full}
    \begin{split}
        \mathcal{I}^{SU(N)}_{\text{BMN }} (y_a)&=\left(\frac{\prod_{a<b}(1-y_ay_b)}{(1-y_1y_2y_3)\prod_{a}(1-y_a)}\right)^{N-1}\\
        &\times\prod^{N-1}_{i=1}\oint_{|s_i|=1} \frac{ds_i}{2\pi i s_i}\prod_{1\leq i\leq j\leq N-1}\left[
         \frac{(1-s_{i,j})\prod_{a<b}(1-y_ay_bs_{i,j}^{\pm})}{(1-y_1y_2y_3s_{i,j}^{\pm})\prod_{a}(1-y_as_{i,j}^{\pm})}\right]\,,
    \end{split}
\end{align}
where, as in Section \ref{Sec:molienweyl}, we use the reduced measure for the gauge integral.
In addition, we use non-standard integration variables $s_i$ with $s_{i,j}=s_i\cdots s_j$, which are related to the $SU(N)$ eigenvalues $u_i$ by\footnote{In Appendix \ref{app:A}, in the context of bosonic multi-matrix models,  a similar change of variables is performed from \eqref{eq:uvar} to \eqref{eq:unr-gauge-sing}. See Appendix A.3 of \cite{deMelloKoch:2025ngs} and Section 2.2 of \cite{vanLeuven:2025gwr}.}
\begin{equation}
    u_i=s_i\cdots s_N\,.
\end{equation}
Finally, we use the notation
\begin{equation}\label{eq:plusminusnotation}
    (1-xs_{i,j}^{\pm})=(1-xs_{i,j})(1-xs_{i,j}^{-1})\,.
\end{equation}
These integrals can be evaluated, for example, through residues \cite{Choi:2023znd,Choi:2023vdm,Gadde:2025yoa}.
The resulting residue sums can be simplified, and the result is a closed form formula which is similar to the Hironaka form of the partition functions studied in Sec.~\ref{Sec:molienweyl}.

The graviton index, on the other hand, has only been evaluated in closed form for $N=2$ \cite{Choi:2023znd}. 
This was done by explicitly constructing the space of states generated by the graviton operators listed above.
For larger values of $N$, this construction has only been performed up to a finite order in the charges using Gr{\"o}bner basis calculations \cite{Choi:2023vdm}.
As a result, the graviton index is also only determined up to some finite order. 

Subtracting the graviton index from the full index yields, by definition, the index for the non-graviton cohomologies, or \emph{non-graviton index}.
This calculation, even if to finite order for $N>2$, allowed the above references to construct explicit representatives of the cohomology classes of non-graviton operators.
A key feature of non-gravitons is their ``fortuity'', that is, they are $Q$-closed at fixed $N$ due to a trace relation specific to that value of $N$, but cease to be $Q$-closed as $N$ increases, since that trace relation no longer holds.
This was proposed as a distinguishing feature of non-graviton cohomologies \cite{Chang:2024zqi}.

In what follows, we study an $SU(3)_R$ singlet projection of the BMN index for $N\leq 4$.
For $N=2,3$ we also study the singlet graviton index and investigate the structure of non-graviton cohomologies. 
The global $SU(3)_R$ singlet projection can be performed similarly to the gauge singlet projection.
In particular, we can express the single letter index in \eqref{eq:bmn-single-letter-index} in terms of $SU(3)_R$ characters as
\begin{equation}
    i^{SU(N)}_{\text{BMN }} (y_a;U,V)=\left(t^6-t^4\chi_{\bar{f}}(V)+t^2\chi_f(V)\right)\chi_{\text{adj}}(U)\,,
\end{equation}
with $|t|<1$ and where $\chi_f(V)=v_1+v_2+(v_1v_2)^{-1}$ is the $SU(3)_R$ character in the fundamental representation, $\chi_{\bar{f}}(V)=\chi_f(V^{-1})$ the character in the anti-fundamental representation, and the contribution from the field strength $f$ is an $SU(3)_R$ singlet.
Note that this rewriting simply amounts to the replacement:
\begin{equation}
    y_1\to t^2v_1\,,\;y_2\to t^2v_2\,,\;y_3\to t^2(v_1v_2)^{-1}\,.
\end{equation}
Comparing for example with \eqref{eq:ref-index-defn}, one can see that the $R_{1,2,3}$ charges of $SU(3)_R$ singlet states satisfy:
\begin{equation}
    R_1=R_2=R_3\,.
\end{equation}
We can then write the $SU(3)_R$ singlet projection through the usual Molien-Weyl formula (see Appendix \ref{app:A}) as an integral over $SU(3)_R$:
\begin{align}\label{Eq:molweylSU3BMN}
\begin{split}
        \mathcal{I}_{[0,0]} (t) =& \prod^{2}_{i=1}\oint \frac{dv_i}{2\pi i v_i}(1-v_1v_2^{-1}) (1-v_{1}^2 v_{2}) (1-v_{1} v_{2}^2) \,\mathcal{I} (t^2 v_{1},t^2 v_{2}, t^2 (v_{1} v_{2})^{-1})\,,
\end{split}
\end{align}
where $\mathcal{I}(y_1,y_2,y_3)$ is the (graviton or BMN) index we wish to project and $[0,0]$ is the Dynkin label for the $SU(3)_R$ singlet representation. 

In what follows, we argue that the global symmetry singlet projection significantly simplifies the construction of the exact graviton index, avoiding the use of numerical Gr{\"o}bner basis calculations. 
This makes the problem of identifying (singlet) non-graviton cohomologies more tractable and sheds light on the structure of their spectrum.

\subsection{\texorpdfstring{$SU(2)$}{SU(2)} gauge theory}

In this section, we study the $SU(3)_R$ singlet projection \eqref{Eq:molweylSU3BMN} for both the BMN and graviton index with the $SU(2)$ gauge group.
To this end, we first collect both indices separately.
The full BMN index is taken from \eqref{eq:bmn-index-full} as
\begin{align}
    \begin{split}
        \mathcal{I}^{SU(2)}_{\text{BMN }} (y_a)&=\frac{\prod_{a<b}(1-y_ay_b)}{(1-y_1y_2y_3)\prod_{a}(1-y_a)}\oint_{|s|=1} \frac{ds}{2\pi i s}
         \frac{(1-s)\prod_{a<b}(1-y_ay_bs^{\pm})}{(1-y_1y_2y_3s^{\pm})\prod_{a}(1-y_as^{\pm})}\,.
    \end{split}
\end{align}
The index can be evaluated through residues by closing the contour inside the unit circle.
Structurally, the expression takes on the form
\begin{equation}
    \mathcal{I}^{SU(2)}_{\text{BMN}}(y_a)=\widetilde{\mathcal{R}}(y_a)+\sum^3_{b=1}\mathcal{R}_b(y_a)\,,
\end{equation}
where $\widetilde{\mathcal{R}}$ corresponds to the residue sum associated with the pole at $s=y_1y_2y_3$ and the $\mathcal{R}_a$ to $s=y_a$.
Explicitly, the functions are given by
\begin{align}
    \begin{split}
        \widetilde{\mathcal{R}}(y_a)&=\frac{y_1y_2y_3(1-y_1^2y_2^2y_3)(1-y_1^2y_2y_3^2)(1-y_1y_2^2y_3^2)}{(1-y_1^2y_2y_3)(1-y_1y_2^2y_3)(1-y_1y_2y_3^2)(1-y_1^2y_2^2y_3^2)}
    \end{split}
\end{align}
and
\begin{align}
    \begin{split}
        \mathcal{R}_1(y_a)&=
    \frac{(1-y_1^2y_2)(1-y_1^2y_3)(1-y_2y_3/y_1)}{(1-y_1^2)(1-y_1^2y_2y_3)(1-y_2/y_1)(1-y_3/y_1)}\,.
    \end{split}
\end{align}
Furthermore, $\mathcal{R}_{2,3}$ are obtained from $\mathcal{R}_1$ through the exchanges $y_1\leftrightarrow y_2$ and $y_1\leftrightarrow y_3$, respectively.
The total expression can be simplified into
\begin{equation}
    \mathcal{I}^{SU(2)}_{\text{BMN}}(y_a)=\frac{P_{23}(y_a)}{(1-y_1^2y_2^2y_3^2)\prod_{b}\left[(1-y_b^2)(1-y_1y_2y_3y_b)\right]}\,,
\end{equation}
where $P_{23}(y_a)$ is a complicated polynomial of total order $23$.

The graviton index in the BMN sector was calculated in \cite[Appendix C]{Choi:2023znd}. The expression is again written in four parts:
\begin{equation}
    \mathcal{I}^{SU(2)}_{\text{BMN grav}}(y_a)=\sum^3_{k=0}I_k(y_a)\,,
\end{equation}
where $k$ labels the number of fermions in the graviton operator.
We have
\begin{align}
    \begin{split}
        I_0(y_a)&=\frac{1+\chi_2+\chi_3(\chi_3-\chi_1\chi_2)+\chi_3^3(\chi_1+\chi_3)}{(1-y_1^2)(1-y_2^2)(1-y_3^2)(1-y_1^2y_2y_3)(1-y_1y_2^2y_3)(1-y_1y_2y_3^2)}\\
        I_1(y_a)&=\chi_3-\chi_2\frac{\chi_1+\chi_3(1-\chi_2-\chi_2^2+\chi_1\chi_3+2\chi_3^2+\chi_2\chi_3^2)}{(1-y_1^2)(1-y_2^2)(1-y_3^2)(1-y_1^2y_2y_3)(1-y_1y_2^2y_3)(1-y_1y_2y_3^2)}\\
        I_2(y_a)&=\chi_1\chi_3\frac{\chi_1^2-\chi_2-\chi_2^2+2\chi_3\chi_1+\chi_3(2\chi_3-\chi_1\chi_2)+\chi_3^3(\chi_1+\chi_3)}{(1-y_1^2)(1-y_2^2)(1-y_3^2)(1-y_1^2y_2y_3)(1-y_1y_2^2y_3)(1-y_1y_2y_3^2)}\\
        I_3(y_a)&=\chi_3\frac{1-\chi_1^2+2\chi_2+\chi_2^2-3\chi_1\chi_3-2\chi_3^2+\chi_3^2(\chi_2+\chi_2^2-\chi_1\chi_3-2\chi_3^2-\chi_2\chi_3^2)}{(1-y_1^2)(1-y_2^2)(1-y_3^2)(1-y_1^2y_2y_3)(1-y_1y_2^2y_3)(1-y_1y_2y_3^2)}\\
        &\quad -\chi_3\,,
    \end{split}
\end{align}
where we defined the symmetric polynomials
\begin{equation}
    \chi_1=y_1+y_2+y_3\,,\quad \chi_2=y_1y_2+y_1y_3+y_2y_3\,,\quad \chi_3=y_1y_2y_3\,.
\end{equation}

\subsubsection{$SU(3)_R$ singlet cohomologies}

As defined above, the non-graviton index is obtained by subtracting the graviton index from the BMN index \cite{Choi:2023znd}. 
Since we know both indices in closed form, we can simply subtract them to obtain
\begin{align}\label{eq:su2-non-gravs}
    \mathcal{I}_{\text{BMN}}^{SU(2)} - \mathcal{I}_{\text{BMN grav}}^{SU(2)}  = - \bigg[ \frac{y_{1}^4 y_{2}^4 y_{3}^4}{1 - y_{1}^2 y_{2}^2 y_{3}^2} \bigg] \bigg[\frac{(1 - y_{1})(1 - y_{2})(1 - y_{3})}{(1 - y_{1}^2 y_{2} y_{3}) (1 - y_{1} y_{2}^2 y_{3}) (1 - y_{1} y_{2} y_{3}^2)} \bigg]\,,
\end{align}
As discussed in \cite{Choi:2023znd,Gadde:2025yoa}, the first term accounts for an infinite tower of ``core black hole cohomologies''.
These are non-trivial $Q$--cohomology classes with representatives
\begin{align}\label{eq:infinitetowerop}
    O_{n} = \tr (f^2)^{n} O_{0} + n \tr (f^2)^{n-1} \tr (f \xi) + \frac{2n^2 + n}{3} \tr (f^2)^{n-1} \chi\,,
\end{align}
where $n \in \mathbb{Z}_{\geq 0}$. 
The threshold cohomology class arises at $n=0$ with representative \cite{Choi:2022caq,Choi:2023znd}
\begin{equation}\label{Eq:24threshold}
    O_{0} = \epsilon^{p_{1} p_{2} p_{3}} \tr (\phi^{m} \psi_{p_{1}}) \tr (\phi^{n} \psi_{p_{2}}) \tr (\psi_{m} \psi_{n} \psi_{p_{3}}) \, ,
\end{equation}
which corresponds to the factor $-y_{1}^4 y_{2}^4 y_{3}^4$ in the numerator. 
For $n > 0$, black hole cohomologies can be identified with the tower generated by the $(1 - y_{1}^2 y_{2}^2 y_{3}^2)$ denominator. Furthermore, $\xi$ and $\chi$ are defined as
\begin{align}
    \xi &= \epsilon^{b_{1} b_{2} b_{3}} \epsilon^{c_{1} c_{2} c_{3}} \psi_{b_{1}} \tr (\phi^{a} \psi_{c_{1}}) \tr (\phi_{b_{2}} \psi_{c_{2}}) \tr (\psi_{a} \psi_{b_{3}} \psi_{c_{3}}),  \\
    \chi &= - \frac{1}{72} \epsilon^{a_{1} a_{2} a_{3}} \epsilon^{b_{1} b_{2} b_{3}} \epsilon^{c_{1} c_{2} c_{3}} \tr (\psi_{a_{1}} \psi_{b_{1}} \psi_{c_{1}}) \tr (\psi_{a_{2}} \psi_{b_{2}} \psi_{c_{2}}) \tr (\psi_{a_{3}} \psi_{b_{3}} \psi_{c_{3}}), 
\end{align}
where we note that $\xi$ transforms in the adjoint of the $SU(2)$ gauge group and is a singlet under $SU(3)_{R}$, whereas $\chi$ is a singlet under both groups.
In particular, each operator in this tower is a singlet under $SU(3)_R$. 

The numerator of the second term in \eqref{eq:su2-non-gravs} incorporates contributions from the superconformal descendents of the core black holes obtained by acting with $Q^{1,2,3}$, where we recall from Table \ref{tab:n=4-letters} that $Q^a$ has index $y_a$.
Finally, the denominator of the second term contains dressings of the core black holes by the $w_{2}$ graviton operators defined in \eqref{Eq:gravlet3}. 
It was found that only this restricted set of graviton operators lead to non-trivial cohomology classes, while all the other dressings correspond to $Q$-exact operators \cite{Choi:2023znd}.
For convenience, we also collect the expression for the unrefined index, setting $y_{i} = t^2$:
\begin{align}\label{eq:indexsubtraction}
    \mathcal{I}_{\text{BMN}}^{SU(2)} (t) - \mathcal{I}_{\text{BMN grav}}^{SU(2)} (t) = -\frac{t^{24}}{1 - t^{12}} \frac{(1 - t^2)^3}{(1 - t^{8})^3}\,.
\end{align}

We now consider the $SU(3)_R$ singlet projection \eqref{Eq:molweylSU3BMN}. 
For the BMN index, we find a substantially simpler expression
\begin{align}
    \mathcal{I}_{\text{BMN } [0,0]}^{SU(2)} (t) = \frac{1 - t^{12} - t^{24} + t^{30} - t^{36}}{1 - t^{12}}\,.
\end{align}
Even simpler, we find that the graviton index completely trivializes, with
\begin{align}\label{eq:su2-bmn-grav-ind}
    \mathcal{I}^{SU(2)}_{\text{BMN grav } [0,0]} (t) = 1\,.
\end{align}
Therefore, either there are no $SU(3)_R$ singlet graviton operators in the $SU(2)$ BMN sector, or they cancel perfectly. 
Below, we will prove the former, stronger statement. 

So, after a (trivial) subtraction we are left with the $SU(3)_R$ singlet non-graviton index
\begin{align}\label{eq:singletsubtractionsu2}
    \mathcal{I}_{\text{BMN } [0,0]}^{SU(2)} (t) - \mathcal{I}^{SU(2)}_{\text{BMN grav } [0,0]} (t) = \frac{-t^{24}+t^{30}-t^{36}}{1 - t^{12}}\,.
\end{align}
In this expression we can identify the first term as the core black hole cohomologies given in \eqref{eq:infinitetowerop}, which were indeed noted to be $SU(3)_R$ singlets.
The second term corresponds to a tower of superconformal descendents of the core black holes 
\begin{align}\label{eq:singlet-sc-descendant}
    Q^{a} Q^{b} Q^{c} \epsilon_{abc} O_{n}\,.
\end{align}
The operators in this tower are $Q$-closed, but not $Q$-exact, simply because they correspond to $SU(3)_R$ invariant linear combinations of the corresponding non-trivial cohomology classes contributing to the unprojected index. 
The last term in the numerator of \eqref{eq:singletsubtractionsu2}, together with denominator, can be identified with the following tower of $SU(3)_R$ singlets
\begin{align}
    \epsilon_{abc} (Q^{a} Q^{b} O_{n}) w_{2}^{c}\,.
\end{align}
This operator, which can be viewed as a graviton dressing of the non-graviton tower, was shown to be $Q$-closed, but not $Q$-exact in \cite{Choi:2023znd}. 
All in all, these operators fully account for the non-graviton index in \eqref{eq:singletsubtractionsu2}.

\subsubsection{Absence of singlet gravitons}

In order to explain the trivialization of the graviton index for the $SU(2)$ theory in the $SU(3)_R$ singlet sector, we consider various (multi-trace) $SU(3)_R$ singlets that can be made from the single-trace graviton generators \eqref{Eq:gravlet1}, \eqref{Eq:gravlet2} and \eqref{Eq:gravlet3}, and argue that all of them vanish. 

The ring of graviton operators is generated by single trace gravitons in the $S_2$ multiplet.
Recall that we can treat all matrices in the graviton generators as simultaneously diagonalizable up to $Q$-exact terms, which for the purpose of enumerating cohomologies suffices. 
For this reason, we can also think about these operators in terms of Weyl invariant polynomials of the eigenvalues.  We denote the eigenvalues of the scalars by $x^a$, the fermions by $\psi_b$ and the field strength by $f$.
Since the Weyl symmetry is simply $\mathbb{Z}_2$ for an $SU(2)$ gauge theory, and acts on the eigenvalues as
\begin{equation}
    x^a, \psi_b, f \to -x^a, -\psi_b, -f, \quad \forall\, a,b \in \{1,2,3\}\,,
\end{equation}
we have to consider even polynomials.
The generators of the graviton sector can be written in terms of the eigenvalues as
\begin{align}
    &x^ax^b, \quad  x^a \psi_b- \frac 13 \delta^a_b x^c \psi_c,\quad f x^a + \frac12\epsilon^{abc}\psi_b\psi_c\,.
\end{align}
The $SU(3)_R$ singlets are polynomials in the above generators where all indices are contracted. 
We can perform two possible contractions: either an (upper) fundamental index is contracted with a (lower) anti-fundamental index, or we may contract three indices with the Levi-Civita tensors $\epsilon_{abc}$ or $\epsilon^{abc}$.
Below, we examine the various $SU(3)_R$ singlets that can be made from the above generators, graded by the number of fermions that appear in the singlet. We note that since there are three flavours of fermions, and thus only three fermionic eigenvalues, any monomial with four or more fermions vanishes.

\paragraph{Zero--fermion sector:} In this sector, the indices have to be contracted with $\epsilon_{abc}$. However, the bosonic eigenvalues $x^a$ commute and therefore any purely bosonic $SU(3)_R$ invariant of the form $x^a x^bx^c \epsilon_{abc}$ vanishes.\footnote{As explained before, whenever an invariant vanishes when expressed using diagonal matrices, it means that the corresponding operator is $Q$-exact \cite{Choi:2023znd,Choi:2023vdm}.} 
It thereby follows that there are no $SU(3)_R$ singlets in the zero--fermion sector.

\paragraph{One--fermion sector:} Gauge invariant operators in this sector are of the form
\begin{equation}
    x^{a_1} \cdots x^{a_n} fx^{a_{n+1}} \cdots fx^{a_m}(x^{b_1}\psi_c-\frac13\delta^{b_1}_c x^{b_2}\psi_{b_2})\,,
\end{equation}
with $n$ even.
The only way to construct a singlet is to contract one of the $a_{1},\ldots, a_m$ indices with the fermion $c$ index and then contract the remaining bosonic indices with a sufficient number of $\epsilon_{abc}$. Any such invariant must involve at least one contraction of the form $x^{a_1}x^{a_2}x^{a_3} \epsilon_{a_1a_2a_3}$, which vanishes. Therefore, the $SU(3)_R$ singlets in the one--fermion sector also vanish.

\paragraph{Two--fermion sector:} Gauge invariant operators in this sector are of two types
\begin{align}
    \begin{split}
         &x^{a_1} \cdots x^{a_n} fx^{a_{n+1}} \cdots fx^{a_m}(fx^a +\frac12 \epsilon^{abc}\psi_b\psi_c)\\
         &x^{a_1} \cdots x^{a_n} fx^{a_{n+1}} \cdots fx^{a_m}(x^{b_1} \psi_{c_1}-\frac13 \delta^{b_1}_{c_1} x^{b_2}\psi_{b_2})(x^{b_3} \psi_{c_2}-\frac13\delta^{b_3}_{c_2} x^{c_3}\psi_{c_3})\,,
    \end{split}
\end{align}
where $n$ is even and only the fermionic part of one $(w_2)^a$ operator is relevant.
There are two ways to construct an $SU(3)_R$ invariant from the first type.
One is given by
\begin{equation}
    x^{a_1}x^{a_2}(fx^a +\frac12 \epsilon^{abc}\psi_b\psi_c)\epsilon_{a_1a_2a},
\end{equation}
which vanishes, since the bosonic eigenvalues commute.
The only other way to obtain a singlet of the first type is through
\begin{equation}
   \epsilon_{abc}fx^afx^b(fx^c +\frac12 \epsilon^{ca_1a_2}\psi_{a_1}\psi_{a_2})\,,
\end{equation}
which again clearly vanishes.

The minimal $SU(3)_R$ invariant of the second type is
\begin{equation}
    (x^{a_1} \psi_{b_1}-\frac13 \delta^{a_1}_{b_1} x^{c_1}\psi_{c_1})(x^{a_2} \psi_{b_2}-\frac13\delta^{a_2}_{b_2} x^{c_2}\psi_{c_2}) \delta^{b_2}_{a_1} \delta^{b_1}_{a_2}.
\end{equation}
This is just the trace of the square of a fermionic matrix $(x^a \psi_b-\frac13 \delta^a_b x^c\psi_c)$, which vanishes due to fermion statistics.
The only other $SU(3)_R$ invariant of the second type is
\begin{equation}
    fx^{a_1} fx^{a_2} fx^{a_3}(x^{a_4} \psi_{b_1}-\frac13 \delta^{a_4}_{b_1} x^{c_1}\psi_{c_1})(x^{a_5} \psi_{b_2}-\frac13\delta^{a_5}_{b_2} x^{c_2}\psi_{c_2}) \delta^{b_1}_{a_1}\delta^{b_2}_{a_2}\epsilon_{a_3a_4a_5}\,,
\end{equation}
which can be seen to vanish due to the anti-symmetrization of bosonic fundamental indices.
We thus find no $SU(3)_R$ singlet gravitons in the two--fermion sector.

\paragraph{Three--fermion sector:} Gauge invariant operators in this sector are again of two types
\begin{align}
    \begin{split}
        &x^{a_1} \cdots x^{a_n} fx^{a_{n+1}} \cdots fx^{a_m}(x^{c_1} \psi_{b_1}-\frac13 \delta^{c_1}_{b_1} x^{c_2}\psi_{c_2})(fx^{c_3} +\frac12 \epsilon^{c_3c_4c_5}\psi_{c_4}\psi_{c_5}) \\
        &(x^{a_1} \psi_{b_1}-\frac13 \delta^{a_1}_{b_1} x^{c_1}\psi_{c_1})(x^{a_2} \psi_{b_2}-\frac13 \delta^{a_2}_{b_2} x^{c_2}\psi_{c_2})(x^{a_3} \psi_{b_3}-\frac13 \delta^{a_3}_{b_3} x^{c_3}\psi_{c_3})
    \end{split}
\end{align}
where $n$ is even and again only the fermionic part of one $(w_2)^a$ operator is relevant. 
We also note that adding any other generators to the second line would necessarily involve anti-symmetrization of bosonic indices, and therefore vanishes.

On the first line, we can ignore the $fx^a$ terms without loss of generality.
There are then two possible $SU(3)_R$ invariant structures. 
The first such structure is
\begin{equation}
    x^{a_1}x^{a_2}(x^{a_3} \psi_{b_1}-\frac13 \delta^{a_3}_{b_1} x^{c_1}\psi_{c_1})(fx^{a_4} +\frac12 \epsilon^{a_4c_2c_3}\psi_{c_2}\psi_{c_3}) \delta^{b_1}_{a_4} \epsilon_{a_1 a_2 a_3},
\end{equation}
which is zero due to anti-symmetrization of the $a_1, a_2$. The other structure is
\begin{equation}
    x^{a_1}x^{a_2}(x^{a_3} \psi_{b}-\frac13 \delta^{a_3}_b x^{c_1}\psi_{c_1})(fx^{a_4} +\frac12 \epsilon^{a_4c_2c_3}\psi_{c_2}\psi_{c_3}) \delta^b_{a_2} \epsilon_{a_1a_3a_4}.
\end{equation}
The only potentially non-vanishing term in this operator is given by
\begin{equation}
    -\frac{1}{6}x^{a_1}x^{a_2}\delta^{a_3}_b x^{c_1}\psi_{c_1}\epsilon^{a_4c_2c_3}\psi_{c_2}\psi_{c_3}\delta^{b}_{a_2} \epsilon_{a_1a_3a_4} = -\frac{1}{6}x^{a_1}x^{a_2}x^{c_1}\psi_{c_1}\epsilon^{a_4c_2c_3}\psi_{c_2}\psi_{c_3} \epsilon_{a_1a_2a_4},
\end{equation}
which vanishes due to anti-symmetrization of $a_1$, $a_2$.

For operators of the second type, the only non-vanishing contraction structure is
\begin{equation}
    (x^{a_1} \psi_{b_1}-\frac13 \delta^{a_1}_{b_1} x^{c_1}\psi_{c_1})(x^{a_2} \psi_{b_2}-\frac13 \delta^{a_2}_{b_2} x^{c_2}\psi_{c_2})(x^{a_3} \psi_{b_3}-\frac13 \delta^{a_3}_{b_3} x^{c_3}\psi_{c_3})\delta^{b_1}_{a_2} \delta^{b_2}_{a_3} \delta^{b_3}_{a_1}\,,
\end{equation}
since any other contraction will involve $\epsilon$ contractions of bosonic indices.
Any term in this product is of the form
\begin{equation}
    x^ax^bx^c \psi_a \psi_b \psi_c,
\end{equation}
which vanishes due to fermion statistics.
We thus see that there are also no $SU(3)_R$ singlets in the three--fermion sector.

All in all, this proves that there are no $SU(3)_R$ singlet graviton operators in the $SU(2)$ BMN sector of the $\mathcal{N}=4$ theory, as suggested by the trivial graviton index \eqref{eq:su2-bmn-grav-ind}.

\subsection{\texorpdfstring{$SU(3)$}{SU(3)} gauge theory}\label{Sec:SU3BMN}

In this section, we study the $SU(3)_R$ singlet projection \eqref{Eq:molweylSU3BMN} for the BMN sector with $SU(3)$ gauge group.
Unlike in the $SU(2)$ case, there is no closed form expression for the graviton index available.
This is essentially due to the complexity of trace relations, see \cite{Choi:2023vdm} for a detailed discussion and a construction of the index up to a finite order.
As we will demonstrate, the structure of the ring of $SU(3)_R$ singlet gravitons only consists of two non-trivial operators.
This leads to a simple closed form formula for the $SU(3)_R$ singlet graviton index.
Since we can also determine the full BMN index in the singlet sector, this enables us to find a closed form expression for the index of singlet non--gravitons.

The full $SU(3)$ BMN index is taken from \eqref{eq:bmn-index-full} and reads
\begin{align}\label{eq:BMNsu3}
    \begin{split}
        \mathcal{I}^{SU(N)}_{\text{BMN }} (y_a)&=\left(\frac{\prod_{a<b}(1-y_ay_b)}{(1-y_1y_2y_3)\prod_{a}(1-y_a)}\right)^{2}\\
        &\times\prod^{2}_{i=1}\oint_{|s_i|=1} \frac{ds_i}{2\pi i s_i}\prod_{1\leq i\leq j\leq 2}\left[
         \frac{(1-s_{i,j})\prod_{a<b}(1-y_ay_bs_{i,j}^{\pm})}{(1-y_1y_2y_3s_{i,j}^{\pm})\prod_{a}(1-y_as_{i,j}^{\pm})}\right]\,,
    \end{split}
\end{align}
which can be evaluated through residues. 
The fully refined expression, which we require to perform the singlet projection, is very long and we avoid writing it here.

Performing the $SU(3)_R$ singlet projection of \eqref{eq:BMNsu3}, using \eqref{Eq:molweylSU3BMN}, we obtain the simpler expression
\begin{align}\label{Eq:BMNfullSU3proj}
    \begin{split}
        &\mathcal{I}^{SU(3)}_{\text{BMN }[0,0]}(t)= \frac{P_{114}(t)}{(1-t^{12})(1-t^{18})},\\
    &P_{114}(t)=1-t^{12}-2 t^{18}+2 t^{30}-2 t^{42}+4 t^{54}-t^{60}-2 t^{66} \\ & \qquad \qquad \quad +t^{78}+t^{84}-t^{90}+3 t^{96}-t^{102}-2 t^{108}+t^{114}  \, .
    \end{split}
\end{align}
In the free theory, we interpret the order $12$ term in the denominator as resulting from the \textit{primary} invariant $\tr (f^2)$. 
We can similarly interpret the order $18$ term in the denominator as another primary invariant $\tr (f^3)$. 
In the interacting theory, these operators are not $Q$-closed since the letter $f$ is not. 
But, we expect that the denominators \emph{can} be interpreted as generating towers of non-graviton cohomologies, similar to the $SU(2)$ tower described in \eqref{eq:infinitetowerop}. 
In particular, these towers should involve the action of arbitrary powers of $\tr (f^2)$ and $\tr (f^3)$, respectively, on a given black hole cohomology in the numerator, alongside additional terms to ensure that the full operator is $Q$-closed.

As in the $SU(2)$ case, we would like to write down a closed form expression for the non-graviton index:
\begin{equation}\label{eq:diff-ind-grav-int-su3}
    \mathcal{I}^{SU(3)}_{\text{BMN }[0,0]}(t)-\mathcal{I}^{SU(3)}_{\text{BMN grav }[0,0]}(t)\,.
\end{equation}
As mentioned above, the graviton index has so far only been constructed up to order $\mathcal{O}(t^{54})$ \cite{Choi:2023vdm}.
We will now determine a closed form expression for the graviton index in the $SU(3)_R$ singlet sector.

\subsubsection{Singlet gravitons}\label{Sec:SU3gravproof}

The $SU(3)_R$ singlet gravitons can be constructed from the graviton generators \eqref{Eq:gravlet1}, \eqref{Eq:gravlet2} and \eqref{Eq:gravlet3}, in the $S_2$ and $S_3$ multiplets:
\begin{align}
    &(u_2)^{a_1a_2}=\tr\lb\bar{\phi}^{(a_1}\bar{\phi}^{a_2)}\rb, \quad (u_3)^{a_1a_2a_3}=\tr\lb\bar{\phi}^{(a_1}\bar{\phi}^{a_2}\bar{\phi}^{a_3)}\rb,\\
    &(v_2)^{a}_b= \tr\lb\bar{\phi}^{a}\psi_b\rb -\frac13 \delta^a_b \tr\lb\bar{\phi}^c\psi_c\rb,\\ &(v_3)^{a_1a_2}_b= \tr\lb\bar{\phi}^{(a_1}\bar{\phi}^{a_2)}\psi_b\rb -\frac14 \delta^{a_1}_b \tr\lb\bar{\phi}^{(a_2}\bar{\phi^{c)}}\psi_c\rb-\frac14 \delta^{a_2}_b \tr\lb\bar{\phi}^{(a_1}\bar{\phi^{c)}}\psi_c\rb, \\
    &(w_2)^{a}= \tr\lb f \bar{\phi}^{a}+\frac 12 \epsilon^{abc} \psi_b\psi_c\rb, \quad (w_3)^{a_1a_2}=\tr\lb f\bar{\phi}^{(a_1}\bar{\phi}^{a_2)}+\epsilon^{bc(a_1}\bar{\phi}^{a_2)}\psi_b\psi_c\rb.
\end{align}
As mentioned before, we may treat all matrices $\bar{\phi}^a, \psi_b$ and $f$ to be diagonal.
These matrices thus have the generic form
\begin{equation}
M=
    \begin{pmatrix}
        m_1&0&0\\
        0&m_2&0\\
        0&0&-m_1-m_2
    \end{pmatrix},
\end{equation}
where $M$ can be any of the letters $\bar{\phi}^a, \psi_b,f$ and contains two independent eigenvalues.

Using the above set of generators, we can explicitly enumerate the $SU(3)_R$ singlet gravitons at any given order. 
Note that each scalar $\bar{\phi}^a$ contributes at order $t^2$, each fermion $\psi_b$ at order $t^4$ and $f$ at order $t^6$.
We then use the diagonal forms of the matrices to explicitly evaluate the singlet gravitons in terms of the eigenvalues. 
The details of this procedure are described in Appendix~\ref{App:gravsinglets}.

Let us summarize the main points.
The first non-vanishing singlet graviton cohomology appears at order $t^{18}$ and is given by
\begin{equation}\label{Eq:t18SU3}
    (v_2)^a_b(v_2)^b_c(v_2)^c_a ,
\end{equation}
Since this invariant involves an odd number of fermions, it contributes $-t^{18}$ to the graviton index.
The second non-vanishing singlet arises at order $t^{24}$ and can be expressed as
\begin{equation}\label{Eq:t24SU3}
     \epsilon_{a_1a_3b_1}\epsilon_{a_2a_4b_2}(u_2)^{a_1a_2}(u_2)^{a_3a_4}(w_2)^{b_1}(w_2)^{b_2}.
\end{equation}
This operator has even fermion number and contributes $t^{24}$ to the index.
Moreover, plugging in the explicit expression for $(w_2)^b$, we find that only the four-fermion term in this operator is non-vanishing:
\begin{align}
    \begin{split}
        \epsilon_{a_1a_3b_1}\epsilon_{a_2a_4b_2}(u_2)^{a_1a_2}(u_2)^{a_3a_4}(w_2)^{b_1}(w_2)^{b_2}=&\frac{1}{4}\epsilon_{a_1a_3b_1}\epsilon_{a_2a_4b_2}(u_2)^{a_1a_2}(u_2)^{a_3a_4}\\& \quad \times\tr(\epsilon^{b_1c_1c_2}\psi_{c_1}\psi_{c_2})\tr(\epsilon^{b_2c_3c_4}\psi_{c_3}\psi_{c_4}).
    \end{split}
\end{align}
Since the $SU(3)$ theory has only six independent fermionic eigenvalues, this implies that both the product of \eqref{Eq:t18SU3} and \eqref{Eq:t24SU3}, and higher powers vanish due to fermion statistics.
 
Let us note an interesting feature of the singlet graviton in \eqref{Eq:t24SU3}. 
Naively, this operator could have also been written down as a singlet graviton for the $SU(2)$ theory.
However, it turns out that in this case it corresponds to a trace relation \cite{Choi:2023znd}\footnote{When the matrices in the graviton letters are taken to be diagonal, this invariant vanishes.}
\begin{equation}
    \epsilon_{a_1a_3b_1}\epsilon_{a_2a_4b_2}(u_2)^{a_1a_2}(u_2)^{a_3a_4} = Q\textrm{-exact}\,,
\end{equation}
which explains why we did not encounter it there.
For the $SU(3)$ theory, this trace relation no longer holds and \eqref{Eq:t24SU3} represents a non-trivial cohomology class. 
In the next section, we will encounter a new graviton singlet cohomology for the $SU(4)$ theory, which is $Q$-exact in the $SU(3)$ theory due to a trace relation. 

We will now show that no more singlet gravitons arise at higher order.
In order to explain this, we first show that invariants involving an $\epsilon$ contraction of three bosonic (scalar) eigenvalues vanish. Note that this does not trivially follow from their bosonic nature, since there are a total of six eigenvalues for the $SU(3)$ theory. 
For instance, the $t^{24}$ operator \eqref{Eq:t24SU3} involves an $\epsilon$ contraction of two bosonic eigenvalues, but is non-vanishing since it involves terms of the form $x_1^2 x_2^3-x_1^3x_2^2 \neq0$, where $x^a_i$ is the $i^{\text{th}}$ eigenvalue of $\bar{\phi}^a$. 
To show that any $SU(3)_R$ singlet involving an $\epsilon$ contraction of three bosonic eigenvalues vanishes, it is useful to first relax the tracelessness condition of the scalar matrices. 
Any invariant of this type takes the form
\begin{equation}
    \epsilon_{abc}x_{i}^ax_j^b x^c_k \times (\text{other bosons and fermions}),
\end{equation}
where $i,j,k\in{1,2,3}$ label the \emph{three} eigenvalues of the matrices. 
For each eigenvalue, we introduce a vector $\vec{x}_{1,2,3}$, whose three components are labelled by the upper $SU(3)_R$ index. The contraction $\epsilon_{abc}x_{i}^ax_j^b x^c_k$ is then nothing but the box product of three vectors $\vec{x}_i$, $\vec{x}_j$ and $\vec{x}_k$. 
The condition that the box product is non-vanishing requires that the $i,j,k$ are distinct.
Therefore, the only non-vanishing component is $\epsilon_{abc}x_{1}^ax_2^b x^c_3$. 
However, the bosonic matrices are traceless, which implies that $\vec{x}_1+\vec{x}_2+\vec{x}_3=0$. 
But, this implies that the vectors are not linearly independent and consequently that the box product vanishes.
This proves our claim that any $SU(3)_R$ singlet involving an $\epsilon$ contraction of three bosonic eigenvalues vanishes. 
This also shows that there are no purely bosonic singlet gravitons for the $SU(3)$ theory, since those singlets must be contracted with $\epsilon$ tensors. 
It follows that the singlet graviton index for the $SU(3)$ theory should truncate at some finite order in $t$. 

We will now argue that this truncation occurs at order $t^{54}$ and there are no $SU(3)$ singlet gravitons between order $24$ and $54$. 
First, we note that any singlets constructed from more than three fermionic matrices $v_2$ will vanish. 
To see this, we note that $v_2$ is traceless and the trace of any even power of $v_2$ vanishes due to Grassmann statistics. 
Furthermore, $\tr(v_2^{5})=0$ due to trace relations\footnote{The $T_3(A,B,C,D)$ trace relation \cite[Appendix C.1]{deMelloKoch:2025ngs} combined with the vanishing of traces of even powers of $v_2$, and the tracelessness of $v_2$ can be used to show that $\tr(v_2^{2n+1})=0$ for $n\geq 2$.} and higher odd powers vanish due to fermion statistics.
Invariants, between orders $24$ and $54$, made solely from combinations of $u_2,u_3,v_2$ and $v_3$ vanish, since they necessarily involve $\epsilon$ contractions of three bosonic eigenvalues, which vanish by our argument above. 

We conclude that the non-vanishing singlets at orders greater than $24$ must involve $w_2$ or $w_3$. 
By explicitly enumerating all the singlet gravitons at any given order, using the vanishing of an $\epsilon$ contraction of three bosonic eigenvalues, and the identity\footnote{This identity follows from the fact that $w_2$ bosonic.}
\begin{equation}\label{Eq:w2w2epsilon}
    \epsilon_{abc}w_2^b w_2^c=0,
\end{equation} 
we find that there are no singlet gravitons beyond $t^{54}$. 
See Appendix~\ref{App:gravsinglets} for more details. 
We note, but cannot explain, the coincidence that the Gr{\"o}bner basis calculation of \cite{Choi:2023vdm} was performed up to order $t^{54}$ for the full set of graviton operators.

In Appendix~\ref{App:gravsinglets}, we also describe how to construct potential singlet gravitons between orders $t^{24}$ and $t^{54}$.
As it turns out, all potential candidates vanish.
The procedure we use is brute force, but tractable for the present $SU(3)$ case. 
However, for the generalization to higher rank gauge groups one would require a more efficient method. 
We believe that the vanishing of \emph{all} singlet gravitons between orders $24$ and $54$ does in fact suggest that there should exist an easier argument for their vanishing, which may well generalize to higher rank gauge groups.
We leave this to future work.

We thus find, by explicit construction, the exceedingly simple form for the singlet graviton index 
\begin{equation}
    \mathcal{I}^{SU(3)}_{\text{BMN grav }[0,0]}(t)=1-t^{18}+t^{24}\,.
\end{equation}
This also leads to a closed-form expression for the non-graviton index: 
\begin{align}\label{Eq:nongravSU3}
    &\mathcal{I}^{SU(3)}_{\text{BMN }[0,0]}(t)-\mathcal{I}^{SU(3)}_{\text{BMN grav }[0,0]}(t)= \frac{-t^{24} P_{90}(t)}{\left(1-t^{12}\right) \left(1-t^{18}\right)},\\
    &P_{90}(t)=\left(1+t^{18}-t^{24}-3 t^{30}+t^{36}+2 t^{42}-t^{54}-t^{60}+t^{66}-3 t^{72}+t^{78}+2 t^{84}-t^{90}\right) \nonumber \, .
\end{align}
One may compare this result to the $SU(2)$ case studied in \eqref{eq:singletsubtractionsu2}.
In particular, we note that the two infinite towers, which also appeared in the singlet projection of the full index \eqref{Eq:BMNfullSU3proj}, remain.
As commented there, we expect these towers to be generated in the free theory by $\text{tr}(f^2)$ and $\text{tr}(f^3)$.
These operators are clearly $SU(3)_R$ singlets and moreover are not part of the $S_n$ multiplets.
It makes sense, therefore, that their contributions remain in the non-graviton index.
In the interacting theory, corrections to these operators are required for them to remain $Q$-closed.
We also observe that terms in the numerator are separated by a power of $t^6$.
This is similar to the $SU(2)$ case, where we noted in \eqref{eq:singlet-sc-descendant} that the action of $\epsilon_{abc}Q^aQ^bQ^c$ creates a singlet descendant for a singlet cohomology, and has index $t^6$ (see Table \ref{tab:n=4-letters}).
Finally, we note that our result is consistent with the finite order result of \cite{Choi:2023vdm} when projected onto singlets.
In particular, these authors construct an explicit representative of the threshold cohomology, which corresponds to a fermionic $SU(3)_R$ singlet at order $t^{24}$.
Clearly, our non-graviton index reflects this cohomology class as its first term when expanded in $t$.

\subsection{\texorpdfstring{$SU(4)$}{SU(4)} gauge theory}\label{Sec:SU4BMN}

In this section, we study the $SU(3)_R$ singlet projection \eqref{Eq:molweylSU3BMN} for the $SU(4)$ BMN index.
Again, there is no closed form expression for the full graviton index available, but, through explicit construction of graviton operators to a finite order, \cite{Choi:2023vdm} have been able to evaluate the first few terms of the expansion.

The full $SU(4)$ BMN index is again taken from \eqref{eq:bmn-index-full} and reads
\begin{align}\label{eq:BMNsu4}
    \begin{split}
        \mathcal{I}^{SU(N)}_{\text{BMN }} (y_a)&=\left(\frac{\prod_{a<b}(1-y_ay_b)}{(1-y_1y_2y_3)\prod_{a}(1-y_a)}\right)^{3}\\
        &\times\prod^{3}_{i=1}\oint_{|s_i|=1} \frac{ds_i}{2\pi i s_i}\prod_{1\leq i\leq j\leq 3}\left[
         \frac{(1-s_{i,j})\prod_{a<b}(1-y_ay_bs_{i,j}^{\pm})}{(1-y_1y_2y_3s_{i,j}^{\pm})\prod_{a}(1-y_as_{i,j}^{\pm})}\right]\,.
    \end{split}
\end{align}
The index can again be evaluated using residues by closing the contours inside the unit circles. 
The $SU(3)_R$ singlet projection can be performed using \eqref{Eq:molweylSU3BMN}.
We then obtain the following formula
\begin{align}
  \mathcal{I}^{SU(4)}_{{\rm BMN}\, [0,0]}(t) =& \frac{\left(1-t^{12}\right)^2 }{\left(1-t^{18}\right) \left(1-t^{24}\right) \left(1-t^{36}\right) \left(1-t^{48}\right) \left(1-t^{60}\right)} P_{300}(t),\\
  P_{300}(t) =&
  1+3 t^{12}-2 t^{18}+4 t^{24}-7 t^{30}+3 t^{36}-10 t^{42}+4 t^{48}- 8 t^{54} +\hdots\,,
\end{align}
where $P_{300}$ is a polynomial of order $\mathcal{O}(t^{300})$, which can be found in Appendix \ref{App:SU4P300}.
Let us make some comments on this expression.
A first thing to note that is that the denominator contains considerably more factors than the $SU(2)$ (1 factor) and $SU(3)$ (2 factors) cases.
Let us include an additional factor of $(1-t^{12})$ in both denominator and numerator, and absorb the factors in the numerator into $P_{300}$ so that $3t^{12}$ term is removed. 
The factors $(1-t^{12})$, $(1-t^{18})$ and $(1-t^{24})$ are likely to correspond, in the free theory, to towers of the $SU(3)_R$ singlets $\text{tr}(f^2)$, $\text{tr}(f^3)$ and $\text{tr}(f^4)$.
This generalizes the denominators seen in the $SU(2)$ and $SU(3)$ cases.
In addition, these towers are expected to stay in the non-graviton index, given that these operators do not correspond to gravitons.

The origin of the other factors is less clear.
Since this expression represents the $SU(3)_R$ singlet sector of the full BMN sector, some of these towers may correspond to singlet gravitons. 
Indeed, we will find an example of a purely bosonic $SU(3)_R$ singlet graviton below, which did not exist in the $SU(2)$ and $SU(3)$ BMN sectors.
We also note that the relatively high powers for $t$ for these primary generators is likely related to observations in the first part of this work.
More specifically, we observed that multi-trace operators in the original matrix model can arise as primary generators in global symmetry singlet sectors.
This results in the relatively high order of these primary generators.

Let us now compare our expression to known results in the literature.
To this end, we first series expand our expression as
\begin{equation}\label{Eq:BMNSU4proj}
    \mathcal{I}^{SU(4)}_{{\rm BMN} \, [0,0]}(t) = 1+t^{12}-t^{18}-2 t^{30}+\mathcal{O}(t^{36})\,.
\end{equation}
In \cite{Choi:2023vdm}, using a Gr{\"o}bner basis calculation, the authors find an expression for the (refined) non-graviton index up to order $t^{30}$:
\begin{align}
    \begin{split}
        \mathcal{I}^{SU(4)}_{{\rm BMN}}-\mathcal{I}^{SU(4)}_{\text{BMN grav}}=& \lb-\chi_{[2,0]}(v_1,v_2)t^{28}-\chi_{[3,0]}(v_1,v_2)t^{30}+\mathcal{O}(t^{32})\rb\\
        &\times (1-t^2v_1)(1-t^2v_2)(1-t^2(v_1v_2)^{-1})\,,
    \end{split}
\end{align}
where $\chi_{[k,l]}$ is the $SU(3)_R$ character in the representation with Dynkin labels $[k,l]$ (see, \textit{e.g.}, Appendix C.4 \cite{Kinney:2005ej} for a general formula).
Furthermore, the factor on the first line can be interpreted as containing the core non-graviton cohomologies, while the remaining factors account for their superconformal descendants. 
It follows that there are no $SU(3)_R$ singlet non-gravitons, at least as seen by the index, up to order $t^{30}$. 
Therefore, the singlets at orders $t^{12}$, $t^{18}$, and $t^{30}$, appearing in the singlet index \eqref{Eq:BMNSU4proj}, must be gravitons.

These singlet gravitons can be constructed from the generators \eqref{Eq:gravlet1}, \eqref{Eq:gravlet2}, and \eqref{Eq:gravlet3} in the $S_{\leq4}$ multiplets. As usual, we treat all letters as simultaneously diagonalizable, so that each letter $\bar{\phi}^a, \psi_b,f$ takes the general form
\begin{equation}
M=
    \begin{pmatrix}
        m_1&0&0&0\\
        0&m_2&0&0\\
        0&0&m_3-m_1&0\\
        0&0&0&-m_2-m_3
    \end{pmatrix}\,,
\end{equation}
The graviton singlet at $+t^{12}$ can be identified to be
\begin{equation}
    \epsilon_{a_1a_3a_5} \epsilon_{a_2a_4a_6}u^{a_1a_2}u^{a_3a_4}u^{a_5a_6}\,.
\end{equation}
As noted in Sec.~\ref{Sec:SU3BMN}, the above invariant is $Q$--exact in the $SU(2)$ and $SU(3)$ theories due to a trace relation. 
However, this trace relation no longer holds in the $SU(4)$ theory, leading to the appearance of a new singlet graviton cohomology. 
Note this cohomology is purely bosonic, which we argued could not arise in the $SU(3)$ BMN sector. 

The singlet at $-t^{18}$ is the same as the one in the $SU(3)$ theory \eqref{Eq:t18SU3}, namely
\begin{equation}
    (v_2)^a_b(v_2)^b_c(v_2)^c_a.
\end{equation}
Furthermore, there are two fermionic singlets at $t^{30}$, one of which is simply the product of the bosonic and fermionic singlets identified above. 
The second fermionic singlet at $t^{30}$ is
\begin{equation}
    (v_2)^a_b(v_2)^b_c(v_2)^c_{d}(v_2)^{d}_{e}(v_2)^{e}_a\,.
\end{equation}
Here, we see again the appearance of a new graviton singlet cohomology, which however vanishes in the $SU(2)$ and $SU(3)$ theories due to a trace relation.

We have identified non-vanishing graviton singlets to explain the non-vanishing terms in the index \eqref{Eq:BMNSU4proj}, but have not performed an exhaustive check to verify that there are no other singlet gravitons whose contributions to the index, from bosonic and fermionic operators, cancel. 
Unlike the $SU(3)$ case, where the graviton index terminates at a finite order, the graviton index in the $SU(4)$ case is non-terminating due to the purely bosonic singlet at $t^{12}$. 
However, we expect the graviton singlet sector of the $SU(4)$ theory to be simple enough to admit an exact enumeration.
As in the $SU(3)$ case, the explicit construction will lead to a closed form for the non-graviton index and, as such, could shed light on the spectrum of non-graviton operators for $N=4$ as well.

\section{Conclusions and future directions}\label{Sec:conclusions}

In this paper, we have demonstrated that the structure of the Hilbert space of both bosonic and supersymmetric matrix models can simplify in the singlet sector of a global symmetry.
This is closely related to previous observations in the literature, which focused on simplification of the dynamics in these sectors, see, \textit{e.g.}, \cite{Sethi:1997pa,Berenstein:2016zgj,Buividovich:2022jgv,Hoppe:2023qem}.
Since the restriction to a singlet sector can be viewed as a gauging of the global symmetry, we have referred to the corresponding matrix model as double--gauged.
Below we provide a brief summary of our main results and point out interesting future directions.

\subsection*{Bosonic matrix models}

In the first part of this paper, we have studied free bosonic matrix models with $d$ matrices transforming in the adjoint of a $U(2)$ or $SU(2)$ gauge group, with an $SO(d)$ global symmetry. We have investigated the structure of the gauge invariant Hilbert spaces of these models in the $SO(d)$ singlet sector.
In particular, we calculated the partition function of the model using two Molien-Weyl projections, one for the gauge symmetry and one for the global symmetry. 
The resulting partition functions were then interpreted in terms of the ring of $SU(2)\times SO(d)$ and $U(2)\times SO(d)$ invariants, taking trace relations into account. 

The main result of this analysis is that the $SO(d)$ singlet sector of $SU(2)$ matrix models is equivalent, for $d\geq 4$, to an $SO(3)$ matrix model of a single $3\times 3$ real symmetric matrix $B$ (see also \cite{Sethi:1997pa,Hoppe:2023qem}).
When $d=3$, the model contains an additional (secondary) invariant, which squares to $\det B$.
Similarly, for arbitrary $d\geq5$, the singlet sector of $U(2)$ models can be described in terms of an $SO(3)$ matrix model of a symmetric and anti-symmetric $3\times 3$ matrix and an additional scalar, the combination of which we denoted by $\mathcal{B}$. 
When $d=4$, there is an additional (secondary) invariant which squares to $\det \mathcal{B}$.
Even though our computations of the partition functions were in the context of free matrix models, the analysis of the ring of invariants was purely ``kinematic'' and therefore extends to interacting models.
For example, as noted in Section \ref{Sec:Btheory}, the commutator squared potential in the Hamiltonian of the $SU(2)$ matrix model can be written in terms of traces of $B$ and $B^2$ in the $SO(d)$ singlet sector.

Thus, an $SO(d)$ singlet projection results in a significant simplification of matrix models. 
For example, the above implies that the bosonic part of the BFSS matrix model, in the $SO(d)$ singlet sector, reduces to the theory of a particle moving in a three-dimensional potential, with the Hamiltonian given in \eqref{Eq:d=3ham}.
This simple Hamiltonian, and its SUSY extension, can be studied more easily numerically \cite{Buividovich:2022jgv}, and are also more amenable to simulation on near-term quantum computers, than the full matrix model \cite{Gharibyan:2020bab,Rinaldi:2021jbg}.
In particular, they may provide interesting alternatives to SYK(-like) models, as studied in \cite{Garcia-Garcia:2020cdo,Xu:2020shn,Hanada:2023rkf,Swingle:2023nvv,Hanada:2025pis}, for the simulation of holographic protocols, including black hole formation \cite{Aoki:2015uha}, black hole evaporation \cite{Berkowitz:2016znt}, and the traversable wormhole teleportation protocol \cite{Brown:2019hmk,Gao:2019nyj,Nezami:2021yaq,Jafferis:2022crx}. 

Matrix models such as the BFSS model are expected to describe semi--classical gravity at large $N$ and strong coupling. It is therefore important to understand how our analysis generalises to matrix models with $SU(N)$, $N\geq3$ gauge groups. As a first step towards this generalization, we provide the global symmetry singlet partition functions for $U(3)$ and $SU(3)$ matrix models, with $d=2$ in Appendix~\ref{App:N3partitionfns}.
We note, however, that the reduction of the multi--matrix models to a single matrix model in the $SU(2)$ case relied on the isomorphism between the $su(2)$ and $so(3)$ Lie algebras. This enabled us to describe $SU(2)$ adjoint matrices as $SO(3)$ vectors and combine the degrees of freedom into the $B$ matrix referred to above. It would be interesting to study the generalization to the $SO(d)$ singlet sectors of $SU(N)$ matrix models. 

\subsection*{Non-graviton cohomologies and the superconformal index}

In the second part of this paper we investigated another global symmetry projection, now in the context of the $1/16$ BPS, so-called BMN subsector of the 4d $\mathcal{N}=4$ SYM theory. 
In particular, the $1/16$ BPS sector has a global $SU(3)_R$ R--symmetry, and we studied the singlet sector under this symmetry.
As reviewed in Section \ref{ssec:rev-bmn}, the BMN sector harbours non-graviton operators, which are expected to be closely related to microstates of BPS $AdS_5$ black holes.

We have shown that in the $SU(3)_R$ singlet sector of the BMN sector, the analysis of the spectrum of non-graviton operators simplifies.
The key reason for this is that, at least for $N=2,3,4$, the spectrum of ``finite $N$ gravitons'' is simple.
For example, for $SU(2)$ gauge group, there are no $SU(3)_R$ singlet gravitons at all!
In this case, the singlet projection of the BMN index automatically isolates the singlet black hole cohomologies. 
For $SU(3)$ gauge group, we have shown that the $SU(3)_R$ singlet graviton sector is extremely simple, consisting of only two non-trivial operators.
By calculating the difference between the full index and the graviton index, we thus obtain closed form expressions for the non-graviton index.
Such closed form expressions manifest various structural features of the non-graviton spectrum, which we hope will help to construct explicit representatives of these cohomologies and eventually lead to a better understanding of this class of operators.

It would be interesting to examine if global symmetry projections also enable the construction of closed form non-graviton indices for $N\geq 4$, since this would offer a way to make progress at larger $N$ without resorting to numerical techniques. 
We have taken the first steps towards this here for $N=4$, but we postpone a complete analysis to a future publication.
In particular, it would be nice to understand if the extreme scarcity of singlet gravitons observed in the low $N$ cases is a feature which extends to larger values of $N$.
Similarly, it would be interesting to understand if there is a reason that key non-gravitons seem to arise as $SU(3)_R$ singlets, such as the threshold cohomologies and the infinite towers on top of them for $N=2,3$. This may be compared with the $R$-charge concentration observed in supersymmetric SYK models \cite{Chang:2024lxt} and the D$1$-D$5$ CFT \cite{Chang:2025wgo}, where the fortuitous states are concentrated in specific charge sectors. 
Furthermore, it will be interesting to study the singlet projections for the full 1/16 BPS spectrum of the $\mathcal{N}=4$ theory.

Finally, we note that the identification of the non-graviton $Q$--cohomologies with black hole microstates in $AdS_5$ relies on the conjecture that the one-loop BPS states remain BPS at strong coupling \cite{Grant:2008sk}. 
This was recently shown to be false in the context of $\mathcal{N}=4$ theories with $SO(2N+1)/USp(2N)$ gauge groups \cite{Gadde:2025yoa,Chang:2025mqp,Choi:2025bhi}. 
But, the status of the conjecture remains unclear for $SU(N)$ gauge groups. 
In this paper, we have focused on investigating how global symmetry projections simplify the space of the one-loop BPS states. 
However, it is important to analyse if and how some of these states lift as we include further corrections. 
The authors in \cite{Chang:2025mqp,Choi:2025bhi} have argued that the $t^{24}$ threshold black hole cohomology in the $SU(2)$ theory, \eqref{Eq:24threshold}, which is an $SU(3)_R$ singlet,  survives higher loop and non-perturbative corrections. 
We also note that the authors in \cite{Choi:2025bhi} have identified a pair of $SU(3)_R$ singlet states, in the $SO(7)$ gauge theory, that lift due to quantum corrections. 
Thus, the $SU(3)_R$ singlet sector could provide a simpler setting to study these issues.

\section*{Acknowledgments}
We thank Jaehyeok Choi for correspondence about an equation in \cite{Choi:2023vdm}. We are especially grateful to Robert de Mello Koch for helpful discussions throughout this project, and particularly for pointing out the numerical algorithm to find relations among invariants. We would like to thank Masanori Hanada for useful discussions at the start of this project and pointing out various important references, and Pratik Roy for collaboration during the initial stages of this work.
SvL acknowledges support from the DSI-NRF Centre of Excellence in Mathematical and Statistical Sciences (CoE-MaSS), South Africa, grant \#2022-59-Phy-Indices.
SvL and KM also gratefully acknowledge generous support from the Wits-IBM Quantum Computing Seed Funding Programme with reference number QCSeed003/2023. 
Opinions expressed and conclusions arrived at are those of the authors and are not necessarily to be attributed to the CoE-MaSS nor NRF.

\begin{appendix}
\section{Molien--Weyl formula for partition functions}\label{app:A}

In this appendix, we review the Molien-Weyl formula for the partition functions of (gauged) matrix models.
This formula allows one to calculate the partition function in three steps.
One first calculates the partition function in the ungauged sector, \textit{i.e.}, one includes contributions from all matrix degrees of freedom.
Secondly, one expands this partition function in characters of the symmetry group $G$.
Finally, one projects onto representations of interest using the orthogonality of characters of representations of a semi--simple Lie group $G$:
\begin{equation}
 \int_G d\mu_G\, \rchi_{R}^{G}(U)\, \rchi_{\widetilde{R}}^{G}(U)=\delta_{R\widetilde{R}}\,, 
\end{equation}
where $d\mu_G$ is the Haar measure on $G$, $U\in G$ and $\rchi^G_{R,\widetilde{R}}(U)$ are characters in the representations $R,\widetilde{R}$.
For example, when $G$ is a gauge symmetry, one would project onto the singlet representation with trivial character.
The usefulness of the formula is that the partition function can be computed without the need to worry about trace relations.

We will be interested specifically in (bosonic) matrix models with $U(N)$ or $SU(N)$ gauge symmetry and a global $SO(d)$ symmetry.
Due to the global symmetry, the gauge--singlet Hilbert space decomposes into a direct sum over irreps of $SO(d)$. 
We will construct the partition functions in these fixed irrep sub-sectors.
This can be done in three parts, which we outline below:
\begin{enumerate}
    \item Compute the partition function for gauge singlets
    \item Refine by the global $SO(d)$ symmetry
    \item Project the refined partition function onto irreps of the global $SO(d)$ symmetry
\end{enumerate}  

The single particle partition function for a bosonic matrix with energy $E_i$ transforming in some representation $R$ of a semi--simple Lie group $G$ is given by
\begin{equation}
    z_i(x_i,U) = x_i \, \rchi^{G}_{R}(U)\,,
\end{equation}
where $x_i = e^{-\beta E_i}$. Then we obtain the multi-particle partition function through plethystic exponentiation of the sum over all single-particle partition functions, as
\begin{equation}
    Z(x_i,U) = \prod_{i=1}^d\exp\left(\sum_{m=1}^{\infty} \frac{1}{m} x_i^m\,  \rchi^{G}_{R}(U^m)\right)\,.
\end{equation}
We now isolate specific irreps $\widetilde{R}$ using the orthogonality of characters:
\begin{equation}\label{eq:formal-proj}
Z(x_i)_{\widetilde{R}} = \int_G d\mu_G\,  Z(x_i,U)\,  \rchi_{\widetilde{R}}^{G}(U)\,, 
\end{equation}
here $d\mu_G$ is the Haar measure of the group $G$. 
In the present work, we are interested in $G = U(N)/SU(N)\times SO(d)$. 
Since the integrand is invariant under $G$, one can convert the integral over $G$ into an integral over eigenvalues.
For the unitary group $U(N)$, this becomes
\begin{align}\label{eq:Umeasure}
    \begin{split}
        \int_{U(N)}\ d\mu_{U(N)} &= \frac{1}{N!}\prod_{j=1}^N\oint_{|u_j|=1} \frac{du_j}{2\pi i u_j}\ \prod_{k<l}(u_l-u_k)(u_l^{-1}-u_k^{-1}) \\
        &=  \prod_{j=1}^N\oint_{|u_j|=1} \frac{du_j}{2\pi i u_j} \prod_{k<l}\left(1-u_{kl}\right)
    \end{split}
\end{align}
where the measure on the first line is comprised of two Vandermonde determinants.
On the second line, we defined $u_{kl}=u_ku_l^{-1}$ and we have fixed the residual $S_N$ Weyl symmetry of the measure.
We refer to the corresponding measure as the reduced measure (see, \textit{e.g.}, \cite{Hanany:2008sb}, Appendix A.3 of \cite{deMelloKoch:2025cec} or Section 2 of \cite{vanLeuven:2025gwr} for more details).
The difference for $SU(N)$ is that one integrates over one eigenvalue less, which is fixed by the unit determinant condition.

In the case of the (special) orthogonal group $SO(d)$, we have instead 
\begin{align}\label{eq:Omeasure}
    \int_{SO(d)} d\mu_{SO(d)} &=\frac{1}{n!\ 2^{(n+\epsilon_0-1)}}\prod_{j=1}^n\oint_{|s_j|=1} \frac{ds_j}{2\pi i s_j}(1-s_j^{\pm})^{\epsilon_0} \prod_{1\leq p<q\leq n} (1-s_p^{\pm} s_q^{\pm})\nonumber \\ &=\prod_{j=1}^n\oint_{|s_j|=1} \frac{ds_j}{2\pi i s_j}(1-s_j)^{\epsilon_0} \prod_{1\leq p<q\leq n} (1-s_p s_q^{\pm})
\end{align}
here $n = {\lfloor d/2 \rfloor}$ and $\epsilon_0 = 1$ or $0$ when $d$ is odd or even, respectively. Above, we use the notation that $(1 - x s^{\pm })=(1 - x s)(1 - x s^{-1 })$.
Again, the expression on the first line corresponds to the Vandermonde determinants and the second line corresponds to the reduced measure \cite{Hanany:2008sb,Hanany:2008kn}.
We will only work with the reduced measures as they are simpler to work with.

\subsection*{Part 1: Gauge singlet partition function}
Let us specialize to the case where we have a multi-matrix model with matrices transforming in the adjoint of the gauge group, which is either $U(N)$ or $SU(N)$. 
The partition function for gauge singlets can be computed by using \eqref{eq:formal-proj} with
\begin{equation}
    \rchi_R^G = \rchi^{SU(N)/U(N)}_{\text{adj}}(u_j) = - \varepsilon + \sum_{k,l=1}^{N} u_k u_l^{-1}, \qquad \rchi_{\widetilde{R}}^G = \rchi_{\text{singlet}}^{SU(N)/U(N)} = 1\,, 
\end{equation}
where $\varepsilon = 0$ or $1$ for $U(N)$ and $SU(N)$, respectively. 
The gauge singlet partition function then takes the form 
\begin{equation}\label{eq:uvar}
    Z^{SU(N)/U(N)}(x_i) = \frac{1}{ \prod_i(1-x_i)^{N-\varepsilon}}\prod_{j=1}^N \oint \ \frac{du_j}{2\pi i u_j}\ \prod_{1\leq k < l \leq N} \frac{\left(1 - u_{kl}\right)}{\prod^d_{i=1}\left(1-x_i u_{kl}^{\pm} \right)}. 
\end{equation}
We make a change of variables $u_j =  t_j\cdots t_N$ and rewrite the above integral as \cite{deMelloKoch:2025ngs,vanLeuven:2025gwr}
\begin{equation}\label{eq:unr-gauge-sing}
    Z^{SU(N)/U(N)}(x_i) = (Z^{U(1)}(x_i))^{N-\varepsilon} \prod_{j=1}^{N-1}\oint \frac{dt_j}{2\pi i t_j}\ \prod_{1\leq k \leq l \leq N-1} \frac{(1-t_{k,l})}{\prod^d_{i=1}(1- x_i t_{k,l}^{\pm})}
\end{equation}
here $Z^{U(1)}(x_i) = \prod^d_{i=1}(1-x_i)^{-1}$ and $t_{k,l} = t_k\cdots t_l$. We note that when $N=2$, this formula reduces to \eqref{eq:mol-weyl-U(2)-d}. 
\subsection*{Part 2: Gauge singlet partition function with $SO(d)$ refinement} 
We now turn to refining the gauge singlet partition functions in terms of irreps of the global $SO(d)$ symmetry. This can be thought of as a partition function for a single bosonic field carrying the adjoint representation of $SU(N)/U(N)$ and the fundamental representation of $SO(d)$, which can be obtained by using \eqref{eq:formal-proj} with
\begin{align}
    \rchi_R^G &=   \rchi^{SU(N)/U(N)}_{\text{adj}}(u_j) \times  \rchi^{SO(d)}_{\text{fund}}(s_k)  = \bigg(- \varepsilon + \sum_{k,r=1}^{N} u_k u_l^{-1} \bigg) \bigg(\epsilon_0 + \sum_{j=1}^{n} (s_j + s_j^{-1} )\bigg),\nonumber \\ \qquad \rchi_{\widetilde{R}}^G&= \rchi_{\text{singlet}}^{SU(N)/U(N)} = 1\,, 
\end{align}
where $n = \lfloor d/2 \rfloor$. The partition function then takes the form
\begin{equation}\label{eq:r-gauge-sing-1}
    Z^{SU(N)/U(N)}(x,s_k) = (Z^{U(1)}(x,s_k))^{N-\varepsilon}\prod_{j=1}^{N-1} \oint \frac{dt_j}{2\pi i t_j}\, I(x,t_i,s_k)\,,
\end{equation}
with 
\begin{eqnarray}\label{eq:r-gauge-sing-2}
    I(x,t_i,s_k) = \prod_{1\leq k \leq l \leq N-1} \frac{(1-t_{k,l})}{(1-x t_{k,l}^{\pm})^{\epsilon_0}} \prod_{j=1}^{d/2} \frac{1}{1 - x s_j^{\pm} t_{k,l}^{\pm}} \, ,
\end{eqnarray}
where the $\pm$ should be understood as the product over all combinations of signs, and finally 
\begin{equation}\label{eq:r-gauge-sing-3}
    Z^{U(1)}(x,s_k) = \frac{1}{\left(1-x\right)^{\epsilon_0}}  \prod_{j=1}^{d/2} \frac{1}{1- x s_j^{\pm}}\,.
\end{equation} 

\subsection*{Part 3: Projected partition function onto different $SO(d)$ irreps} 

Next, we would like to project the refined gauge singlet partition function onto the spin-$J$ irrep of the $SO(d)$ global symmetry. This is done by taking \eqref{eq:r-gauge-sing-1} and further projecting onto irreps of $SO(d)$ using the $SO(d)$ measure in \eqref{eq:Omeasure} and multiplying by the character of interest. In this case, in \eqref{eq:formal-proj} we choose
\begin{equation}
    \rchi_R^G =   \rchi^{SU(N)/U(N)}_{\text{adj}}(u_j) \times \rchi^{SO(d)}_{\text{fund}}(s_k), \qquad \rchi_{\widetilde{R}}^G = \rchi^{SU(N)/U(N)}_{\text{singlet}}  \times \rchi^{SO(d)}_{J}
\end{equation}
with, for $d=2,3$, 
\begin{equation}
    \rchi^{SO(2)}_{J}(s) = \frac{s^J + s^{-J}}{2},\qquad \rchi^{SO(3)}_{J}(s) = \sum_{r=-J}^J s^r.
\end{equation}
Now staring at \eqref{eq:unr-gauge-sing} and (\ref{eq:r-gauge-sing-1}, \ref{eq:r-gauge-sing-2}, \ref{eq:r-gauge-sing-3}), it is easy to see that the refined gauge singlet partition function can be obtained from the gauge singlet partition function by the variable change 
\begin{equation}
    x_{2i-1} = x s_i,\quad x_{2i} = x s_{i}^{-1},\quad \forall\ i \leq \lfloor d/2\rfloor 
\end{equation}
when $d$ is even, and when $d$ is odd $x_d = x$. Combining these when $N=2$ and $d=2,3$, we obtain \eqref{eq:mol-weyl-so(even)}, \eqref{eq:mol-weyl-so(odd)}, \eqref{eq:mol-weyl-so(2)Jneq0}, and \eqref{eq:mol-weyl-so(3)Jneq0}. 

\subsection{\texorpdfstring{$U(2)$}{U(2)} and \texorpdfstring{$SU(2)$}{SU(2)} matrix models} \label{App:partfnexpr}
This section is a compendium of various partition functions for bosonic multi-matrix models with $U(2)$ and $SU(2)$ gauge group. These can be evaluated using the Molien-Weyl integral formulas defined above. The contour integrals that appear therein are evaluated using residue sums, by closing the contours inside the unit circles.

\subsection*{$\bm{d=3}$}
We first recall the evaluation of \eqref{eq:unr-gauge-sing} for $d=3$ \cite{deMelloKoch:2025ngs}.
        The residue sum can be simplified into
        \begin{equation}\label{eq:res-sum-U(2)-3}
        Z^{U(2)}_{d=3}(x_i)=\frac{1+x_1x_2x_3}{\prod^3_{i=1}(1-x_i)(1-x_i^2)\prod_{i<j}(1-x_ix_j)}\,.
        \end{equation}
        The $SU(2)$ version corresponds to
        \begin{equation}\label{eq:res-sum-SU(2)-3}
            Z^{SU(2)}_{d=3}(x_i)=\frac{1+x_1x_2x_3}{(1-x_1^2)(1-x_2^2)(1-x_3^2)(1-x_1x_2)(1-x_1x_3)(1-x_2x_3)}\,,
        \end{equation}
        and can also be obtained by simply removing the linear $1-x_{1,2,3}$ terms in the $U(2)$ answer.
\subsection*{$\bm{d=4}$}
        Similarly, for $d=4$ \cite{deMelloKoch:2025ngs} the residue sum can be simplified into
        \begin{align}
            \begin{aligned}
                Z_{d=4}^{U(2)}(x_{i}) &= \frac{P_{4}(x_{1}, x_{2},x_{3}, x_{4})}{\left(1 - x_1 \right) \left( 1 - x_2 \right) \left( 1 - x_3 \right) \left( 1 - x_4 \right) \left(1-x_1^2\right)  \left(1-x_2^2\right) \left(1-x_3^2\right) \left(1-x_4^2\right)} \\ &\quad \times \frac{1}{\left(1-x_1 x_2\right)  \left(1-x_1 x_3\right)\left(1-x_2 x_3\right) \left(1-x_1 x_4\right) \left(1-x_2 x_4\right) \left(1-x_3 x_4\right)},
            \end{aligned}
    \end{align}
    with
    \begin{align}
        \begin{aligned}
            P_{4} (x_{1},x_{2},x_{3},x_{4}) &= 1 + x_1 x_2 x_3 + x_1 x_2 x_4 + x_1 x_3 x_4 + x_2 x_3 x_4 - x_1^2 x_2 x_3 x_4 \\ &\quad - x_1 x_2^2 x_3 x_4 - x_1 x_2 x_3^2 x_4 - x_1 x_2 x_3 x_4^2 - x_1^2 x_2^2 x_3^2 x_4^2,
        \end{aligned}
    \end{align}
    and
    \begin{align}\label{eq:res-sum-SU(2)-4}
        \begin{aligned}
        Z^{SU(2)}_{d=4}(x_{i}) &= \frac{P_{4}(x_{1},x_{2},x_{3},x_{4})}{\left(1-x_1^2\right)  \left(1-x_2^2\right) \left(1-x_3^2\right) \left(1-x_4^2\right) \left(1-x_1 x_2\right)  \left(1-x_1 x_3\right) \left(1-x_2 x_3\right)} \\
        &\quad \times \frac{1}{\left(1-x_2 x_3\right) \left(1-x_1 x_4\right) \left(1-x_2 x_4\right) \left(1-x_3 x_4\right)}.
        \end{aligned}
    \end{align}
\subsection*{$\bm{d=5}$}
    The partition functions for $d=5$ $U(2)$ and $SU(2)$ adjoint matrices are listed below, where $P_{5} (x_{1}, x_{2}, x_{3}, x_{4}, x_{5})$ contains 94 terms, given by
    \begin{align}
    \begin{aligned}
    &Z^{U(2)}_{d=5}(x_{i}) =\frac{1}{(1 - x_{1} x_{2})(1 - x_{1} x_{3})(1 - x_{2} x_{3})(1 - x_{1} x_{4})(1 - x_{2} x_{4})(1 - x_{3} x_{4})}  \\
    &\qquad \times \frac{P_5(x_{1},x_{2},x_{3},x_{4},x_{5})}{(1 - x_{1}) (1 - x_{2}) (1 - x_{3}) (1 - x_{4}) (1 - x_{5}) (1 - x_{1}^2) (1 - x_{2}^2) (1 - x_{3}^2) (1 - x_{4}^2) (1 - x_{5}^2)} \\
    &\qquad \times \frac{1}{(1 - x_{1} x_{5})(1 - x_{2} x_{5})(1 - x_{3} x_{5})(1 - x_{4} x_{5})},
    \end{aligned}
\end{align}
with
\begin{align}
    \begin{aligned}
    P_5(x_{1},x_{2},x_{3},x_{4},x_{5}) &= 1 + x_1 x_2 x_3 + x_1 x_2 x_4 + x_1 x_2 x_5 + x_1 x_3 x_4 + x_1 x_3 x_5 + x_1 x_4 x_5 \\ &\qquad+ x_2 x_3 x_4  + x_2 x_3 x_5 + x_2 x_4 x_5 + x_3 x_4 x_5 - 4 x_1 x_2 x_3 x_4 x_5\\ &\qquad - x_1^2 x_2 x_3 x_4 - x_1 x_2^2 x_3 x_4  - x_1 x_2 x_3^2 x_4 - x_1 x_2 x_3 x_4^2 - x_1^2  x_2 x_3 x_5 \\ &\qquad- x_1 x_2^2 x_3 x_5 - x_1 x_2 x_3^2 x_5 - x_1 x_2 x_3 x_5^2  - x_1^2 x_2 x_4 x_5 - x_1 x_2^2 x_4 x_5\\ &\qquad - x_1 x_2 x_4^2 x_5 - x_1 x_2 x_4 x_5^2 - x_1^2  x_3 x_4 x_5 - x_1 x_3^2 x_4 x_5  - x_1 x_3 x_4^2 x_5 \\ &\qquad- x_1 x_3 x_4 x_5^2 - x_2^2 x_3 x_4 x_5 - x_2 x_3^2 x_4 x_5 - x_2 x_3 x_4^2 x_5 - x_2 x_3 x_4 x_5^2  \\ &\qquad- x_1^2 x_2^2 x_3^2 x_4^2 - x_1^2 x_2^2 x_3^2 x_5^2 - x_1^2 x_2^2 x_4^2 x_5^2 - x_1^2 x_3^2 x_4^2 x_5^2 - x_2^2 x_3^2 x_4^2 x_5^2\\ &\qquad + x_1 x_2^2 x_3^2 x_4 x_5  + x_1 x_2 x_3^2 x_4^2 x_5 + x_1 x_2^2 x_3 x_4^2 x_5 + x_1 x_2^2 x_3 x_4 x_5^2 \\ &\qquad+ x_1 x_2 x_3^2 x_4 x_5^2 + x_1 x_2 x_3 x_4^2 x_5^2  +  x_1^2 x_2 x_3^2 x_4 x_5 + x_1^2 x_2^2 x_3 x_4 x_5\\ &\qquad + x_1^2  x_2 x_3 x_4^2 x_5 + x_1^2 x_2 x_3 x_4 x_5^2 - x_1 x_2^2 x_3^2 x_4 x_5^2    - x_1^2 x_2^2 x_3^2 x_4 x_5\\ &\qquad - x_1 x_2^2 x_3 x_4^2 x_5^2 -  x_1 x_2^2 x_3^2 x_4^2 x_5 -x_2^2 x_3 x_4^2 x_5 x_1^2 - x_1 x_2 x_3^2 x_4^2 x_5^2 \\ &\qquad - x_1^2 x_2 x_3^2 x_4^2 x_5 -x_1^2 x_2^2 x_3 x_4 x_5^2 - x_1^2 x_2 x_3^2 x_4 x_5^2 - x_1^2 x_2 x_3 x_4^2 x_5^2\\ &\qquad  + 4 x_2^2 x_3^2 x_4^2 x_5^2 x_1^2  + x_1^3 x_2 x_3 x_4 x_5  + x_1 x_2^3 x_3 x_4 x_5 + x_1 x_2 x_3^3 x_4 x_5\\ &\qquad  + x_1 x_2 x_3 x_4^3 x_5 + x_1 x_2 x_3 x_4 x_5^3  + x_1 x_2^2 x_3^2 x_4^3 x_5^2 + x_1 x_2^2 x_3^3 x_4^2 x_5^2\\ &\qquad  + x_1 x_2^3 x_3^2 x_4^2 x_5^2 + x_1^2 x_2 x_3^3 x_4^2 x_5^2 + x_1^2 x_2^3 x_3 x_4^2 x_5^2 \\ &\qquad  + x_1^2 x_2^2 x_3^3 x_4 x_5^2 + x_1^2 x_2^3 x_3^2 x_4 x_5^2 + x_1^2 x_2^2 x_3^2 x_4^3 x_5 + x_1^2 x_2^2 x_3^3 x_4^2 x_5 \\ &\qquad + x_1^2 x_2^3 x_3^2 x_4^2 x_5  + x_1 x_2^2 x_3^2 x_4^2 x_5^3 + x_1^2 x_2^2 x_3 x_4^3 x_5^2 + x_1^2 x_2 x_3^2 x_4^3 x_5^2 \\ &\qquad+ x_1^2 x_2^2 x_3^2 x_4 x_5^3 + x_1^2 x_2 x_3^2 x_4^2 x_5^3  + x_1^2 x_2^2 x_3 x_4^2 x_5^3 + x_1^3 x_2 x_3^2 x_4^2 x_5^2 \\ &\qquad+ x_1^3 x_2^2 x_3 x_4^2 x_5^2 + x_1^3 x_2^2 x_3^2 x_4 x_5^2 + x_1^3 x_2^2 x_3^2 x_4^2 x_5  -x_1^3 x_2^2 x_3^2 x_4^2 x_5^3 \\ &\qquad-x_1^3 x_2^2 x_3^2 x_4^3 x_5^2 -x_1^3 x_2^2 x_3^3 x_4^2 x_5^2 -x_1^3 x_2^3 x_3^2 x_4^2 x_5^2 - x_1^2 x_2^2 x_3^3 x_4^3 x_5^2  \\ &\qquad - x_1^2 x_2^3 x_3^2 x_4^3 x_5^2 - x_1^2 x_2^3 x_3^3 x_4^2 x_5^2 - x_1^2 x_2^2 x_3^2 x_4^3 x_5^3 - x_1^2 x_2^2 x_3^3 x_4^2 x_5^3 \\ &\qquad- x_1^2 x_2^3 x_3^2 x_4^2 x_5^3 - x_1^3 x_2^3 x_3^3 x_4^3 x_5^3, 
    \end{aligned}
\end{align}
and
\begin{align}
    \begin{aligned}
    &Z^{SU(2)}_{d=5}(x_{i}) = \frac{1}{(1 - x_{2} x_{4})(1 - x_{3} x_{4})(1 - x_{1} x_{5})(1 - x_{2} x_{5})(1 - x_{3} x_{5})(1 - x_{4} x_{5})} \\
    &\quad\times\frac{P_5(x_{1},x_{2},x_{3},x_{4},x_{5})}{(1 - x_{1}^2) (1 - x_{2}^2) (1 - x_{3}^2) (1 - x_{4}^2) (1 - x_{5}^2)(1 - x_{1} x_{2})(1 - x_{1} x_{3})(1 - x_{2} x_{3})(1 - x_{1} x_{4})} .
    \end{aligned}
\end{align}
We do not provide the explicit expressions for the partition functions for $d\geq6$ since they are very unwieldy, with the numerator $P_{6} (x_{1}, x_{2}, x_{3}, x_{4}, x_{5}, x_{6})$ containing 1301 terms and $P_{7} (x_{1}, x_{2}, x_{3}, x_{4}, x_{5}, x_{6}, x_{7})$ containing 23512 terms, respectively. These can be obtained in a straightforward way from \eqref{eq:unr-gauge-sing}.

\subsection{\texorpdfstring{$U(3)$}{U(3)} and \texorpdfstring{$SU(3)$}{SU(3)} matrix models}\label{App:N3partitionfns}
Here, we compute the partition functions for the $SO(d)$ singlet sector of matrix models with $2$ matrices transforming in the adjoint of the $U(3)$ and $SU(3)$ gauge groups. As above, these can be evaluated using the Molien-Weyl formulas in terms of residue sums.

\subsection*{$SU(3)$ gauge group}
The partition function of the $SU(3)$ matrix model with $d=2$ can be evaluated with \eqref{eq:unr-gauge-sing} as
\begin{equation}\label{eq:res-sum-SU(3)-2}
    Z^{SU(3)}_{d=2}(x_i)=\frac{1+x_1^3x_2^3}{(1-x_1^2)(1-x_2^2)(1-x_1x_2)(1-x_1^3)(1-x_2^3)(1-x_1^2x_2)(1-x_1x_2^2)(1-x_1^2x_2^2)}\,.
\end{equation}
This formula reflects the presence of 8 primary invariants and one secondary invariant. 
The projection onto $SO(2)$ singlets is performed by applying \eqref{eq:Omeasure} to \eqref{eq:res-sum-SU(3)-2} to find
\begin{align}
    \begin{split}
        Z^{SU(3),J=0}_{d=2}(x)&=\frac{1+x^6}{(1-x^2)(1-x^4)}\oint\frac{ds}{2\pi i s}\frac{1}{(1-x^3s^{\pm1})(1-x^2s^{\pm2})(1-x^3s^{\pm3})} \\
        &=\frac{1+3x^8+2x^{10}+3x^{12}+x^{20}}{(1-x^2)(1-x^4)^2(1-x^6)^3(1-x^8)}
    \end{split}
\end{align}
We see that the number of primary invariants reduces by one, as expected since fixing the $SO(2)$ angular momentum reduces the degrees of freedom by one.
\subsection*{$U(3)$ gauge group}

In the case of the $U(3)$ gauge group, we obtain the following partition function in the $J=0$ sector
\begin{align}
    \begin{aligned} 
    Z^{U(3),J=0}(x) &= \frac{1 + x^{6}}{(1 - x^2)(1 - x)^4} \oint\frac{ds}{2 \pi i s} \frac{1}{ (1 - x s^{\pm})(1-x^2s^{\pm2})(1-x^3s^{\pm3})(1-x^3s^{\pm})}  \\
    &= \frac{1 + 3 x^4 + 6 x^6 + 9 x^8 + 6 x^{10} + 12 x^{12} + 6 x^{14} + 9 x^{16} + 
 6 x^{18} + 3 x^{20} + x^{24}}{(1 - x^2)^2 (1 - x^4)^3 (1 - x^6)^3 (1 - x^8)}.
    \end{aligned}
\end{align}
This features 9 primary invariants, which is one less than the number of invariants in the original $U(3)$ partition function \cite{deMelloKoch:2025ngs}, as expected from the $SO(2)$ constraints.

\section{\texorpdfstring{$SO(d)$}{SO(d)} non-singlet sectors of bosonic matrix models}\label{Sec:loopspaceJneq0}

In this section, we generalize the analyses of sections \ref{Sec:molienweyl} and \ref{sec:struct-loop-singlet} to non-singlet or $SO(d)$ spin $J\neq 0$ sectors of the $U(2)$ and $SU(2)$ multi-matrix models.

\subsection{\texorpdfstring{$d=2$}{d=2} \texorpdfstring{$U(2)$}{U(2)} matrix model}
 Partition functions for $d=2$ matrices in the adjoint of $U(2)$, in fixed $SO(2)$ spin $J\neq0$ sectors can be constructed using \eqref{eq:mol-weyl-so(2)Jneq0}, as
\begin{align}
     &Z^{U(2), J=2n}_{d=2} (x) = \frac{x^{2n}+x^{2n+4}}{(1-x^2)^2(1-x^4)^2}+\frac{nx^{2n}}{(1-x^2)^2(1-x^4)}, \qquad n \in \mathbb{Z}^{+},\\
     &Z^{U(2), J=2n+1}_{d=2} (x) = \frac{x^{2n+1}+x^{2n+3}}{(1-x^2)^2(1-x^4)^2}+\frac{nx^{2n+1}}{(1-x^2)^2(1-x^4)}, \qquad n \in \mathbb{Z}^{+}\,.
\end{align}
Firstly, we note that in any fixed spin sector, invariants with non-zero spin necessarily act as secondaries, since their products will not preserve spin. Conversely, primary invariants must have $J=0$. Thus, the denominators in the above correspond to the spin-$0$ primary invariants $p_1,p_2,q_1,q_2$ identified in Sec.~\ref{Sec:invd2j0}, whereas the numerators are associated  with $n+2$ spin-$J$ secondary invariants. Note that we do not include invariants that are anti-symmetric under the exchange $X_1\leftrightarrow X_2$ of the two matrices, since they are projected out by the Molien-Weyl integral for $J\neq0$, as explained below equation~\eqref{eq:mol-weyl-so(2)Jneq0}.

A noteworthy feature of these partition functions is that for $J\geq3$  they are expressed as a sum of two modules, with one less primary invariant appearing in the second module. This indicates that the invariant ring has the following decomposition for even $J=2n\geq4$
\begin{equation}
\mathcal{R}^{U(2),J=2n} = s_{2n,0} K[p_1,p_2,q_1,q_2] +s_{2n+4} K[p_1,p_2,q_1,q_2] +\sum_{k=1}^{n} s_{2n,k} K[p_1,p_2,q_1]\,,
\end{equation}
where $s_{2n,k}$, for $k=0,\hdots,n$, are the $n+1$ spin-$2n$ secondaries at order $2n$ and $s_{2n+4}$ is the spin-$2n$ secondary at order $2n+4$. A similar structure holds for odd spins $J\geq3$.
This indicates that the ring \emph{does not} have a Hironaka decomposition, \textit{i.e.}, it is not a free module over a \textit{single} sub-ring. Note that we are guaranteed to find a free module over a single sub-ring for the invariant ring of a linearly reductive group due to the Hochster–Roberts theorem \cite{HOCHSTER1974115}. However, the fixed $J\neq0$ sectors are not $SO(d)$ invariants, but correspond to fixed irreps of $SO(d)$. There is no theorem that guarantees the same structure for the $J\neq0$ sectors as the $J=0$ sector. See \cite{VandenBergh:1991} for counter examples to Cohen-Macauly-ness for covariants (non-singlets).

 We further explain the structure of these partition functions below, by explicitly identifying the invariants for the first few spins $J=1,2,3$. For $J\neq0$, we do not provide analytic proofs of completeness of the invariant generators; however, we expect that a similar strategy as that for $J=0$ in Sec.~\ref{Sec:completed2j0} can be applied.
 
\subsubsection{Invariants for $J=1$}\label{Sec:invd2j1}
The partition function in the $J=1$ sector is
\begin{equation}
    Z^{U(2),J=1}_{d=2}(x)=\frac{x+x^3}{(1-x^2)^2(1-x^4)^2}\,,
\end{equation}
which indicates two secondary invariants, which  can be identified as
\begin{equation}
    S^{(1,1)}_i = \tr(X_i), \quad S^{(3,1)}_i=\tr(X_i X_jX_j).
\end{equation}
Here, the superscripts ${(\alpha,\beta)}$ denote an order $\alpha$ invariant, and $\beta$ distinguishes the different invariants at the same order $\alpha$. The spin of the invariant is clear from the number of free indices it carries. 

The only other $J=1$ invariant at order $3$ is $S^{(3,2)}_i=\tr(X_j)\tr(X_iX_j)$, which is reducible due to the following trace relation
\begin{equation}
    2 S^{(3,1)}_i-S^{(1,1)}_ip_1 -2 S^{(3,2)}_i+ S^{(1,1)}_i p_2=0\,.
\end{equation}
We have numerically verified, up to order $20$, that there are no other relations among the primary and the secondary invariants listed above.

\subsubsection{Invariants for $J=2$}\label{Sec:invd2j2}
The partition function in the $J=2$ sector is
\begin{equation}
    Z^{U(2),J=2}_{d=2}(x)= \frac{2 x^2}{(1-x^2)^2(1-x^4)^2}\,,
\end{equation}
which indicates two independent $J=2$ invariants at order $2$. These are
\begin{equation}
    S^{(2,1)}_{ab}= \tr(X_a X_b)- \frac12\delta_{ab} p_1\,,\quad S^{(2,2)}_{ab}=\tr(X_a)\tr(X_b)-\frac12\delta_{ab} p_2\,,
\end{equation}
where we have subtracted out the trace (spin-$0$) parts. There are no other $J=2$ invariants at order $2$. Moreover, these two invariants are algebraically independent, since there are no trace relations at order two for $2\times2$ matrices. We have numerically verified, up to order $20$, that there are no relations among these primary and secondary generators.

\subsubsection{Invariants for $J=3$}\label{Sec:invd2j3}
The first non-trivial partition function, with two modules, arises for $J=3$ and is
\begin{equation}\label{Eq:U2J3d2}
    Z^{U(2),J=3}_{d=2}(x) = \frac{x^3+x^5}{(1-x^2)^2(1-x^4)^2}+\frac{x^3}{(1-x^2)^2(1-x^4)}\,.
\end{equation}
This indicates two $J=3$ invariants at order $3$ and one $J=3$ invariant at order $5$. We identify these below. 

A spin-$3$ invariant can be made from multi-traces of the matrices with three symmetrized free indices. The spin-$3$ invariants at order $3$ are
\begin{subequations}
\begin{align} 
    S^{(3,1)}_{abc} &= \tr\left(X_{(a}\ X_b\ X_{c)}\right)- {\rm trace}\\ 
    S^{(3,2)}_{abc} &= \tr\left(X_{(a}\ X_b\right)\ \tr\left(X_{c)}\right)-{\rm trace}\\
    S^{(3,3)}_{abc} &= \tr\left(X_a\right)\ \tr\left(X_b\right)\ \tr\left(X_c\right)-{\rm trace}\,,
\end{align}
\end{subequations}
where the brackets around the indices denote complete symmetrisation under any pair-wise exchange, and the spin of the invariant is clear from the number of indices it carries. Here, we have subtracted out the trace (spin-$1$) parts of these invariants to ensure that they transform in the spin-$3$ irrep.  It is easy to see that only two out of $S^{(3,1)}_{abc},S^{(3,2)}_{abc}$ and $S^{(3,3)}_{abc}$ are independent, using the following trace relation:
\begin{multline}
    \tr(X_a)\tr(X_b)\tr(X_c)-\tr(X_aX_b)\tr(X_c)-\tr(X_aX_c)\tr(X_b)-\tr(X_bX_c)\tr(X_a)\\+\tr(X_aX_bX_c)+\tr(X_aX_cX_b)=0\,.
\end{multline}
We will choose $S^{(3,1)}_{abc}$ and $ S^{(3,2)}_{abc}$ as the independent invariants. 

Next, we consider spin-$3$ invariants at order $5$. There are 12 order 5 invariants with three symmetrized free indices and two indices contracted with a Kronecker delta $\delta_{ij}$, which cannot be written in terms of lower order invariants. These are
\begin{subequations}
    \begin{align}
        O^{(1)}_{5,3}&=\tr(X_iX_jX_{(k}X_lX_{m)})\delta_{ij},\\
        O^{(2)}_{5,3}&=\tr(X_iX_{(j}X_{k}X_lX_{m)}) \delta_{ik},\\
        O^{(3)}_{5,3}&=\tr(X_i)\tr(X_jX_{(k}X_lX_{m)})\delta_{ij},\\
        O^{(4)}_{5,3}&=\tr(X_{(i})\tr(X_jX_{k}X_lX_{m)})\delta_{jk},\\
        O^{(5)}_{5,3}&=\tr(X_{(i})\tr(X_jX_{k}X_lX_{m)})\delta_{jl},\\
        O^{(6)}_{5,3}&=\tr(X_{(i}X_j)\tr(X_kX_lX_{m)})\delta_{jk},\\
        O^{(7)}_{5,3}&=\tr(X_{(i}X_j)\tr(X_kX_lX_{m)})\delta_{kl},\\
        O^{(8)}_{5,3}&=\tr(X_i)\tr(X_jX_{(k})\tr(X_lX_{m)})\delta_{ij},\\
        O^{(9)}_{5,3}&=\tr(X_{(i})\tr(X_jX_k)\tr(X_lX_{m)})\delta_{jl},\\
        O^{(10)}_{5,3}&=\tr(X_i)\tr(X_{(j})\tr(X_kX_lX_{m)})\delta_{ik},\\
        O^{(11)}_{5,3}&=\tr(X_{(i})\tr(X_j)\tr(X_kX_lX_{m)})\delta_{kl},\\
        O^{(12)}_{5,3}&=\tr(X_i)\tr(X_{(j})\tr(X_k)\tr(X_lX_{m)})\delta_{il}.
    \end{align}
\end{subequations}
To make the trace relations easier to read we have suppressed the indices from the left hand sides of the above. The subscript $5,3$ indicates that these are order $5$ invariants with $J=3$.
Only one of these 12 invariants is independent since we have 11 in-equivalent trace relations, which are listed below
\begin{subequations}
    \begin{align}
       &O^{(12)}_{5,3}-O_{5,3}^{(8)}-2O_{5,3}^{(10)}+2O_{5,3}^{(3)}=0,\\
       &O_{5,3}^{(11)}-O_{5,3}^{(7)}-2O_{5,3}^{(4)}+2O_{5,3}^{(1)}=0,\\
       &O^{(10)}_{5,3}-O^{(6)}_{5,3}-O^{(4)}_{5,3}-O^{(3)}_{5,3}+O^{(2)}_{5,3}+O^{(1)}_{5,3}=0,\\
       &O^{(12)}_{5,3}-2O^{(9)}_{5,3}-O^{(8)}_{5,3}+2O^{(6)}_{5,3}=0,\\
       &O^{(8)}_{5,3}-O^{(7)}_{5,3}-O^{(6)}_{5,3}-O^{(3)}_{5,3}+O^{(1)}_{5,3}+O^{(2)}_{5,3}=0,\\
       &O^{(9)}_{5,3}-O^{(4)}_{5,3}-2O^{(6)}_{5,3}+O^{(1)}_{5,3}+O^{(2)}_{5,3}=0,\\
       &O^{(11)}_{5,3}-2O^{(5)}_{5,3}-O^{(7)}_{5,3}+2O^{(2)}_{5,3}=0,\\
       &O^{(2)}_{3,3}p_1-O^{(7)}_{5,3}-O^{(1)}_{3,3}p_1-O^{(4)}_{5,3}+2O^{(1)}_{5,3}=0,\\
       &p_2O^{(1)}_{3,3}-p_1 O^{(1)}_{3,3}-2O^{(3)}_{5,3}+2O^{(1)}_{5,3}=0,\\
       &p_2 O^{(2)}_{3,3}-p_1 O^{(2)}_{3,3} -2O^{(8)}_{5,3}+2O^{(7)}_{5,3}=0,\\
       &O^{(9)}_{5,3}-2O^{(6)}_{5,3}-O^{(5)}_{5,3}+2O^{(2)}_{5,3}=0.
    \end{align}
\end{subequations}
In the above list, the trace relations that involve four traces have been generated by multiplying the $T_2$ relation at order 4 by a single-trace invariant. All other trace relations have been generated with a straightforward use of $T_2(A,B,C)$, at order five, with appropriate choices for the matrices $A,B,C$, as explained in Sec.~\ref{Sec:invd2j0}. We will choose $O^{(1)}_{5,3}$ to be the independent invariant from the above list. 

Thus, we have the following $J=3$ invariants as the secondary generators
\begin{align}
    S^{(3,1)}_{abc} &= \tr(X_{(a}X_bX_{c)})-\frac12\lb\tr(X_{(a} X_dX_d)\delta_{bc)}\rb,\\
    S^{(3,2)}_{abc}&= \tr(X_{(a}X_b)\tr(X_{c)}) - \frac14 \Bigg(2\tr(X_{(a}X_d)\tr(X_d)\delta_{bc)}+\tr(X_dX_d)\tr(X_{(a})\delta_{bc)}\Bigg),\\
    S^{(5,1)}_{abc} &= \tr(X_{(a} X_bX_{c)} X_dX_d)\\ & \quad -\frac12 \Bigg(\tr(X_{(a} X_e X_e X_d X_d) \delta_{bc)}+\tr(X_eX_{(a} X_e X_d X_d) \delta_{bc)}+\tr(X_e X_e X_{(a} X_d X_d) \delta_{bc)}\Bigg) \nonumber,
\end{align}
where we have defined $S_{abc}^{(5,1)}$ using $O^{(1)}_{5,3}$ and subtracting its trace.
These three secondary invariants, together with the $J=0$ primary  generators $p_1,p_2,q_1,q_2$ defined in Sec.~\ref{Sec:invd2j0}, generate the $J=3$ invariant ring and account for all the terms in the partition function.

We note that in addition to the above, there are 5 non-zero invariants at order $5$ with three free symmetrized indices and two indices contracted with the $\epsilon_{ij}$ tensor.  We do not include these invariants, since they are parity odd under the $X_1 \leftrightarrow X_2$ exchange and are projected out by the Molien-Weyl integral (see \eqref{eq:mol-weyl-so(2)Jneq0}).

The reduction of an order four primary in the second module in the partition function \eqref{Eq:U2J3d2} is due to the following order $7$ relation
\begin{multline}\label{Eq:j3ord7reln}
    \frac23 S^{(3,2)}_{ijk} q_1 +S^{(3,1)}_{ijk} q_2 - S^{(3,2)}_{ijk} q_2 +S^{(3,1)}_{ijk}p_2^2 -S^{(3,1)}_{ijk} p_1p_2 \\ -\frac13 S^{(3,2)}_{ijk} p_1^2 -\frac23 S^{(3,2)}_{ijk}p_2^2 +\frac43 S^{(3,2)}_{ijk}p_1p_2 -\frac13 S^{(5,1)}_{ijk}p_2 =0\,,
\end{multline}
which can be checked numerically. Furthermore, we have verified numerically, up to order $50$, that \eqref{Eq:j3ord7reln} is the only relation among the generators. The invariant ring therefore has the following structure
\begin{equation}
    \mathcal{R}^{U(2),J=3}_{d=2} = S^{(3,1)}_{abc} K[p_1,p_2,q_1,q_2]+S^{(5,1)}_{abc} K[p_1,p_2,q_1,q_2]+S^{(3,2)}_{abc} K[p_1,p_2,q_1]\,,
\end{equation}
where $K[p_1,p_2,q_1,q_2]$ and $K[p_1,p_2,q_1]$ are the rings freely generated by the primary invariants in the respective parenthesis.

\subsubsection{$SU(2)$ theory}
The $SO(2)$ spin$-J$ projection of the $SU(2)$ singlet partition function is 
\begin{equation}
    Z^{SU(2),J}_{d=2}(x) = \frac{1}{(1-x^2)}\oint \frac{ds}{2\pi i}\ \frac{s^{1+J}}{(1-x^2s^{2})(s^2-x^2)} = \left(\frac{(-1)^J+1}{2}\right)\ \frac{x^J}{(1-x^2)(1-x^4)}\,.
\end{equation}
When $J$ is odd the projection vanishes, since there are no odd spin invariants that can be constructed in the $SU(2)$ theory. For example, the only possible spin-$3$ invariant is of the form $\tr(X_{(i} X_j X_{k)})- {\rm traces}$, which vanishes due to the $T_2$ trace relation and tracelessness of the $SU(2)$ matrices. Higher odd-spin invariants can similarly be shown to vanish. 

On the other hand, for $J=2n$ we obtain 
\begin{equation}
    Z^{SU(2),2n}_{d=2}(x) =  \frac{x^{2n}}{(1-x^2)(1-x^4)}\qquad n \in \mathbb{Z}^+.
\end{equation}
This partition function is in the Hironaka form and reflects that the spin-$J$ sector of $SO(2)$ (symmetric traceless tensors with $J$ indices) when $J$ is even, has one secondary operator. This operator can be constructed using $O_{ij} = $ tr $X_iX_j$. The operator for $J=2n$ is
\begin{equation}
   S_{a_1,b_1,\cdots,a_n,b_n} = O_{(a_1b_1}\cdots O_{a_{n}b_n)} - {\rm traces}.
\end{equation}
The primary invariants here are the $p_1$ and $q_2$ spin-$0$ invariants identified in Sec.~\ref{Sec:invd2j0}.
\subsection{\texorpdfstring{$d=3$}{d=3} \texorpdfstring{$U(2)$}{U(2)} matrix model}

The fixed spin partition function for even $J=2n$ is
\begin{equation}
    Z_{d=3}^{U(2),J=2n}(x)= x^{2n}(1+x^3)\frac{n(1 + x^2) (1 + x + x^2) (1 + x^3)+Q_n(x)}{(1-x^2)^2(1-x^3)(1-x^4)^2(1-x^6)},
\end{equation}
with
\begin{equation}
    Q_n(x)= \begin{cases}
                1+x^9,& n=0,\\
                1-2x^3-2x^4-2x^5-x^6-x^7 & n =1,\\
               1-2x^3-2x^4-3 x^5-3x^6-4x^7\lb\frac{1-x^{2n-4}}{1-x}\rb-3x^{2n+3}-2x^{2n+4}-x^{2n+5},& n\geq2.
            \end{cases}
\end{equation}
For odd $J=2n+1$, we get
\begin{equation}
    Z_{d=3}^{U(2),J=2n+1}(x)=x^{2n+1}(1+x^3)\frac{n\lb1 + x + 2 x^2 + 2 x^3 + 2 x^4 + 2 x^5 + x^6 + x^7\rb+P_n(x)}{(1-x^2)^2(1-x^3)(1-x^4)^2(1-x^6)},
\end{equation}
with
\begin{equation}
P_n(x)=
    \begin{cases}
    1+x^2+x^4,& n=0,\\
    1+x^2-2x^5-2x^6-2x^7-x^8, &n=1,\\
    1+x^2-2x^5-3x^6-4 x^7 \lb\frac{1-x^{2n-3}}{1-x}\rb-3x^{2n+4}-2x^{2n+5}-x^{2n+6},& n\geq2.
    \end{cases}
\end{equation}
The two partition functions above are not in the Hironaka form for $J\geq3$ due to the presence of minus signs in the numerator in the $Q_n(x)$ and $P_n(x)$ terms. It is non-trivial to write down a general Hironaka form for arbitrary $J$. But, when a choice of $J$ is made, the partition function does indeed take the Hironaka form. We illustrate this for the first few spins below.

For $J=3$ the partition function is
\begin{equation}
    Z^{U(2),J=3}_{d=3}=\frac{x^{10}+x^9+2 x^8+2 x^7+3 x^6+2 x^5+x^4+2 x^3}{\left(1-x^2\right)^2 \left(1-x^3\right) \left(1-x^4\right)^2 \left(1-x^6\right)}+\frac{x^8+x^7+x^6+x^5}{\left(1-x^2\right)^2 \left(1-x^3\right) \left(1-x^4\right)^2},
\end{equation}
which features two modules as seen in the $d=2$ model before. The second module has one less primary invariant at order $6$. Similarly for $J=4$ we get
\begin{equation}
    Z^{U(2),J=4}_{d=3}=\frac{x^{11}+3 x^9+3 x^8+3 x^7+3 x^6+2 x^5+3 x^4}{\left(1-x^2\right)^2 \left(1-x^3\right) \left(1-x^4\right)^2 \left(1-x^6\right)}+\frac{x^{10}+2 x^9+x^8+2 x^7+x^6}{\left(1-x^2\right)^2 \left(1-x^3\right) \left(1-x^4\right)^2},
\end{equation}
where the second module again has one less primary at order $6$. This reduction of the order $6$ primary in the second module also occurs for $J=5$ as follows
\begin{multline}
   Z^{U(2),J=5}_{d=3}= \frac{x^{12}+x^{11}+4 x^{10}+3 x^9+4 x^8+4 x^7+2 x^6+3 x^5}{\left(1-x^2\right)^2 \left(1-x^3\right) \left(1-x^4\right)^2 \left(1-x^6\right)}\\+\frac{x^{12}+2 x^{11}+3 x^{10}+3 x^9+3 x^8+x^7}{\left(1-x^2\right)^2 \left(1-x^3\right) \left(1-x^4\right)^2}.
\end{multline}

We have checked that all the partition functions up to and including $J=9$ can be written as a sum of two modules, where the second module has a reduction of the order $6$ primary invariant. The set of primary invariants in the first module are the same as those for the $J=0$ sector \eqref{Eq:U2d3j0}, as expected.

It is very cumbersome to explicitly construct the invariants for non-zero spin in this case and verify the completeness of the invariant generators using trace relations. We expect that the reduction of the primary invariant in the second module occurs through a set of syzygies, as in the $d=2$, $J=3$ case in Sec.~\ref{Sec:invd2j3}.

\subsubsection{$SU(2)$ theory}
The fixed spin partition function for a model with $d=3$ matrices in the adjoint of $SU(2)$, with $J=2n$ is
\begin{align}
     Z_{d=3}^{SU(2),J=2n}(x) &= \frac{x^{J}-x^{2J+2}}{(1-x^2)^2(1-x^3)(1-x^4)}=  \frac{1}{(1-x^2)(1-x^3)(1-x^4)}\ \left(\sum_{n=J/2}^{J} x^{2n}\right)\nonumber\\
     & =  \frac{1+x^3}{(1-x^2)(1-x^4)(1-x^6)} \left(x^J + x^{J+2} + \cdots + x^{2J-2} + x^{2J}\right).
\end{align}
The primary invariants are the same as the $J=0$ sector in Sec.~\ref{Sec:BtheorySU2}, and  we see that the order $3$, $J=0$ secondary invariant is retained. The second factor in the numerator indicates that there are $(1 + J/2)=n+1$ secondary invariants, with spin $2n$, all of even orders ranging from $J$ to $2J$. 

We find that the $J=1$ projection vanishes, and for odd $J >1$ we have 
\begin{align}
     Z_{d=3}^{SU(2),J=2n+1}(x) &= \frac{x^{J+3}-x^{2J+2}}{(1-x^2)^2(1-x^3)(1-x^4)}\\
     & = \frac{1+x^3
     }{(1-x^2)(1-x^6)(1-x^4)}\left(x^{J+3} + x^{J+5}+\cdots + x^{2J-2} + x^{2J} \right).
\end{align}
Here, the numerator indicates that there are $(J-1)/2=n$ secondary invariants, with spin $2n+1$, all of even orders ranging from $J+3$ to $2J$. These secondary invariants are \textit{dressed} by the ring of $J=0$ invariants identified before.


\section{Enumerating graviton singlets}\label{App:gravsinglets}

The $SU(3)_R$ singlet gravitons in the BMN sector of the $\mathcal{N}=4$ SYM theory, with the $SU(3)$ gauge group, were analysed in Sec.~\ref{Sec:SU3BMN}. In this section, we explain the procedure we followed to enumerate these gravitons. 

As a start, the multi--trace gravitons are generated by the following single--trace gravitons
\begin{align}
    &(u_2)^{i_1i_2}=\tr\lb\bar{\phi}^{i_1}\bar{\phi}^{i_2}\rb, \quad (u_3)^{i_1i_2i_3}=\tr\lb\bar{\phi}^{(i_1}\bar{\phi}^{i_2}\bar{\phi}^{i_3)}\rb,\\
    &(v_2)^{i}_j= \tr\lb\bar{\phi}^{i}\psi_j\rb -\frac13 \delta^i_j \tr\lb\bar{\phi}^k\psi_k\rb,\\ &(v_3)^{i_1i_2}_j= \tr\lb\bar{\phi}^{(i_1}\bar{\phi}^{i_2)}\psi_j\rb -\frac14 \delta^{i_1}_j \tr\lb\bar{\phi}^{(i_2}\bar{\phi^{k)}}\psi_k\rb-\frac14 \delta^{i_2}_j \tr\lb\bar{\phi}^{(i_1}\bar{\phi^{k)}}\psi_k\rb, \\
    &(w_2)^{i}= \tr\lb f \bar{\phi}^{i}+\frac 12 \epsilon^{ijk} \psi_j\psi_k\rb, \quad (w_3)^{i_1i_2}=\tr\lb f\bar{\phi}^{(i_1}\bar{\phi}^{i_2)}+\epsilon^{jk(i_1}\bar{\phi}^{i_2)}\psi_j\psi_k\rb\,.
\end{align}
We will refer to the single--trace gravitons below as graviton letters, not to be confused with the matrix letters inside the traces.

Each boson $\bar{\phi}^i$ contributes at order $t^2$, each fermion $\psi_j$ contributes at $t^4$, and $f$ at order $t^6$ (see Section \ref{ssec:rev-bmn} and, in particular, Table \ref{tab:n=4-letters}). Using the above generating letters, we can write a formula to enumerate all graviton \textit{words} in the theory as
\begin{multline}
   \frac{1}{1-U_2 x^2 t^4 \alpha^2} \frac{1}{1-U_3 x^3 t^6 \alpha^3} \frac{1}{1-V_2 x t^6 z \alpha \beta}\frac{1}{1-V_3 x^2 t^8 z \alpha^2 \beta}\\ \times \frac{1}{1-t^{8}\alpha(x W_{2b}+z^2W_{2f})}\frac{1}{1-t^{10}\alpha^2(x^2 W_{3b}+xz^2W_{3f})},
\end{multline}
where the variables $U_2,U_3,V_2,V_3,W_2,W_3$, keep track of the graviton letters. The powers of $t$ in each term keep track of the order of the generator, $x$ counts the number of bosons in the generator, $z$ counts the number of fermions, $\alpha$ counts the number of upper indices, and $\beta$ counts the number of lower indices.  For the $w_2$ and $w_3$ generators, we separate the purely bosonic term ($W_{2b},W_{3b}$) from the term containing fermions ($W_{2f},W_{3f}$) for reasons that will become clear below. We series expand the above to the desired order in $t$ to obtain a polynomial in the graviton letters. Each monomial therein is a graviton word at that particular order. Since the theory features $6$ independent fermionic eigenvalues, we can set any monomial with powers of $z$ greater than six to zero. Note that the above formula overcounts the gravitons since it does not account for trace relations. This is sufficient for us, since we find that only two graviton singlets exist in this theory, which are \eqref{Eq:t18SU3} at order $t^{18}$ and \eqref{Eq:t24SU3} at order $t^{24}$. All other singlet gravitons either vanish or are proportional to the two identified in \eqref{Eq:t18SU3} and \eqref{Eq:t24SU3}.

After obtaining a list of gravitons at a given order in $t$ as a sum of monomials (words) of the form
\begin{equation}
    U_2^aU_3^bV_2^cV_3^d W_{2b}^eW_{2f}^pW_{3b}^qW_{3f}^r x^{i}z^j\alpha^k \beta^l,
\end{equation}
we then select the potential global $SU(3)$ singlet words as the ones with
\begin{equation}
    k-l= 0 \mod3.
\end{equation}
The $\text{mod } 3$ enforces the fact that bosonic indices can only be contracted with the Levi-Civita tensor, and therefore we need them to occur in multiples of three to be able to form a singlet. We subtract off the number of lower indices $l$, since upper and lower indices can be contracted using the Kronecker delta. This leads us to a list of graviton words that have the correct index structure to construct $SU(3)_R$ singlets.

As explained in Sec.~\ref{Sec:SU3gravproof}, singlets involving an $\epsilon$ contraction of three bosonic eigenvalues vanish. We can remove these singlets from our list by examining the partitions of the exponent of $x$ in the graviton word into $\frac{k-l}{3}$ parts. Each partition corresponds to a specific index contraction structure in the graviton. If any of these parts contain three $x$'s, then that partition (contraction structure) vanishes, and any word that has only such partitions is therefore zero. We need to separate out the $w_2$ and $w_3$ generators into their bosonic and fermionic parts to ensure that this bosonic contraction constraint is imposed correctly, since only the $\epsilon$ contractions of three bosonic eigenvalues vanish.

Using the fact that traces of even powers of the matrix $v_2$ vanish, and \begin{equation}
    \tr(v_2^{2n+1})=0, \, n\geq2,
\end{equation} 
we find, from the above procedure, that all graviton singlet words beyond order $t^{72}$ vanish.

We can further reduce the set of potential singlet graviton words by accounting for the identity \eqref{Eq:w2w2epsilon}
\begin{equation}
    \epsilon_{ijk}w_2^j w_2^k=0,
\end{equation}
in a similar way. We examine the partitions of the power of $W_2$ in a word into $\frac{k-l}{3}$ parts. Any partition involving two or more $W_2$'s vanishes and therefore any monomial that has only such partitions can be excluded from our list. After imposing this constraint, we find that there are no singlet graviton words beyond $t^{54}$. Secondly, we find that the first set of non-vanishing monomials arises at $t^{18}$.

Having identified the list of potential graviton singlet words, we then proceed to explicitly construct the singlet gravitons, with all possible contraction structures. This is done using the graviton letters and treating all matrices inside the trace to be simultaneously diagonal, as
\begin{equation}
M=
    \begin{pmatrix}
        m_1&0&0\\
        0&m_2&0\\
        0&0&-m_1-m_2
    \end{pmatrix}\,,
\end{equation}
where $M$ is any of $\bar{\phi}^i, \psi_j$ or $f$. Fermion statistics for the Grassmann variables are imposed by ordering the fermions in a fixed manner in each invariant and then adding all possible permutations of the fermions, with minus signs for odd permutations. 

Using the above procedure, we identify the independent singlet gravitons at $t^{18}$ and $t^{24}$ as \eqref{Eq:t18SU3} and \eqref{Eq:t24SU3}, respectively. All other non-vanishing gravitons at order $t^{18}$ and $t^{24}$ are proportional to those identified in \eqref{Eq:t18SU3}  and \eqref{Eq:t24SU3}. The gravitons beyond order $t^{24}$ are considered in the following section.

\subsection{Vanishing singlet gravitons beyond \texorpdfstring{$t^{24}$}{t24}}
In this section, we list the identities that can be used to show that all singlet gravitons beyond order $t^{24}$ vanish. Using the procedure explained above, we can enumerate all potential singlet graviton words at any given order in $t$, and then explicitly evaluate the singlets using diagonal matrices. This leads to the identities listed below.

We first note the following identities, which are not global symmetry singlets:
\begin{equation}
   w_3^{il}w_3^{jm}w_3^{kn}\epsilon_{ijk}=0, \quad w_3^{il}w_3^{jm}w_3^{kn}\epsilon_{ija}\epsilon_{lkb}\epsilon_{mnc}=0.
\end{equation}
The above identities imply that any singlets with three or more $w_3$'s, which have the above contraction structures, vanish.

Next, we list the non-trivial singlet identities that arise at a given order in $t$. The left hand side of each identity is a potential singlet graviton that we obtain from the procedure explained above. We then explicitly check that each such singlet graviton vanishes, leading to the identities below.

There is a single identity at $t^{54}$, which is
\begin{equation}
    \epsilon_{ikm}\epsilon_{jln}(v_2)^i_a(v_2)^j_b w_2^k w_2^l w_2^aw_2^b w_3^{mn}=0.
\end{equation}

Then the identities at order $t^{48}$ are:
\begin{align}
    \begin{aligned}
        &\epsilon_{ikn}\epsilon_{jlb}\epsilon_{mac} u_2^{ij} w_2^k w_2^l w_2^m w_3^{na}w_3^{bc}=0, \quad \epsilon_{ikm}\epsilon_{jln}u_2^{ij}(v_2)^k_c(v_2)^l_d w_2^m w_2^nw_2^c w_2^d=0,\\
        &\epsilon_{ikm}\epsilon_{jln}(v_3)^{jk}_c(v_3)^{il}_dw_2^m w_2^nw_2^c w_2^d=0, \quad \epsilon_{ikm}\epsilon_{jln}(v_2)^{k}_c(v_3)^{il}_dw_3^{jm} w_2^nw_2^c w_2^d=0,\\
        &\epsilon_{ikm}\epsilon_{jln}(v_2)^{k}_c(v_2)^{l}_dw_3^{jm} w_3^{in}w_2^c w_2^d=0, \quad \epsilon_{imn}w_2^iw_2^cw_2^d (v_2)^k_c(v_2)^l_d(v_2)^m_k(v_2)^n_l=0,\\
        &\epsilon_{imn}w_2^iw_2^cw_2^d (v_2)^n_c(v_2)^l_d(v_2)^m_k(v_2)^k_l=0.
    \end{aligned}
\end{align}
Some of the identities listed above (also below for lower orders) are equivalent due to the following trace relations:
\begin{equation}
    \epsilon_{ijk}(4 u_2^{jb} (v_3)_b^{kl}+3 u_3^{jbl} (v_2)_b^k)=0, \quad \epsilon_{ab(j} \left((v_2)^a_{k)}w_3^{bi}+(v_3)^{ai}_{k)}w_2^{b}\right)=0,
\end{equation}
which are listed in \cite{Choi:2023vdm}, and were found using Gr{\"o}bner basis calculations. Note, however, that we do not need to invoke these relations. Instead we can explicitly evaluate all the graviton singlets listed above and find that they vanish.

Next, the identities at order $t^{42}$ are:
\begin{align}
    \begin{aligned}
         &\epsilon_{ikn}\epsilon_{jld}\epsilon_{mce}u_2^{ij}w_2^m w_3^{kc}w_3^{ln}w_3^{de} =0, \quad 
            \epsilon_{ikn}\epsilon_{jld}\epsilon_{mce}u_3^{ijc}w_2^m w_2^{k}w_3^{ln}w_3^{de} =0,\\
            &\epsilon_{ikn}\epsilon_{jld}\epsilon_{mce}u_2^{ij} u_2^{nc}w_2^m w_2^{k}w_2^{l}w_3^{de} =0,\quad
            \epsilon_{iln} \epsilon_{jmc} (v_3)^{ij}_k w_2^k w_2^l w_2^m w_3^{nc}=0,\\
            & \epsilon_{ikm}\epsilon_{lnc} (v_2)^i_j w_2^{k} w_2^l w_3^{mn} w_3^{cj}=0 ,\quad            \epsilon_{ilc}\epsilon_{jmd}u_3^{ijk}(v_2)^{l}_k (v_2)^m_n w_2^c w_2^d w_2^n=0,\\
            &\epsilon_{ilc}\epsilon_{jmd}u_2^{ik}(v_3)^{jl}_k (v_2)^m_n w_2^c w_2^d w_2^n=0, \quad 
            \epsilon_{ilc}\epsilon_{jmd}u_2^{jk}(v_2)^{l}_k (v_3)^{im}_n w_2^c w_2^d w_2^n=0,\\
            &\epsilon_{ilc}\epsilon_{jkd}u_2^{ik}(v_3)^{jl}_m (v_2)^m_n w_2^c w_2^d w_2^n=0, \quad 
            \epsilon_{ilc}\epsilon_{jmd}u_2^{ik}(v_2)^{j}_k (v_2)^m_n w_2^c w_3^{ld} w_2^n=0,\\
            &\epsilon_{ilc}\epsilon_{jmd}u_2^{jk}(v_2)^{l}_k (v_2)^{i}_n w_3^{mc} w_2^d w_2^n=0, \quad
            \epsilon_{ilc}\epsilon_{jkd}u_2^{ik}(v_2)^{j}_m (v_2)^m_n w_2^c w_3^{dl} w_2^n=0,\\
            &\epsilon_{mil}\epsilon_{nkj} (v_3)^{id}_c(v_3)^{kc}_d w_2^m w_2^n w_3^{jl}=0, \quad
            \epsilon_{mik}\epsilon_{njl} (v_3)^{ij}_c(v_3)^{kl}_d w_2^m w_2^n w_3^{cd}=0,\\
            &\epsilon_{mik}\epsilon_{njl} (v_3)^{ij}_c(v_3)^{kl}_d w_2^c w_2^n w_3^{md}=0, \quad 
            \epsilon_{mik}\epsilon_{njl} (v_3)^{ij}_c(v_3)^{kc}_d w_2^m w_2^n w_3^{ld}=0,\\
            &\epsilon_{mik}\epsilon_{njl} (v_3)^{id}_c(v_3)^{kl}_d w_2^c w_2^n w_3^{mj}=0, \quad
            \epsilon_{mil}\epsilon_{nkj} (v_2)^{i}_c(v_3)^{kc}_d w_3^{md} w_2^n w_3^{jl}=0, \\
            &\epsilon_{mik}\epsilon_{njl} (v_2)^{i}_c(v_3)^{kl}_d w_2^{mj} w_2^n w_3^{cd}=0, \quad
            \epsilon_{mik}\epsilon_{njl} (v_2)^{j}_c(v_3)^{kl}_d w_2^c w_3^{ni} w_3^{md}=0, \\
            &\epsilon_{mik}\epsilon_{njl} (v_3)^{ij}_c(v_2)^{k}_d w_2^c w_3^{nl} w_3^{md}=0,\quad
             \epsilon_{mik}\epsilon_{njl} (v_2)^{i}_c(v_3)^{kc}_d w_3^{mj} w_2^n w_3^{ld}=0,\\
            &\epsilon_{mik}\epsilon_{njl} (v_3)^{ij}_c(v_2)^{k}_d w_3^{mc} w_2^n w_3^{ld}=0, \quad
            \epsilon_{mik}\epsilon_{njl} (v_3)^{id}_c(v_2)^{l}_d w_2^c w_3^{nk} w_3^{mj}=0,\\
            &\epsilon_{mil}\epsilon_{nkj} (v_2)^{i}_c(v_2)^{k}_d w_3^{md} w_3^{nc} w_3^{jl}=0, \quad
            \epsilon_{mik}\epsilon_{njl} (v_2)^{i}_c(v_2)^{l}_d w_2^{mj} w_3^{kn} w_3^{cd}=0,\\
            &\epsilon_{mik}\epsilon_{njl} (v_2)^{j}_c(v_2)^{k}_d w_3^{mc} w_3^{in} w_3^{ld}=0, \quad
            \epsilon_{ijk}(v_2)^i_c(v_2)^m_d(v_2)^j_m w_2^c w_2^d w_2^k=0,\\
            &\epsilon_{ijk} (v_2)^i_l (v_2)^l_d (v_2)^d_m (v_3)^{jm}_c w_2^c w_2^k=0, \quad
            \epsilon_{ijk} (v_2)^i_c (v_2)^l_d (v_2)^d_m (v_3)^{jm}_l w_2^c w_2^k=0,\\
            &\epsilon_{ijk} (v_2)^l_c (v_2)^i_d (v_2)^d_m (v_3)^{jm}_l w_2^c w_2^k=0, \quad
            \epsilon_{ijk} (v_2)^d_c (v_2)^l_d (v_2)^i_m (v_3)^{jm}_l w_2^c w_2^k=0,\\
            &\epsilon_{ijk} (v_2)^i_l (v_2)^j_d (v_2)^d_m (v_3)^{lm}_c w_2^c w_2^k=0, \quad
            \epsilon_{ijk} (v_2)^i_c (v_2)^l_d (v_2)^j_m (v_3)^{dm}_l w_2^c w_2^k=0, \\
            &\epsilon_{ijk} (v_2)^l_c (v_2)^i_d (v_2)^j_m (v_3)^{dm}_l w_2^c w_2^k=0,\quad
            \epsilon_{ijk} (v_2)^j_c (v_2)^l_d (v_2)^i_m (v_3)^{dm}_l w_2^c w_2^k=0,\\
            &\epsilon_{ijk} (v_2)^i_l (v_2)^l_d (v_2)^d_m (v_2)^{j}_c w_2^c w_3^{mk}=0, \quad
            \epsilon_{ijk} (v_2)^l_c (v_2)^i_d (v_2)^d_m (v_2)^{j}_l w_2^c w_3^{mk}=0,\\
            &\epsilon_{ijk} (v_2)^i_l (v_2)^l_d (v_2)^d_m (v_2)^{j}_c w_2^k w_3^{mc}=0,\quad
            \epsilon_{ijk} (v_2)^i_l (v_2)^l_d (v_2)^d_m (v_2)^{m}_c w_2^k w_3^{jc}=0.
    \end{aligned}
\end{align}

Then the identities at order $t^{36}$ are:
\begin{align}
    \begin{aligned}
        &\epsilon_{ikm}\epsilon_{jln} w_2^i w_2^j w_3^{kl}w_3^{mn}=0, \quad \epsilon_{ilc}\epsilon_{jmg}\epsilon_{kdh} w_2^i w_2^j w_2^k u_2^{lm} u_2^{cd} u_2^{gh}=0,\\
        &\epsilon_{ihc}\epsilon_{jdl}\epsilon_{kgm}u_2^{lm} u_3^{cdg} w_2^i w_2^j w_3^{kh} = 0, \quad  \epsilon_{ihm}\epsilon_{jcd}\epsilon_{klg}u_2^{mc} u_2^{dg} w_2^i w_3^{jl} w_3^{kh} = 0,\\
        & \epsilon_{ikm}\epsilon_{jnc} u_2^{mn}(v_2)^c_l w_2^l w_2^i w_3^{jk}=0, \quad \epsilon_{ikm}\epsilon_{jnc} u_2^{ml}(v_2)^c_l w_2^n w_2^i w_3^{jk}=0,\\
        &\epsilon_{ikm}\epsilon_{jnc} u_2^{mn}(v_2)^c_l w_2^l w_2^i w_3^{jk}=0, \quad \epsilon_{ikm}\epsilon_{jnc} u_2^{ml}(v_2)^c_l w_2^n w_2^i w_3^{jk}=0,\\
        &\epsilon_{kni}\epsilon_{mcj}(v_3)_l^{ij} w_2^l w_3^{km} w_3^{nc}=0, \quad \epsilon_{kni}\epsilon_{mcj}(v_3)_l^{ij} w_2^l w_3^{lm} w_3^{nc}=0,\\
        &\epsilon_{ijk}(v_2)^j_n(v_2)^k_m w_2^n w_2^m w_2^i=0, \quad \epsilon_{ikm} \epsilon_{jln} u_2^{md}u_2^{nc}(v_2)^k_d (v_2)^l_c w_2^i w_2^j=0,\\
        &\epsilon_{ikm} \epsilon_{jln} u_2^{md}u_2^{nk} (v_2)^c_d (v_2)^l_c w_2^i w_2^j=0,\quad \epsilon_{iln} \epsilon_{jmk} u_3^{kdc} (v_2)^n_c (v_3)^{lm}_d w_2^i w_2^j=0,\\
        &\epsilon_{iln} \epsilon_{jmk} u_3^{kdl} (v_2)^n_c (v_3)^{cm}_d w_2^i w_2^j=0, \quad \epsilon_{iln} \epsilon_{jmk} u_3^{knc} (v_2)^d_c (v_3)^{lm}_d w_2^i w_2^j=0,\\
        &\epsilon_{iln} \epsilon_{jmk} u_2^{kl} (v_3)^{dn}_c (v_3)^{cm}_d w_2^i w_2^j=0, \quad \epsilon_{iln} \epsilon_{jmk} u_2^{dl} (v_3)^{kn}_c (v_3)^{cm}_d w_2^i w_2^j=0,\\
        & \epsilon_{iln} \epsilon_{jmk} u_2^{cd} (v_3)^{kn}_c (v_3)^{lm}_d w_2^i w_2^j=0,\quad  \epsilon_{iln} \epsilon_{jmk} u_3^{kdc} (v_2)^n_c (v_2)^{l}_d w_3^{im} w_2^j=0,\\
        &\epsilon_{iln} \epsilon_{jmk} u_3^{kdl} (v_2)^n_c (v_2)^{m}_d w_2^i w_3^{cj}=0, \quad  \epsilon_{iln} \epsilon_{jmk} u_3^{kdc} (v_2)^n_c (v_2)^{m}_d w_2^i w_3^{lj}=0,\\
        &\epsilon_{iln} \epsilon_{jmk} u_3^{knc} (v_2)^d_c (v_2)^{l}_d w_3^{im} w_2^j=0, \quad \epsilon_{iln} \epsilon_{jmk} u_3^{knc} (v_2)^d_c (v_2)^{m}_d w_2^i w_3^{lj}=0,\\
        &\epsilon_{iln} \epsilon_{jmk} u_3^{kdc} (v_2)^n_c (v_2)^{l}_d w_3^{im} w_2^j=0,\quad \epsilon_{iln} \epsilon_{jmk} u_3^{kdl} (v_2)^n_c (v_2)^{m}_d w_2^i w_3^{cj}=0,\\
        & \epsilon_{iln} \epsilon_{jmk} u_2^{kd} (v_2)^n_c (v_3)^{ml}_d w_2^i w_3^{cj}=0,\quad \epsilon_{iln} \epsilon_{jmk} u_2^{kd} (v_2)^n_c (v_3)^{mc}_d w_2^i w_3^{lj}=0,\\
        &\epsilon_{iln} \epsilon_{jmk} u_2^{kn} (v_2)^d_c (v_3)^{cl}_d w_3^{im} w_2^j=0,\quad \epsilon_{iln} \epsilon_{jmk} u_2^{kn} (v_2)^d_c (v_3)^{mc}_d w_2^i w_3^{lj}=0,\\
        &\epsilon_{iln} \epsilon_{jmk} u_2^{nc} (v_2)^d_c (v_3)^{kl}_d w_3^{im} w_2^j=0, \quad \epsilon_{iln} \epsilon_{jmk} u_2^{dl} (v_3)^{kn}_c (v_2)^{m}_d w_2^i w_3^{cj}=0,\\
        &\epsilon_{iln} \epsilon_{jmk} u_2^{cd}(v_2)^m_c(v_2)^n_d w_3^{ik} w_3^{jl}=0, \quad \epsilon_{iln} \epsilon_{jmk} u_2^{nd}(v_2)^m_c(v_2)^c_d w_3^{ik} w_3^{jl}=0,\\
        & \epsilon_{ilm} \epsilon_{jnk} (v_3)^{md}_c (v_3)^{nc}_d w_3^{ik} w_3^{jl}=0,\quad \epsilon_{ijk} (v_2)^l_c (v_2)^j_d (v_3)^{kd}_l w_2^c w_2^i=0,\\
        &\epsilon_{ijk} (v_2)^j_c (v_2)^l_d (v_3)^{kd}_l w_2^c w_2^i=0, \quad  \epsilon_{ijk} (v_2)^j_c (v_2)^k_d (v_3)^{cd}_l w_2^l w_2^i=0, \\
        &\epsilon_{ijk} (v_2)^d_c (v_2)^k_d (v_3)^{cj}_l w_2^l w_2^i=0, \quad \epsilon_{ijk} (v_2)^l_c (v_2)^j_d (v_2)^{k}_l w_2^c w_3^{id}=0,\\
        &\epsilon_{ijk} (v_2)^j_c (v_2)^l_d (v_2)^{k}_l w_2^c w_3^{id}=0, \quad \epsilon_{ijk} (v_2)^k_c (v_2)^l_d (v_2)^{c}_l w_2^j w_3^{id}=0,\\
        &\epsilon_{ijk} (v_2)^k_c (v_2)^i_d (v_2)^{c}_l w_2^j w_3^{ld}=0, \quad \epsilon_{ijk} (v_2)^k_c (v_2)^l_d (v_2)^{i}_l w_2^j w_3^{cd}=0,\\
        &\epsilon_{ijk}u_2^{cd} (v_2)^l_c(v_2)^n_d(v_2)^j_l (v_2)^k_n w_2^i=0, \quad \epsilon_{ijk}u_2^{cj} (v_2)^l_c(v_2)^n_d(v_2)^d_l (v_2)^k_n w_2^i=0,\\
        &\epsilon_{ijk}u_2^{cd} (v_2)^l_c(v_2)^j_d(v_2)^n_l (v_2)^k_n w_2^i=0, \quad \epsilon_{ijk}(v_2)^l_c (v_2)^n_d (v_3)^{jc}_l (v_3)^{kd}_n w_2^i=0,\\
        &\epsilon_{ijk}(v_2)^n_c (v_2)^l_d (v_3)^{jc}_l (v_3)^{kd}_n w_2^i=0, \quad \epsilon_{ijk}(v_2)^k_c (v_2)^n_d (v_3)^{jc}_l (v_3)^{ld}_n w_2^i=0,\\
        &\epsilon_{ijk}(v_2)^k_d (v_2)^n_c (v_3)^{jc}_l (v_3)^{ld}_n w_2^i=0, \quad \epsilon_{ijk}(v_2)^k_c (v_2)^j_d (v_3)^{nc}_l (v_3)^{ld}_n w_2^i=0,\\
        &\epsilon_{ijk}(v_2)^l_c (v_2)^n_d (v_2)^{j}_l (v_3)^{kd}_n w_3^{ic}=0,\quad \epsilon_{ijk}(v_2)^n_c (v_2)^l_d (v_2)^{j}_l (v_3)^{kd}_n w_3^{ic}=0,\\
        &\epsilon_{ijk}(v_2)^k_c (v_2)^n_d (v_2)^{j}_l (v_3)^{ld}_n w_3^{ic}=0, \quad  \epsilon_{ijk}(v_2)^k_d (v_2)^n_c (v_2)^{j}_l (v_3)^{ld}_n w_3^{ic}=0,\\
        &  \epsilon_{ijk}(v_2)^l_d (v_2)^n_l (v_2)^{j}_c (v_3)^{kd}_n w_3^{ic}=0.
    \end{aligned}
\end{align}

Lastly, the identities at order $t^{30}$ are:
\begin{align}
    \begin{aligned}
        &\epsilon_{imk} \epsilon_{jln} u_2^{mn} w_2^i w_2^j w_3^{kl}=0, \quad \epsilon_{imk} \epsilon_{jln} u_2^{mn} u_2^{kc} (v_2)^l_c w_2^i w_2^j=0,\\
        &\epsilon_{imk} \epsilon_{jln} u_3^{klc} (v_3)^{mn}_c w_2^i w_2^j=0,\quad \epsilon_{imk} \epsilon_{jln} u_3^{klc} (v_2)^{m}_c w_3^{in} w_2^j=0,\\
        &\epsilon_{imk} \epsilon_{jln} u_2^{kc} (v_3)^{ml}_c w_3^{in} w_2^j=0, \quad  \epsilon_{imk} \epsilon_{jln} u_2^{kn} (v_3)^{ml}_c w_3^{ic} w_2^j=0,\\
        &\epsilon_{imk} \epsilon_{jln} u_2^{kc} (v_2)^{l}_c w_3^{in} w_3^{mj}=0, \quad  \epsilon_{imk} \epsilon_{jln} u_2^{kn} (v_2)^{l}_c w_3^{ic} w_3^{mj}=0,\\
        &\epsilon_{ijk} (v_2)^k_c (v_3)^{jc}_l w_2^l w_2^i=0, \quad \epsilon_{ijk} (v_2)^k_c (v_2)^{j}_l w_2^l w_3^{ic}=0,\\
        &\epsilon_{ijk} u_2^{cn} (v_2)^m_c (v_2)^j_n (v_2)^k_m w_2^i=0, \quad \epsilon_{ijk} u_2^{ck} (v_2)^m_c (v_2)^j_n (v_2)^n_m w_2^i=0,\\
        &\epsilon_{ijk}(v_2)^j_c (v_3)^{cm}_n (v_3)^{kn}_m w_2^i=0,\quad \epsilon_{ijk}(v_2)^m_c (v_3)^{cj}_n (v_3)^{kn}_m w_2^i=0,\\
        &\epsilon_{ijk}(v_2)^j_c (v_3)^{cm}_n (v_2)^{k}_m w_3^{in}=0,\quad \epsilon_{ijk}(v_2)^m_c (v_2)^{j}_n (v_3)^{kn}_m w_3^{ic}=0,\\
        &\epsilon_{ijk}(v_2)^m_c (v_3)^{cj}_n (v_2)^{k}_m w_3^{in}=0,\quad \epsilon_{ijk}(v_2)^m_n (v_2)^{j}_c (v_3)^{kn}_m w_3^{ic}=0.
    \end{aligned}
\end{align}

\section{BMN index for singlets in the \texorpdfstring{$SU(4)$}{SU(4)} theory}\label{App:SU4P300}
The $SU(3)_R$ singlet projection of the BMN index for the $SU(4)$ theory is:
\begin{align}
  \mathcal{I}^{SU(4)}_{{\rm BMN} \, [0,0]}(t) =& \frac{\left(1-t^{12}\right)^2 }{\left(1-t^{18}\right) \left(1-t^{24}\right) \left(1-t^{36}\right) \left(1-t^{48}\right) \left(1-t^{60}\right)} P_{300}(t),\\
  P_{300}(t) &= 1 + 3 t^{12} - 2 t^{18} + 4 t^{24} - 7 t^{30} + 3 t^{36} - 10 t^{42} + 4 t^{48} - 8 t^{54} - 4 t^{60} \nonumber\\&\quad - 6 t^{66} + t^{72} + 15 t^{78} - 14 t^{84} + 21 t^{90} - 14 t^{96} + 42 t^{102} + 2 t^{108} + 2 t^{114} \nonumber\\&\quad + 17 t^{120} - 20 t^{126} + 48 t^{132} - 35 t^{138} + 20 t^{144} - 14 t^{150} - 9 t^{156} + 32 t^{162} \nonumber\\&\quad - 58 t^{168} + 47 t^{174} - 77 t^{180} + 46 t^{186} - 59 t^{192} + 18 t^{198} - 22 t^{204} - 14 t^{210} \nonumber\\&\quad + t^{216} - 24 t^{222} + 20 t^{228} - 19 t^{234} + 10 t^{240} - 13 t^{246} + 3 t^{252} + t^{258} + 3 t^{264} \nonumber\\&\quad + 6 t^{270} + 2 t^{282} - 2 t^{288} -t^{300}.
\end{align}

\end{appendix}

\bibliographystyle{JHEP}
\bibliography{sample}
\end{document}